\documentclass[twocolumn]{aastex631}

\shorttitle{Blackbody temperature of 200+ stellar flares}

\shortauthors{Rabello Soares et al.}

\begin{document}

\title{Blackbody temperature of 200+ stellar flares observed with the CoRoT satellite}

\correspondingauthor{M. Cristina Rabello Soares}
\email{cristina.rabello.soares@stanford.edu}

\author[0000-0003-0172-3713]{M. Cristina Rabello Soares}
\altaffiliation{Physics Department \\
Universidade Federal de Minas Gerais\\
Belo Horizonte 31270-901, Brazil}
\affiliation{W. W. Hansen Experimental Physics Laboratory\\
Stanford University \\
Stanford, CA, 94305, USA}

\author[0000-0001-6879-5872]{Marcia C. de Freitas}
\affiliation{Physics Department \\
Universidade Federal de Minas Gerais\\
Belo Horizonte 31270-901, Brazil}

\author[0000-0002-7552-3063]{Bernardo P. L. Ferreira}
\affiliation{Physics Department \\
Universidade Federal de Minas Gerais\\
Belo Horizonte 31270-901, Brazil}




\begin{abstract}

We estimated blackbody temperature for 209 flares observed at 69 F-K stars,
significantly increasing the number of flare temperature determinations.
We used the Blue and Red channels obtained by the 27 cm telescope of the CoRoT satellite at high cadence and long duration.
The wavelength limits of the channels were estimated using spectra from the Pickles library for the spectral type and luminosity class of each star, provided by the Exodat Database.
The temperatures were obtained
from the flare energy Blue-to-Red ratio,
using the flare equivalent duration and stellar flux in both channels.
The expected value of the analyzed flares is equal to 6,400 K with a standard deviation of 2,800 K, where the mean stellar spectral type, weighted by the number of flares in each spectral subclass, is equal to G6.
%
%
Contrary to our results, a stellar white-light flare is often assumed to emit as a blackbody with a temperature of 9,000 K or 10,000 K. Our estimates agree, however, with values obtained for solar flares.
The GAIA G-band transmissivity is comparable to that of the CoRoT White channel, which allows us to calibrate the flares to the Gaia photometric system.
The energy in the G band of the analyzed flares varies between $10^{32}$ and $10^{37}$ erg
and 
the flare area ranges from 30\,$\mu$sh to 3\,sh (solar hemisphere).
The energy release per area in a flare is proportional to $T_{\rm flare}^{2.6}$, at least up to 10,000 K.

\end{abstract}





\section{Introduction}

Although solar flares are more noticeable 
in UV and X-ray observations, they emit in a broad range of electromagnetic radiation.
In fact, most of the radiated flare energy emerges at visible and UV wavelengths 
\citep{woods2006,kretzschmar2011}.
Flares observed in the visible spectrum are called white-light flares (WLFs).
Solar WLFs are generally rare events to observe, mainly due to the low contrast with the solar disk \citep{lin1976,jess2008}.
%

High-cadence, high-precision continuous photometry observations from space (e.g., CoRoT, Kepler, Tess)
have made possible extensive observations of stellar flares in the visible
on various types of stars.  
These stellar flares can be up to $\sim$10,000 
\citep[or even a million; see][]{schaefer2000}
times larger than the largest solar flares.
Despite this, several statistical studies show that stellar flares have solar-like properties, indicating that they may also be the result of magnetic reconnection events
\citep[][and references within]{namekata2017}.
Fast-rotating stars have higher magnetic activity,
as is to be expected from the dynamo theory of magnetic field generation.
In fact, it is observed that they have 
a higher occurrence rate of flares than slow rotators \citep{shibayama2013}
but not more energetic flares \citep{maehara2012}.
\citet{notsu2013} found that there is a correlation between spot coverage and flare energy.

Determining the flare temperature provides valuable information that can help to understand the physical processes responsible for the flare emission \citep[e.g.][]{kowalski2016}.
Unfortunately, these space missions only provide information in a single filter passband, not allowing to estimate the temperature of the flare.
There are some papers where it was estimated.
\citet{hawley1992} analyzed a large flare at the M-dwarf star AD Leo. They
obtained a blackbody temperature of 8500$-$9500 K
fitting
the average surface flux observed at six passbands
(Johnson U, B, V, and R and two in the near UV) during the flare.
\citet{hawley2003} analyzed four other flares on AD Leo
and obtained similar results ($\sim$9000 K).
\citet{kowalski2013} 
analyzed 20 flares in 5 M-dwarf stars
using 
high-cadence near-ultraviolet and visible spectra.
They
observed that all flares present a linear decrease in flux at
wavelengths from 4000 to 4800\,\AA\, corresponding to a
blackbody temperature
between 9000\,K and 14000\,K.
They reported variations in the Balmer jump during the flare
that
indicate
an additional component to the blackbody emission in the near UV range
\citep{kowalski2019}.
\citet{howard2020}
used simultaneous observations in the narrow g'-band obtained by 
a series of small telescopes 
\citep[Evryscope]{law2016} 
and the broadband filter of NASA's TESS satellite \citep{ricker2014}
to estimate the temperature and energy of 47 flares on 27 K5-M5 dwarfs.
They found that nearly half 
of the flares emit above 14,000\,K during their peak.
However, the temperature of the entire flare varies widely, from 2,600\,K to 50,000\,K.

Based on these stellar temperature estimations,
the WLF spectrum is
often described as a blackbody with a temperature between 9,000$-$10,000\,K
and
used to estimate the bolometric energy of the flare
\citep[e.g.][]{maehara2012, shibayama2013, davenport2020, gunther2020, maehara2021}.
A change in the assumed flare temperature may significantly affect the energy-frequency distribution estimate, which is used to model the atmospheric conditions of the planets orbiting the flaring star, that allows to assess the possibility of life and the probability of becoming a candidate for a biosignature survey \citep[e.g.][]{gunther2020}.

For solar WLFs, there is an earlier estimation of a temperature similar to that of the M-dwarf stellar flares, equal to $\sim$9000\,K by \citet{kretzschmar2011}.
Later,
\citet{kerr2014} estimated temperatures between 5,000 and 6,000\,K,
\citet{kleint2016} between $\sim$ 6000 and 6300\,K, and
\citet{namekata2017} between 5500 and 7000\,K. 


In this work, we used the CoRoT color channels to estimate the temperature of more than 200 stellar flares observed in more than 50 stars with different spectral types.
We describe in Section 2 the data used and the fitting and equivalent duration estimation of the flares.
Section 3 presents the estimation of the limits of the CoRoT color channels
that we used in the flare temperature calculation, described in Section 4.
As stellar rotation has an effect on stellar activity, 
the rotation estimate is presented in Section 5, to be compared with the flare temperatures.
Finally, in Section 6, we took advantage of the similarity of the CoRoT White channel to the Gaia G band to calibrate the analyzed flares into the Gaia photometric system.
In Section 7, we present our conclusions.

\section{Data and Methodology}\label{sec:fitting}

CoRoT (Convection, Rotation and planetary Transits) was a space-based mission that obtained precise photometric observations of hundreds of thousands of stars from 2007 to 2012 in two regions of the sky 
(one toward the galactic center and another toward the Galactic anticenter) $-$ \citet{baglin2016}. We studied stars observed through the so-called Faint Stars or Exoplanet channel \citep{ollivier2016}.
The stellar images were dispersed by a prism that was in the optical path before the CCDs, producing a low-dispersion spectrum of each star. For about 6000 stars, CoRoT provided not only a light curve integrating the entire photometric mask (called White light curve) but also three light curves dividing the mask into three regions corresponding to different spectral bands whose fluxes were recorded separately \citep{auvergne2009}.
The boundaries within the mask were chosen so that 
each flux 
was close to a given fraction of the total flux
and 
corresponds to an integer number of columns on the CCD as the dispersion was done in rows \citep[Fig.10 and 11 in][]{leger2009}.
These color light curves were named Red, Green, and Blue, 
although its spectral content varies with the shape of the photometric mask used, the star's apparent magnitude, spectral type, etc \citep{borde2010}.
The flux in the Green channel is low compared to the others \citep{borsa2013} 
and was not used here due its low signal-to-noise ratio.


\cite{drabent2012} published a list of 111 flaring stars 
after analyzing all CoRoT light curves observed until 2010 December (including observational sequence LRa04).
We used the light curves for these stars
provided by the CoRoT public archive
\footnote{http://idoc-corot.ias.u-psud.fr}.
%
Three data correction levels for each light curve
are provided.
We used the most basic (provided by the BAR extension in the FITS files), which does not include correction of jumps or replacement of invalid and missing data \citep{chaintreuil2016},
as a flare could be misinterpreted as a jump or invalid data.
The housekeeping temperature data show jumps 
\citep{ollivier2016}
that can affect the light curve and mimic a flare.
Thus, we disregard the corresponding time intervals in the analyzed light curves (provided by the keyword "statusfil" in the BARFILL extension\footnote{see Table II.4.15 in \citet{chaintreuil2016} for a description}).
Although CoRoT observations last 32\,s, in most stars, 16 observations are gathered at a cadence of 512\,s \citep{ollivier2016}.
Approximately one-third of the analyzed stars have a cadence of 512\,s and the remaining have a cadence of 32\,s.

\begin{figure}
	\includegraphics[width=\columnwidth]{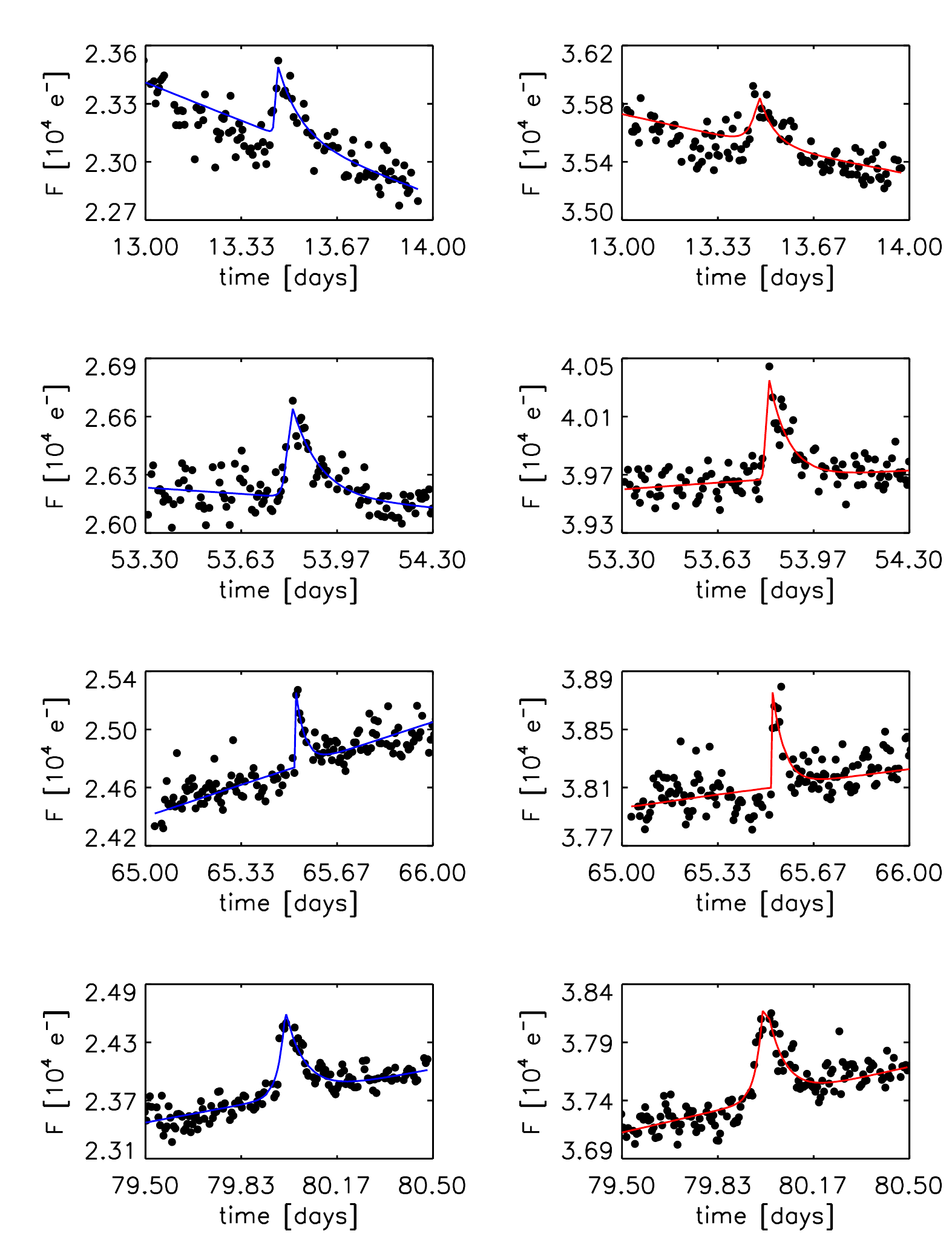}
    \caption{
    Flare light curves for CoRoT 106048015 obtained in the Blue (left) and Red (right) channels.
    The solid line shows the fitted model.
    }
    \label{fig:fitting}
\end{figure}

We visually inspected each light curve
and manually selected the approximate beginning and end of each flare,
using FBEYE software developed by \citet{davenport2014}.
We fitted each flare and its background using a two-exponential function
plus a straight line:
\begin{eqnarray}
  F_{\rm local}(t)
  = F_{\rm bk}(t) + F_{\rm flare}(t) = \nonumber 
  \\
  a_4 + a_5 t + a_0 \left\{
  \begin{array}{c}
  e^{a_2(t-a_1)} \,\,\, , \,\,\, t\,<\,a_1\\
  e^{-a_3(t-a_1)} \,\, , \,\,\, t\,\geq\,a_1\\
\end{array}
\right.
\label{eq:flarefitting}
\end{eqnarray}
observed in each color channel.
We applied a 
least-squares Levenberg–Marquardt minimization algorithm 
to obtain the best-fit flare parameters
and a
Maximum likelihood via Monte-Carlo Markov Chain algorithm 
\citep{foreman2013}
to estimate the uncertainties of the fitted parameters, using the LMFIT Python package.
Figure~\ref{fig:fitting} shows a few examples of the fitting.
The flare fitted parameters are the flare peak flux ($a_0$), the peak flare occurrence time ($a_1$), the exponential growth rate ($a_2$) and the decay rate ($a_3$).

\begin{figure}
	\includegraphics[width=\columnwidth]{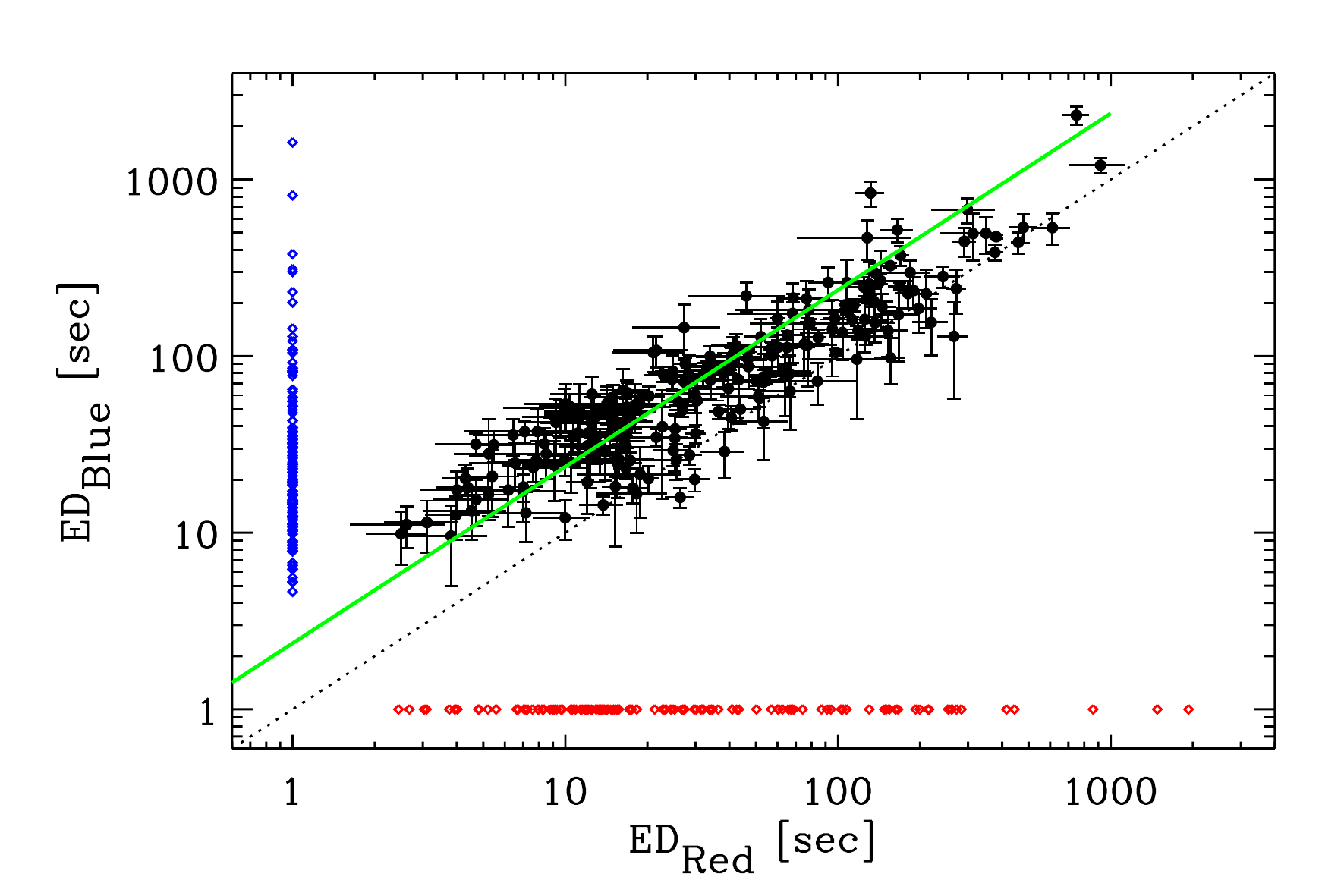}
    \caption{
Equivalent duration observed in the Blue channel versus in the Red channel.
There are 209 flares observed in both the Red and Blue channels, shown as black circles.
A similar number of flares is only seen in the Blue or Red channel,
which are shown as small blue and red diamonds, respectively, with
the equivalent duration on the other channel set to one.
The solid green line is given by 
${\rm ED}_{\rm Blue} = 2.36 \, {\rm ED}_{\rm Red}$.
}
    \label{fig:equiv_duration}
\end{figure}

The equivalent duration, ED, of a flare \citep{gershberg1972}
is given by:
\begin{equation}
{\rm ED} = \int_0^\infty \frac{F_{\rm flare}(t)}{F_{{\rm qu}}} \, dt =
\frac{a_0}{F_{\rm qu}} \left(
\frac{1}{a_2} + \frac{1}{a_3}
\right)
\label{eq:ed}
\end{equation}
where $F_{\rm flare}(t)$ is the flux due only to the flare (after removing the background)
and $F_{\rm qu}$ is the flux in the quiescent state.
As the background varies throughout the flare, we used the background value at the flare peak as the quiescent flux:
$F_{\rm qu} = a_4 + a_5 \, a_1$.
Figure~\ref{fig:equiv_duration} shows the equivalent duration obtained for each flare 
observed in the Blue and Red channels.
Only flares for which we obtained an ED 
greater than 1.5$\sigma$
(where $\sigma$ is the uncertainty in ED)
were used here and are shown in the Figure.
The ED varies from 2 to 2,000 seconds.
There are 209 flares observed in both the Blue and Red channels 
and 
a similar number of flares that are visible only in the Red or Blue channel.
The mean value of 
the Blue-to-Red ED ratio 
and its standard error is equal to
2.36$\pm$0.09
(shown as a green line in Figure~\ref{fig:equiv_duration}).

\begin{figure}
	\includegraphics[width=\columnwidth]{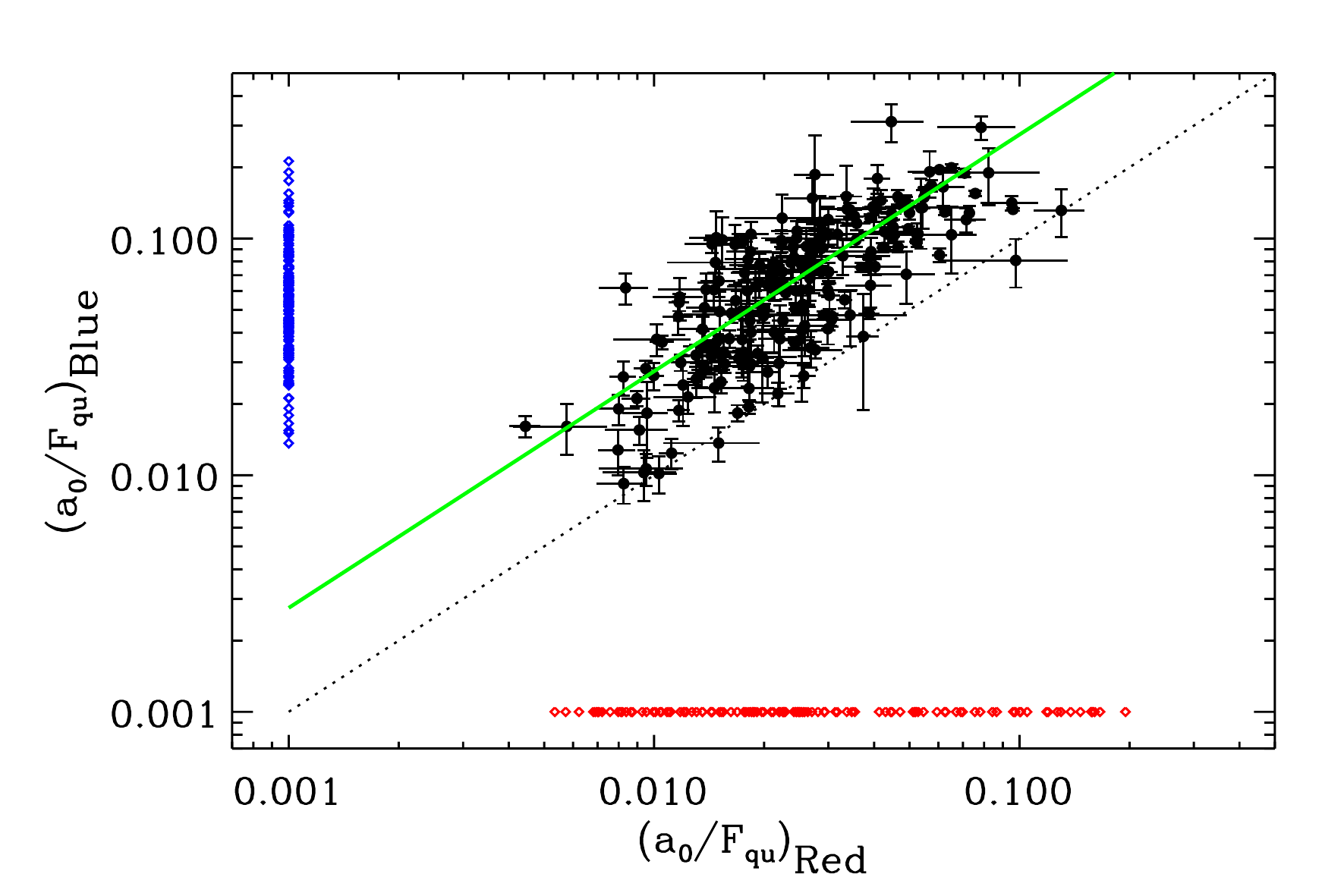}
    \caption{
    Blue-versus-Red flare peak amplitude normalized by its quiescent state
    shown as black circles.
    Flares that are only seen in the Blue or Red channel are shown as small blue and red diamonds, respectively, with
    the peak amplitude ratio on the other channel set to 0.001.
    The solid green line corresponds to: 
    (a$_0$/F$_{\rm qu}$)$_{\rm Blue}$ = 2.75
    (a$_0$/F$_{\rm qu}$)$_{\rm Red}$. 
    }
    \label{fig:a0param}
\end{figure}

The flare peak flux ($a_0$ in Eq.~\ref{eq:flarefitting})
divided by the flux in the quiet state
observed
in the Blue channel versus in the Red channel is shown in Figure~\ref{fig:a0param}. 
The mean and its standard error of the
Blue-to-Red flare peak amplitude ratio 
normalized by its respective quiet state 
is equal to 
2.75$\pm$0.09 
(shown as a green line in Figure~\ref{fig:a0param}).
The Blue-to-Red full width at half-maximum (FWHM) ratio is close to 1.
The flare FWHM ranges from 2 minutes to 2.5 hours.

In our analysis, besides ruling out several flares 
where the equivalent duration was not well determined (i.e., ${\rm ED} < 1.5\sigma$) 
due to its small duration or amplitude,
we also discarded those within a thermal jump or with missing data during the flare.
In the end, we accepted 209 flares observed both in the Blue and Red channels. 
These flares belong to 69 stars, 
reducing the number of stars in our initial sample.

\section{Blue and Red channel wavelength limits}
\label{sec:wavelength}

To obtain the chromatic light curves,
the observed low-resolution stellar spectrum
multiplied by the CoRoT response function
was divided into three parts:
the Blue band corresponds to the shorter wavelengths, 
the Red band to the longer ones, and 
the Green band to intermediate values.
Unfortunately, the boundaries that separate the channels vary for each observed star and
for each observation period (since the star may be on a different CCD) - \citet{borde2010}.
In this section, we estimate these limits for each light curve: 
the upper wavelength limit of the Blue channel, $\lambda_{\rm Blue}$, and
the lower wavelength limit of the Red channel, $\lambda_{\rm Red}$.

For each star,  the spectral type and luminosity class were retrieved from the Exodat Database\footnote{http://cesam.lam.fr/exodat} 
\citep{deleuil2009} - later shown in Table~\ref{tab:temperature}.
The stellar classification in the Exodat Database was estimated by fitting the apparent magnitude of CoRoT stars (obtained by ground-based multi-color broadband photometric observations) 
to the spectral flux from the 
Pickles Stellar Spectral Flux Library \citep{pickles1998}.
\cite{damiani2016A&A} reported that
the luminosity class is estimated with great significance and that the typical uncertainty on the spectral type 
is of half a spectral class.
We used the spectra from the Pickles 
Library for the spectral type and luminosity class of the star $s$, $f^s_P(\lambda)$, to estimate the wavelength limits:
\begin{eqnarray}
\frac{F^{\rm obs}_{{\rm Blue}}}{F^{\rm obs}_{{\rm White}}}(l) =
\frac{\int_0^{\lambda_{\rm Blue}(l,s)}{\Phi(\lambda)\,\, f_{P}^{s}(\lambda) \, \lambda \, d\lambda}}{\int_0^\infty{\Phi(\lambda)\,\, f_{P}^{s}(\lambda) \, \lambda \, d\lambda}} \nonumber\\
\frac{F^{\rm obs}_{{\rm Red}}}{F^{\rm obs}_{{\rm White}}}(l) = 
\frac{\int_{\lambda_{\rm Red}(l,s)}^\infty{\Phi(\lambda)\,\, f_{P}^{s}(\lambda) \, \lambda \, d\lambda}}{\int_0^\infty{\Phi(\lambda)\,\, f_{P}^{s}(\lambda) \, \lambda \, d\lambda}} \label{eq:fluxratio}
\end{eqnarray}
where 
the terms on the left-hand side
are
the observed average stellar flux ratios for each light curve $l$.
$\Phi$($\lambda$) is the CoRoT Response Function
for the Faint Stars channel
\citep{auvergne2009}.
Most stars $s$ have only one light curve $l$, except for eight of them, which have two light curves observed four years apart. 
We used the Pickles library
UVKLIB\footnote{https://archive.stsci.edu/hlsps/reference-atlases/cdbs/grid/pickles/dat$\_$uvk/}, which covers the spectral range 1150$-$25,000 \AA.
In the library, the fluxes are in units of
erg\,cm$^{-2}$~s$^{-1}$\AA$^{-1}$
and the original fluxes in the Pickles V band were normalized to a 0 magnitude in the Vega magnitude system\footnote{https://www.stsci.edu/hst/instrumentation/reference-data-for-calibration-and-tools/astronomical-catalogs/pickles-atlas}. 
As the functions on the right-hand side of equations~\ref{eq:fluxratio}
vary smoothly with $\lambda_{\rm Blue}(l,s)$ and $\lambda_{\rm Red}(l,s)$,
we can use the error propagation equation \citep{bevington2003}.
The uncertainties of the wavelength limits, 
$e_{\lambda_{\rm Blue}}(l,s)$ and
$e_{\lambda_{\rm Red}}(l,s)$,
were estimated
dividing
the standard error of the mean 
observed flux ratio by 
the derivative of the right-hand side expression
at the estimated wavelength limit.
The uncertainties 
are very small, they are less than 0.2\,\AA.

\begin{figure}
	\includegraphics[width=\columnwidth]{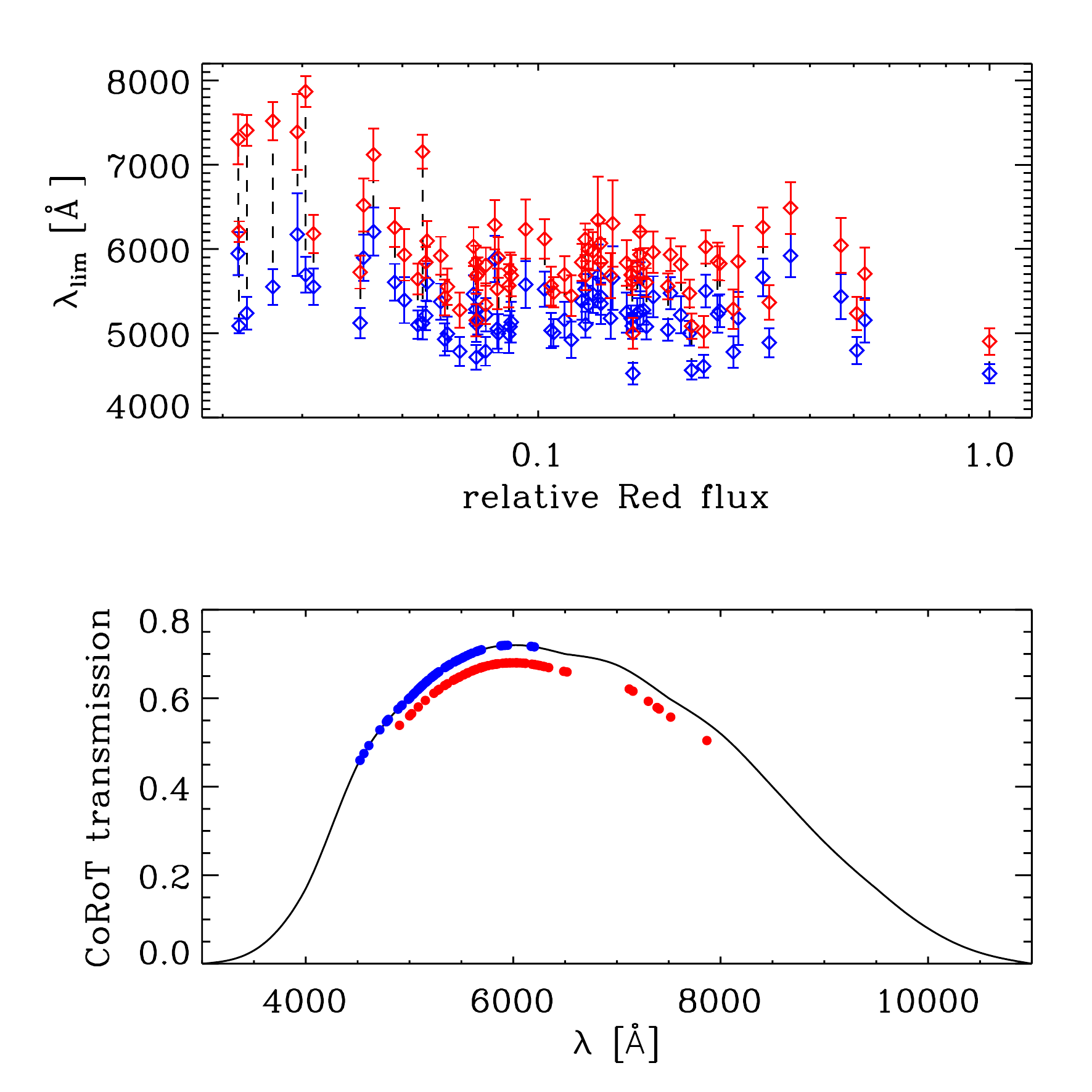}
    \caption{
    Top panel. 
    Weighted mean of the wavelength limits for each light curve:
    $\bar{\lambda}_{\rm Blue}(l)$ (blue symbols) and $\bar{\lambda}_{\rm Red}(l)$ (red symbols),
    as a function of the observed mean stellar flux in the Red channel,
    $F^{\rm obs}_{{\rm Red}}(l)$,
    divided by its maximum value among the analyzed light curves.
    The error bars correspond to the uncertainties in the weighted averages.
    The limits of a given light curve in the Blue and Red channels are connected by a vertical dashed black line for better visualization.
    We observe an increase in $\bar{\lambda}_{\rm Red}$ at small values of the stellar Red flux, indicating a narrowing of this channel. We do not observe any variation in the wavelength limits as a function of the Blue or White stellar flux.
    Bottom panel.
    The CoRoT Response Function for the Faint Stars channel \citep[Figure 14 in][]{auvergne2009} is shown in black. 
    The wavelength limits (from the top panel) are shown in the color of their respective band. The ones for the Red channel are displaced vertically by -0.04 for a better visualization. 
    }
    \label{fig:wavelimits}
\end{figure}

We also included 
the uncertainty in the stellar classification
in the wavelength limit estimation.
To be on the safe side,
we assumed twice the uncertainty in stellar classification given in \cite{damiani2016A&A}.
For each light curve $l$, we estimated the limit of each channel, using all spectra in the Pickles library within plus/minus a spectral class and within plus/minus two luminosity classes of the star $s$:
$(\lambda_{\rm Blue} \pm e_{\lambda_{\rm Blue}})(l,s_i)$
and
$(\lambda_{\rm Red} \pm e_{\lambda_{\rm Red}})(l,s_i)$, 
$i = 1, ..., N$.
For our stars, there are between 18 and 68 spectra within these uncertainties 
(18 $\leq N \leq$ 68)
and
only 7 of the 69 stars have less than 30 spectra.
Next we calculated the weighted average limit of the channels for each light curve $l$,
$\bar{\lambda}_{\rm Blue} 
(l)$
and
$\bar{\lambda}_{\rm Red} 
(l)$,
where the weight is given by the inverse square of the uncertainty.
We estimated the uncertainty in the weighted averages (\citealt{bevington2003} and in more detail in the supplement of \citealt{kirchner2020}):
$\sigma_{\lambda_{\rm Blue}}(l)$
and
$\sigma_{\lambda_{\rm Red}}(l)$.
They have a percent uncertainty ranging from
2\%$-$5\%.
The results are shown in Figure\,\ref{fig:wavelimits} (top panel).
Later in this paper, 
we compare our results using these limits
(Fig.\,\ref{fig:wavelimits})
with limits
calculated as described above, but assuming
a smaller uncertainty in stellar classification:
within plus/minus half a spectral class and within plus/minus one luminosity class. 
These weighted average limits agree with the previous ones within their uncertainty. 
As expected, their uncertainties are half of the former ones.

According to \cite{rouan1999}, 
the fraction in each channel should not be far from
20\% of the bluest photons for the Blue channel and 65\% of the reddest photons for the Red channel.
The averages over the analyzed stars of the Blue-to-White and Red-to-White observed flux ratio,
measured in electrons,
are equal to 0.16$\pm$0.01 and 0.72$\pm$0.01, respectively, where the uncertainties are given by the error of the mean.
However, 
if we calculate the stellar fluxes in units of energy 
using the right-hand side of Equations~\ref{eq:fluxratio}, but without multiplying by the wavelength, the average ratios are 0.22$\pm$0.01 and 0.63$\pm$0.01, respectively,
which agree with the fractions given by \cite{rouan1999}.

\section{Effective flare temperatures}\label{sec:temp}

We assumed that the spectrum of white-light flares 
can be described by a blackbody, $B_\lambda$, with 
temperature, $T_{\rm flare}$
\citep[][and others]{mochnacki1980, shibayama2013, gunther2020}.
Following \citet{shibayama2013}, 
we can estimate the flare luminosity ($L_{\rm flare}$) 
observed by CoRoT in the White channel, in photons,
as:
\begin{equation}
L_{\rm flare}(t) = A_{\rm flare}(t) \,\,\, \int_0^\infty \Phi(\lambda) B_\lambda(T_{\rm flare}) \lambda \, d\lambda
\label{eq:lumflare}
\end{equation}
where $A_{\rm flare}(t)$ is the area of the flare
as a function of time and 
$\Phi(\lambda)$ is CoRoT Response Function
for the Faint Stars channel (Fig.~\ref{fig:wavelimits}).

The equivalent duration (Equation~\ref{eq:ed})
can be written as:
\begin{equation}
{\rm ED} = 
\frac{1}{L_\star} \int_{\rm flare} L_{\rm flare}(t) \, dt.
\label{eq:ed_tflare}
\end{equation}
where $L_\star$ is the luminosity of the star observed with the CoRoT White channel.
Assuming that the flare area and temperature are constant during the flare
\citep[as in][]{shibayama2013,gunther2020}
and 
substituting Equation~\ref{eq:lumflare}
into Equation~\ref{eq:ed_tflare}, we have:
\begin{equation}
{\rm ED} = 
\frac{A_{{\rm flare}}}
{L_\star}
\,\, \Delta t
\,\,\, \int_0^\infty \Phi(\lambda) B_\lambda(T_{{\rm flare}}) \, \lambda \, d\lambda
\label{eq:ed_area}
\end{equation}
where $\Delta t$ 
is the flare duration.
The ratio between the equivalent duration observed in the Blue and Red channels is:
\begin{equation}
\frac{{\rm ED}_{{\rm Blue}}}{{\rm ED}_{{\rm Red}}} =
\frac{F_{\rm Red}}{F_{\rm Blue}}
\,
\frac{\Delta t_{\rm Blue}}{\Delta t_{\rm Red}}
\,\,
\frac{\int_0^{\lambda_{\rm Blue}}{\Phi(\lambda)\,\, B_{\lambda}(T_{{\rm flare}}) \, \lambda \, d\lambda}} 
{\int_{\lambda_{\rm Red}}^\infty{\Phi(\lambda)\,\, B_{\lambda}(T_{{\rm flare}}) \, \lambda \, d\lambda}}.
\label{eq:tflare}
\end{equation}
We also assumed that the area of the flare is the same in observations using the Blue or the Red channel.
The energy of a flare, $E$, is defined as the product of its equivalent duration (ED) and the quiescent luminosity of the star $L_\star$
\citep[e.g.][]{shibayama2013,hawley2014,martinez2020,gunther2020}.
Using Equation~\ref{eq:tflare}, the energy ratio is given by:
\begin{eqnarray}
\frac{E_{{\rm Blue}}}{E_{{\rm Red}}} =
\frac{{\rm ED}_{{\rm Blue}}}{{\rm ED}_{{\rm Red}}} \,
\frac{F_{\rm Blue}}{F_{\rm Red}}
= \nonumber \\
\,
\frac{\Delta t_{\rm Blue}}{\Delta t_{\rm Red}}
\,\,
\frac{\int_0^{\lambda_{\rm Blue}}{\Phi(\lambda)\,\, B_{\lambda}(T_{{\rm flare}}) \, \lambda \, d\lambda}} 
{\int_{\lambda_{\rm Red}}^\infty{\Phi(\lambda)\,\, B_{\lambda}(T_{{\rm flare}}) \, \lambda \, d\lambda}}.
\label{eq:energy_ratio}
\end{eqnarray}

\begin{figure}
	\includegraphics[width=\columnwidth]{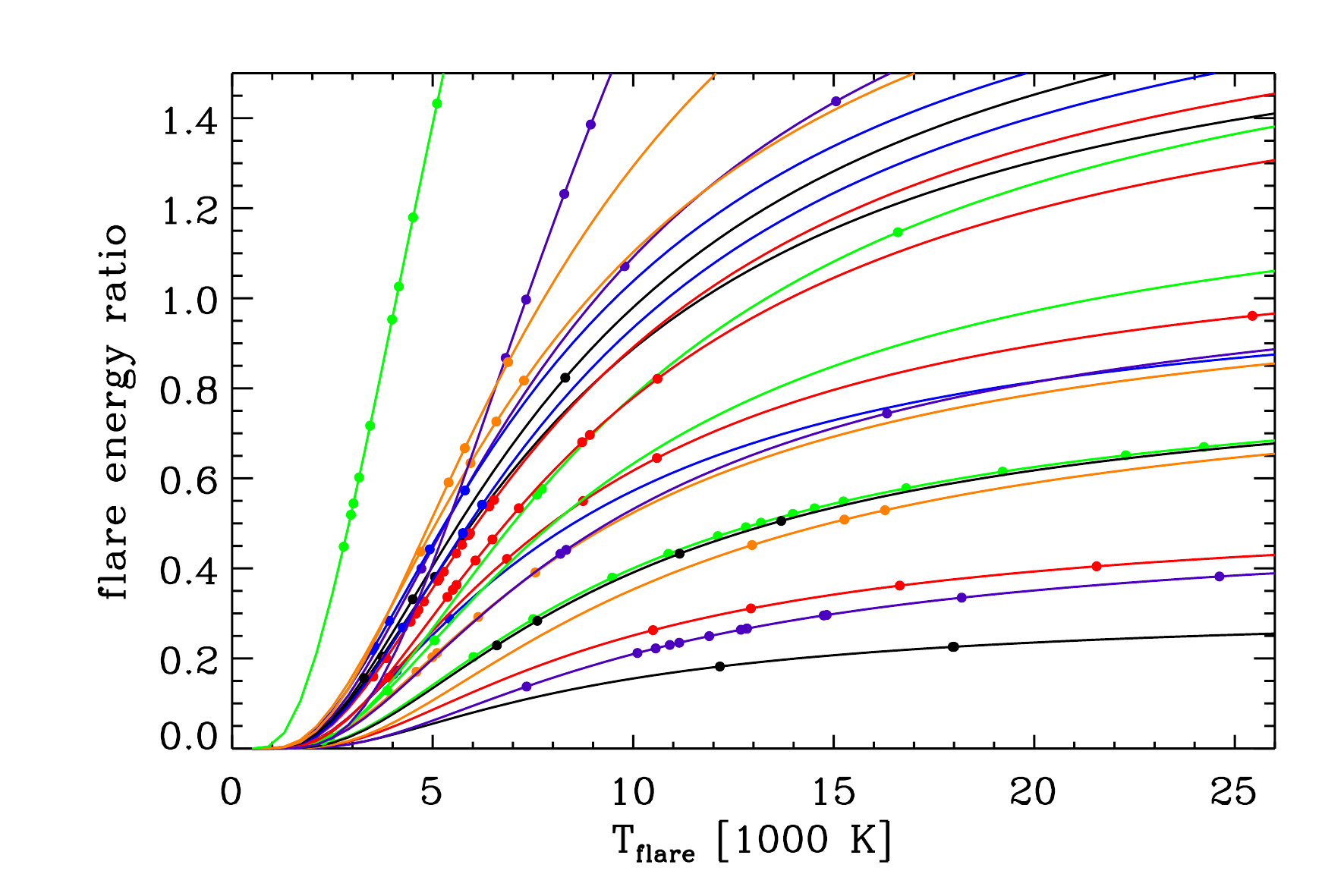}
    \caption{Flare energy ratio, $E_{\rm Blue}/E_{\rm Red}$, 
    as a function of flare temperature, $T_{\rm flare}$, 
    given by the rightmost term in Equation~\ref{eq:energy_ratio_ik}. Each color is for a given light curve $l$. For a better visualization, only light curves with 3 or more observed flares are plotted.
    The full circles are the observed flare energy ratio 
    and the correspondent estimated flare temperature, $T_{\rm flare}$(k).
    }
    \label{fig:ratio}
\end{figure}

The flare duration was calculated using the flare fitted parameters (Equation~\ref{eq:flarefitting}).
The analyzed flares have, on average, the same duration
in the Blue and Red channels. 
We used Equation~\ref{eq:energy_ratio} to estimate $T_{\rm flare}(k)$, for each flare $k$ observed in a given light curve $l$,
since it is the only unknown parameter in the equation:
\begin{eqnarray}
\frac{E_{{\rm Blue}}}{E_{{\rm Red}}}(l,k) =
\frac{{\rm ED}_{{\rm Blue}}}{{\rm ED}_{{\rm Red}}}(k) \, \cdot \,
\frac{F^{\rm obs}_{{\rm Blue}}}{F^{\rm obs}_{{\rm Red}}}(l)
= \nonumber \\
\frac{\int_0^{\bar{\lambda}_{\rm Blue}(l)}{\Phi(\lambda)\,\, B_{\lambda}[T_{{\rm flare}}(k)] \,\, \lambda \, d\lambda}}
{\int_{\bar{\lambda}_{\rm Red}(l)}^\infty{\Phi(\lambda)\,\, B_{\lambda}[T_{{\rm flare}}(k)] \,\, \lambda \, d\lambda}}.
\label{eq:energy_ratio_ik}
\end{eqnarray}
The flare energy ratio as a function of $T_{\rm flare}$, 
given by the ratio of integrals on the right-hand side of Equation~\ref{eq:energy_ratio_ik}, 
is shown in Figure~\ref{fig:ratio}.
It varies smoothly with $T_{\rm flare}$ 
for a given pair of wavelength limits: 
$\bar{\lambda}_{\rm Blue}(l)$ and $\bar{\lambda}_{\rm Red}(l)$.
Using the error propagation equation, we estimated the flare temperature uncertainty, $\sigma_{T_{\rm flare}}(k)$,
by dividing the uncertainty of the observed flare energy ratio,
ED$_{{\rm Blue}}$/ED$_{{\rm Red}}(k)\,\cdot F_{{\rm Blue}}/F_{{\rm Red}}(l)$,
by the derivative of the right-hand side expression
at the estimated flare temperature.
As seen in Fig.~\ref{fig:ratio}, the slope of the flare energy rate
tends to zero as the flare temperature increases.
Therefore, the estimated  uncertainty in  the flare temperature increases with the flare temperature.
This is a consequence of using observations in the visible range to estimate high temperatures.
According to Wien's law, a blackbody with a temperature of 10,000\,K will have its maximum emission at 2900\,\AA \, and outside the CoRoT Response Function (Fig.~\ref{fig:wavelimits}).

\begin{figure*}
	\includegraphics[width=\textwidth]{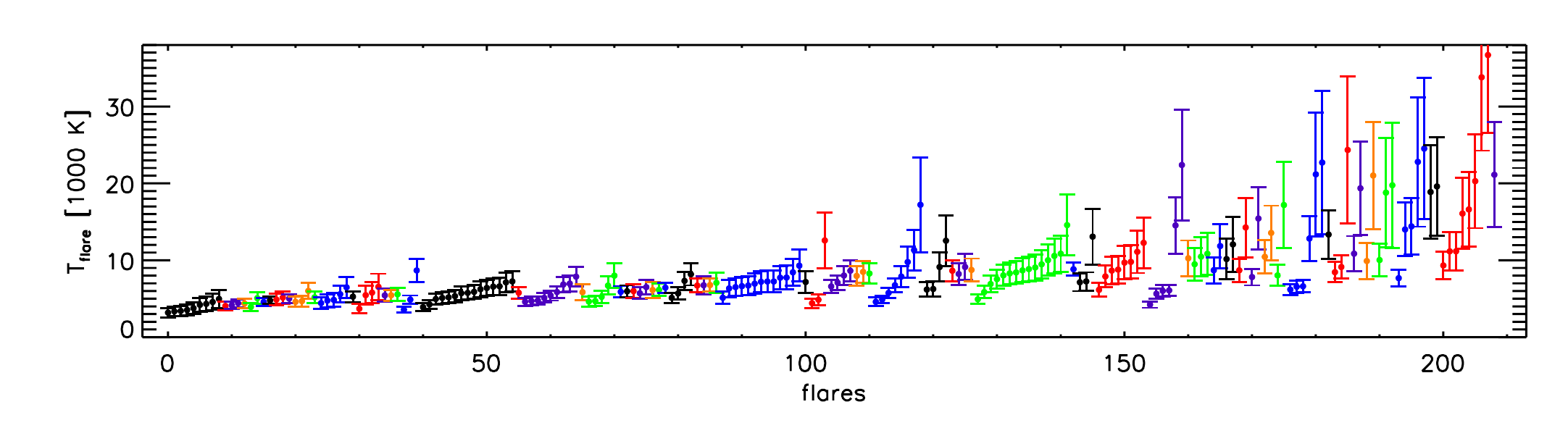}
	\includegraphics[width=\textwidth]
	{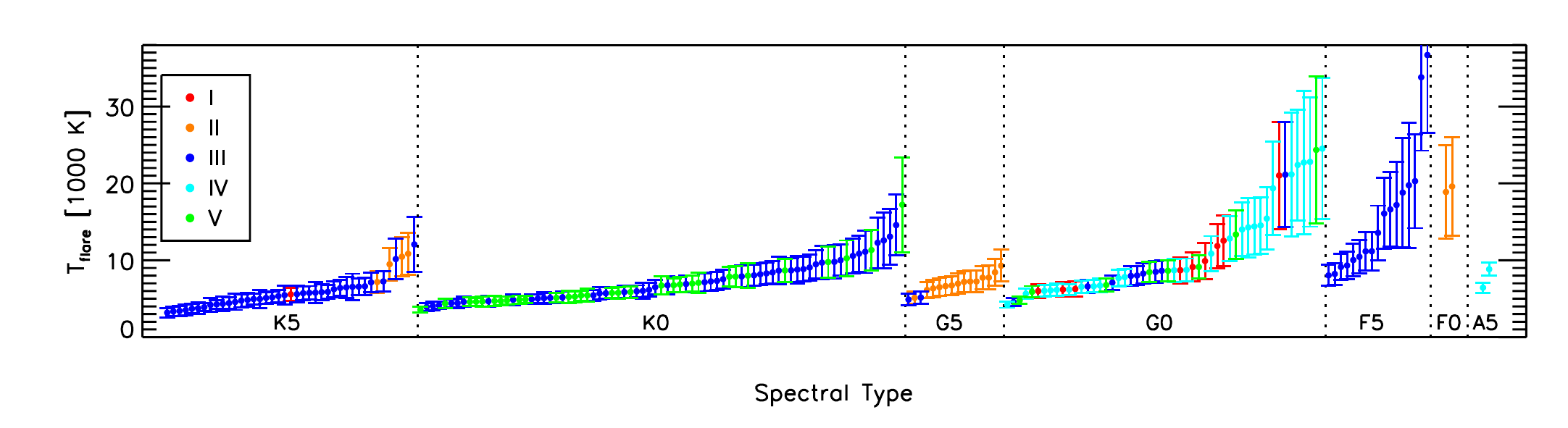}
	\includegraphics[width=\textwidth]
	{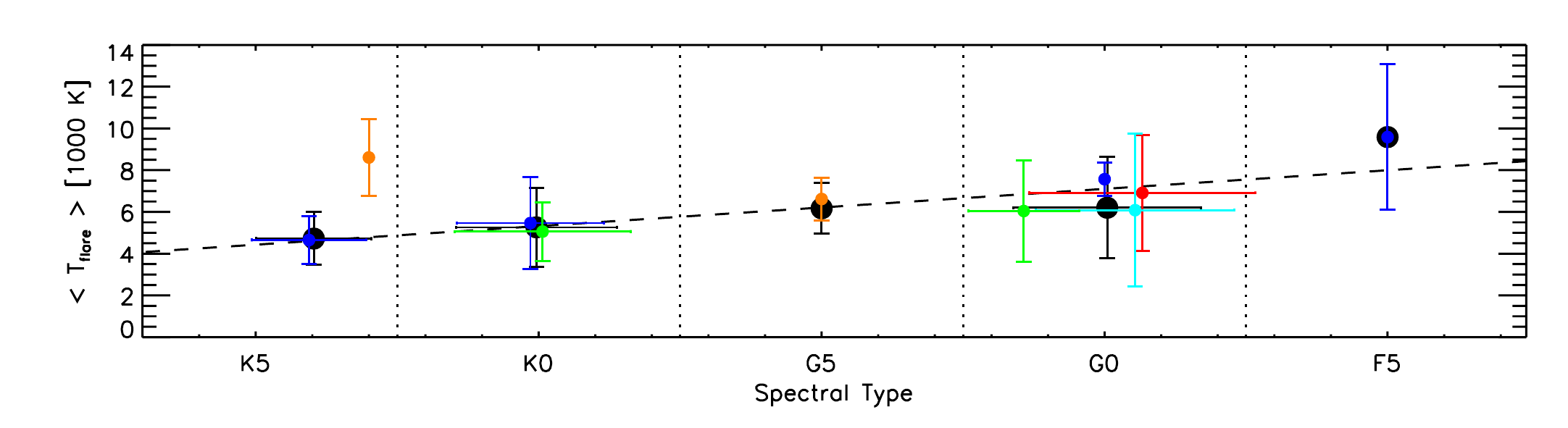}
    \caption{
    1st row: Weighted mean flare temperatures, $\bar{T}_{\rm flare}$(k), 
    of the 209 flares and their standard deviation,
$\sigma_{T_{\rm flare}}(k)$.
The flares in a given star are shown in the same color. They are ordered with increasing mean temperature of all flares in a given star.
2nd row: Same as the first row plot, but as function of spectral type, going from late to early type stars,
in intervals of half a spectral class. 
The spectral type at the center of each interval is shown at the bottom of the plot.
Inside each spectral class interval, the flare temperatures are in ascending order.
The different colors are for the different luminosity classes.
3rd row: 
Weighted average of the temperature of the flares
versus the average of the spectral subclass corresponding to the flares within each interval of half a spectral class.  
The averages for a given luminosity class, when there are 4 or more flares, are shown in different colors (as in the 2nd-row panel). The averages including all luminosity classes are shown in black and their linear regression is given by the black dashed line.
The vertical error bars correspond to the standard deviation of the flare temperature weighted average
and
the horizontal error bars to the standard deviation of the average of the spectral subclass.
The absence of a horizontal bar means that only one spectral type was used in the temperature averaged.
    }
    \label{fig:tmean}
\end{figure*}

\begin{figure}
    \includegraphics[width=\columnwidth]{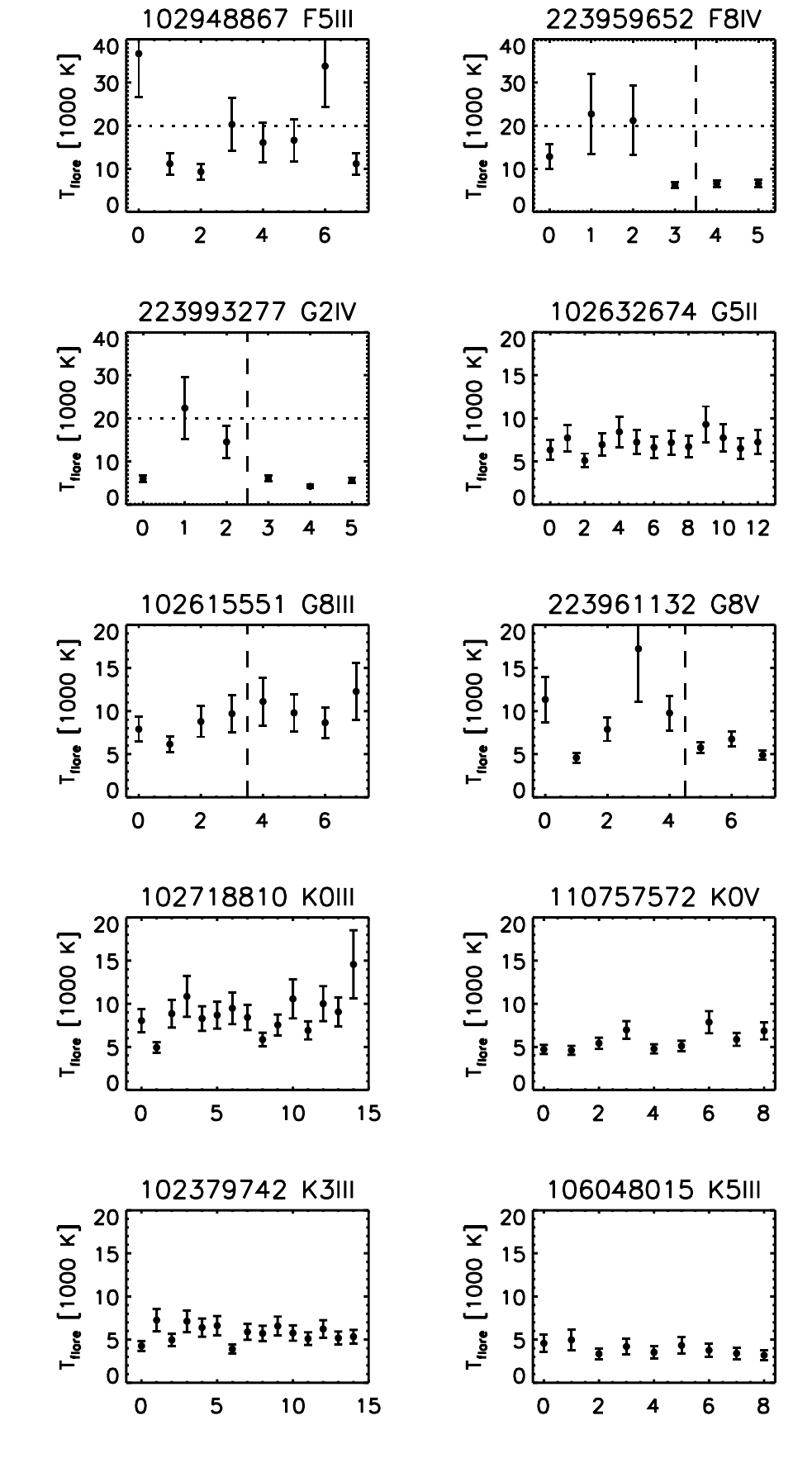}
    \caption{Weighted mean flare temperature, $\bar{T}_{\rm flare}$(k), as in Figure~\ref{fig:tmean} (1st-row panel), for stars with 6 or more flares.
    The CoRoT ID and stellar spectral type are shown in each panel.
    The vertical dashed lines on four of the stars separate the flares obtained in different observation sequences: during 2008 and 2012.
    The vertical axis range is twice as large for the first three stars as for the others.
   }
   \label{fig:many}
\end{figure}

To take into account the uncertainty in the estimated wavelength limits, we added noise to the flare temperature calculation.
For each pair 
$\bar{\lambda}_{\rm Blue}(k)$ and $\bar{\lambda}_{\rm Red}(k)$,
we generated 1000 input pairs by adding different random realizations of Gaussian noise 
with a standard deviation given by
$\sigma_{\lambda_{\rm Blue}}(l)$
and
$\sigma_{\lambda_{\rm Red}}(l)$, respectively.
The final estimate of the flare temperature is the
weighted average of 1000 estimates obtained using perturbed input wavelength limits:
$(\bar{T}_{\rm flare} \pm \sigma_{T_{\rm flare}})(k)$ $-$ Table~\ref{tab:temperature}.
The percent uncertainty of the weighted mean ranges from 10\%$-$40\%.
Figure~\ref{fig:tmean} (first row) shows the average flare temperatures for each star (in the same color) in ascending order of the average temperature of the flares in a given star.
A total of 70\% of the analyzed stars have only one or two flares.
Those with two 
or more flares often have similar flare temperatures.
Figure~\ref{fig:many} shows the temperatures for stars with 6 or more flares.


We checked whether there is any dependence of the flare temperature with stellar spectral type.
Figure\,\ref{fig:tmean} (2nd row) shows the flare temperatures in intervals of half-spectral type, where within each interval the temperatures are in ascending order and the flares in each luminosity class are in a different color.
We calculated the weighted average of the flare temperature over each interval which is shown in the 3rd-row panel. The average over all luminosity classes are shown in black.
The average flare temperature, including all luminosity classes (black circles), ranges from
(4700 $\pm$ 1300)\,K for spectral type K4
to 
(9600 $\pm$ 3500)\,K for F5 type
and
the slope of the linear regression is equal to (1800$\pm$1400)\,K 
per spectral type.
The Pearson correlation coefficient (of the black circles) is 0.9 which has only a 5\% probability of not being correlated \citep{bevington2003}. 
We calculated the correlation coefficient 
after adding 
normally distributed random numbers with the same  
standard deviation of the mean flare temperature (shown in the 3rd-row panel) 
to examine the effect of their uncertainty on the correlation. The average Pearson coefficient and its standard deviation of 10,000 instances is: (0.58$\pm$0.37) and the probability of not being correlated is 30\%. 

\begin{figure}
    \includegraphics[width=\columnwidth]{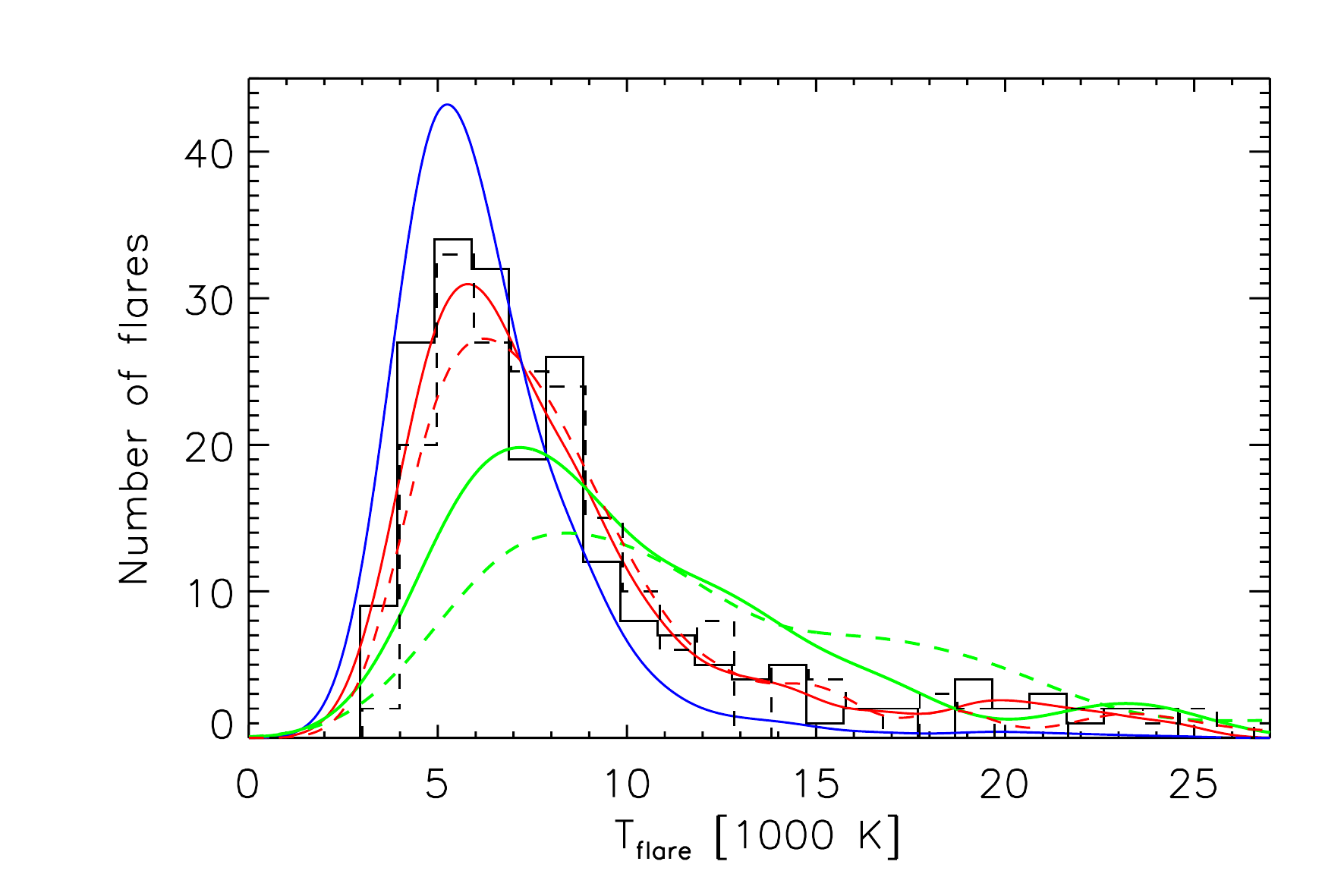}
    \caption{
    Histogram (solid black line)
    of the mean flare temperature, $\bar{T}_{\rm flare}$, for all 209 flares in intervals of 983\,K, which is close to the median of flare temperature uncertainty.
    The PDF obtained with the kernel density estimation \citep{silverman1986} (using a Gaussian kernel) 
    with and without taking into account flare temperature uncertainty are shown as blue and red lines, respectively.
    The histogram shown with dashed black line is for mean flare temperatures obtained using
    wavelength limits estimated assuming a smaller uncertainty 
    in the stellar classification used
    (i.e., half a spectral type and one luminosity class).
    It is slightly displaced horizontally for a better visualization.
    Its PDF estimated,
    without taking into account the uncertainty, is shown by the dashed red line.
    The green solid and dashed lines show the PDF 
    (without taking into account flare temperature uncertainty) obtained by increasing the input ED by 50\% and 100\%, respectively.
    The probability density functions plotted here were normalized to be compared with the histograms, they were multiplied by the number of flares (209) and size of histogram bin (983\,K).
    }
   \label{fig:histo}
\end{figure}

Next, we analyzed the frequency distribution of the estimated flare temperatures. Figure\,\ref{fig:histo} shows the histogram (solid black line) of the 209 $\bar{T}_{\rm flare}(k)$ plotted in Fig.~\ref{fig:tmean} (1st row).
We estimated the probability density function (PDF) using the kernel density estimator (KDE) method \citep{silverman1986} - with a Gaussian kernel - without taking into account flare temperature uncertainties.
It agrees well with the histogram after the PDF was multiplied by the number of flares and size of histogram bin (full red line). 
The kernel has an optimal window width 
 \citep[given by equations 3.30 and 3.31 in][]{silverman1986} equal to 983\,K,
 which is used as the histogram bin width (solid black line).
However, when including the flare temperature uncertainties in the estimation, the normalized PDF (blue line) has its maximum shifted toward a smaller flare temperature, with a mode equal to 5,300\,K. The distribution no longer has a long tail, and the width of the highest density interval (HDI) is narrower by 33\%.
This is to be expected, since flares with high temperatures have greater uncertainties, as mentioned earlier.
In Table~\ref{tab:kde}, we summarize the characteristics of the distributions (called "A"). They give us some indication of the upper and lower limits of the flare temperature distribution.
The average spectral type of all flaring stars, weighted by the number of flares observed in stars in each spectral subclass, is equal to G6.

As mentioned in Section~\ref{sec:wavelength}, we also calculated 
wavelength limits using a smaller uncertainty in the stellar classification (i.e., within plus/minus half a spectral class and within plus/minus one luminosity class).
Reestimating $\bar{T}_{\rm flare}$ in the same way as described above, but with these wavelength limits, we find that the new $\bar{T}_{\rm flare}$
are systematically larger than before by a small amount, which is, on average, equal to 1~$\sigma_{T_{\rm flare}}$.
Their histogram is shown in Fig.~\ref{fig:histo} by a black dashed line.
Their normalized PDF calculated without including the uncertainties (dashed red line) and including the uncertainties (not shown in the figure) have modes and mean values that are a few percent larger than before (labeled "B" in Table~\ref{tab:kde}).

We did a simple exercise to see how higher-than-observed flare temperatures would show in the results of our analysis.
We artificially increased the value of the observed equivalent duration ratio given as input to Eq.~\ref{eq:energy_ratio_ik} to analyze the corresponding increase in $\bar{T}_{\rm flare}(k)$.
After increasing the equivalent duration ratio by 50\% and 100\%
and repeating the calculation of  $\bar{T}_{\rm flare}(k)$ as described above,
the estimated PDF is shifted by approx. 20\% (green solid line) and $\sim$40\% (green dashed line) in $T_{\rm flare}$, respectively.
While the width of 75\% HDI increases by $\sim$40 \% and $\sim$80 \%, respectively.
A wide HDI width indicates a greater uncertainty in the parameter determination, which could be explained by the difficulty of observations in the visible range to measure large flare temperatures.

\begin{table}
\begin{center}
 \caption{
 PDF properties of our main results (`A').
 The table also shows for comparison the PDF properties of the results obtained assuming a smaller uncertainty in the stellar classification (`B') and those obtained by the arbitrary multiplication of the equivalent duration of the flare by 1.5 (`1.5$\times$') and 2 ( `2.0$\times$').
The columns are as follows: 
(1) PDF used;
(2) y (yes) or no to indicate whether or not the flare temperature uncertainties were included in the PDF estimate;
(3) colored lines as plotted in Figure~\ref{fig:histo} (if dashed, it is labeled $-$);
(4) mode; (5) mean value (or expected value); (6) standard deviation of the mean; (7) lower and (8) upper limits in the 75\% HDI.
All values are given in Kelvin.
 }
 \label{tab:kde}
 \begin{tabular}{lcccRcRc}
 &  & line & mode & mean & $\sigma$ & lower & upper\\
 \hline
\hline
                     A & y & blue &   5300 &  6400 &  2800 &  3400 &  8100 \\
                      \vspace{5pt}
                         B & y &  none  &   5400 &  6600 &  2700 &  3500 &  8200 \\
   A & no & red &   5800 &  8400 &  4600 &  3300 & 10300 \\
          B & no & $-$red &   6200 &  8700 &  4700 &  3700 & 10700 \\
                 \\
        1.5$\times$ & no & green &   7200 & 10200 &  5000 &  3900 & 13600 \\
         2.0$\times$ & no & $-$green &   8400 & 12100 &  5700 &  4600 & 17300 \\
         \hline
                          1.5$\times$ & y &    none &   6300 &  7700 &  3600 &  3400 & 10000 \\
                   2.0$\times$ & y & none  &   7100 &  9100 &  4500 &  3500 & 11900 \\
 \end{tabular}
 \end{center}
\end{table}

\section{Rotation}

As suggested by dynamo theories and obtained from observations of cool stars, of any spectral type, the stellar activity increases with the rate of rotation and even more so with the Rossby Ro number, where small values of Ro indicate very active stars \citep{donati2009}.
On the other hand, intermediate- and high-mass stars exhibit virtually no intrinsic variability.
We calculated the stellar rotation to search for a correlation with flare temperature.

The presence of active regions on a star's surface produces periodic variations in stellar observed brightness, due to the region's motion relative to the observer as it follows the star's rotation. 
However, other phenomena can also result in periodic variations in the light curve, such as eclipsing binaries, planetary transits, and pulsations. To distinguish them, we need not only a more detailed analysis but especially radial velocity measurements, both of which are outside the scope of this work.
We analyzed the autocorrelation function (ACF) as described in \cite{mcquillan2013}, 
after smoothing the White channel light curves using a four-hour Gaussian filter (Table~\ref{tab:temperature}).
When the light curve is longer than two rotation periods,
several maxima of the ACF 
will be present. 
In this case, 
the period is defined as the gradient of a straight-line fit to the ACF peak positions as a function of peak number,
to obtain a more robust determination of the period and its uncertainty
\citep[see][for details]{mcquillan2014}.
In some cases, the estimated errors are very small and were arbitrarily made equal to 32\,s. 
We visually inspected the light curves and their periodograms to check the estimated period.
One of the advantages of the ACF method is its potential to give the correct rotation period even when there are two diametrically opposite active regions or, in the case of a close contact binary, when the two minima have similar depths. This is the case for CoRoT 102854684, 102718810 and 223993277, where the first two are probably eclipsing binary systems.
\cite{klagyivik2017} obtained orbital periods for them that agree with ours within 2$\sigma$, by removing the signature of an eclipsing binary from the light curve to look for transit-like features.

Two of the analyzed stars (CoRoT 102615551 and CoRoT 223983509) are in the eclipsing binary catalog of the CoRoT/Exoplanet program \citep{deleuil2018}.
The two aforementioned likely binaries reported by \citet{klagyivik2017} are not among them. 
As CoRoT 102615551 has very deep and narrow eclipses \citep{carone2012}, we report in Table~\ref{tab:temperature} the value given by \citet{deleuil2018}. 
The other object, CoRoT 223983509, identified as a contact binary by 
\citet{deleuil2018} has an orbital period that agrees with ours within 1$\sigma$ and it has also been identified as a T Tauri star \citep{venuti2015}.
Ten of our stars are members of the star-forming region NGC\,2264 ($\sim$3 Myr) and were classified as weak-lined T Tauri stars \citep{venuti2015}. 
In young stellar objects and binary stars, flare events can originate from magnetic reconnection in the field that connects the star to the accretion disk or to the other star, as the case may be.
There is only one case (CoRoT 110741064) in which no periodic variation was detected. 

\begin{figure}
	\includegraphics[width=\columnwidth]{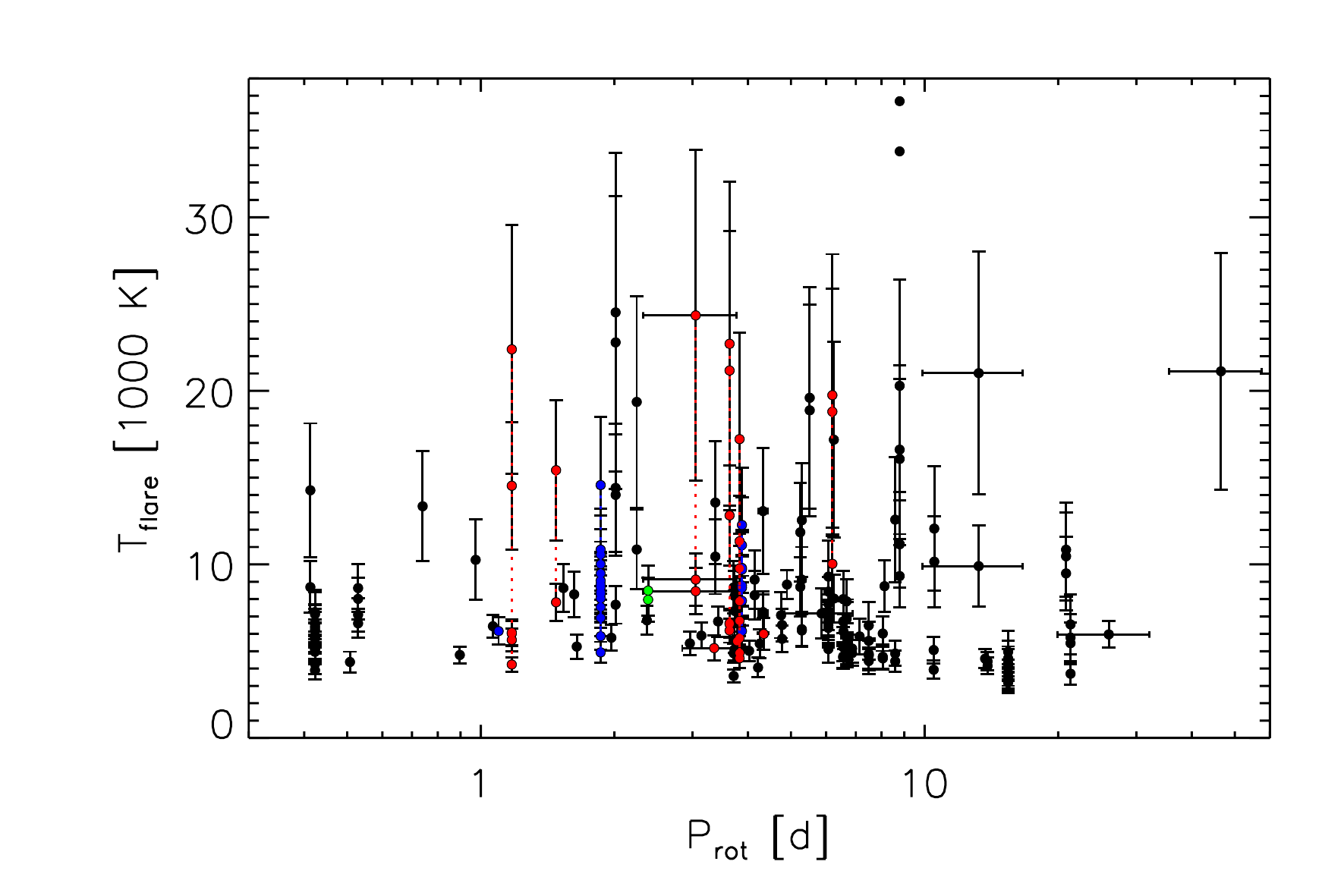}
	\caption{Flare temperature $\bar{T}_{\rm flare}(k)$ (as in top panel of Fig.~\ref{fig:tmean}) versus stellar rotation period. 
	The horizontal error bars correspond to the uncertainty in the estimated rotation period, which is very small in most cases.
	Two flares with temperatures larger than 30,000\,K have uncertainties around 10,000\,K
	and their error bars 
	are not shown for a better visualization. 
	Stars identified as T Tauri are shown in red, and eclipsing binaries in blue, and their flare temperatures are connected by a dotted line of the same color. CoRoT 223983509 is shown in green as it has been classified as T Tauri and eclipsing binary.
	In the case of binaries, the period shown is their orbital period.
    }
    \label{fig:rotation}
\end{figure}

Figure~\ref{fig:rotation} shows the flare temperature $\bar{T}_{\rm flare}(k)$
versus stellar rotation period. We did not find any correlation between them. 
The T Tauri stars and eclipsing binaries are shown in color.
The mean rotation period of the analyzed flaring stars, weighted by the number of flares, is
equal to (5.5$\pm$0.4) days. 
Only 10 stars
have a rotation period larger than 10 days.

\footnotesize{
\begin{center}
\startlongtable
\begin{deluxetable*}{cccccc}
\tablecaption{
The spectral type, rotation period, flare temperature, Gaia flare energy, and flare area for the 69 flaring stars analyzed. 
Their spectral type is given by the Exodat Database.
Flares observed at different epochs (taken $\sim$4 years apart) are marked with a star next to their temperature.
 A small letter after the rotation period indicates: an eclipsing binary identified by \citet{deleuil2018}$^a$, or by \citet{klagyivik2017}$^b$, and a T Tauri analysed by \citet{venuti2015}$^t$.
 The value in the $P_{\rm rot}$ column for eclipsing binaries is actually their orbital period. 
 In the case of the eclipsing binary
 CoRoT 102615551, the value obtained by \citet{deleuil2018} $-$ marked with the symbol $^\dagger$ is given.
Due to the large uncertainties, there are flares with the same temperature and uncertainty for some stars.
\label{tab:temperature}
}
\tablewidth{700pt}
\tabletypesize{\scriptsize}
\tablehead{
\colhead{CoRoT ID} & \colhead{Type} & \colhead{Prot [d]} &
\colhead{$T_{\rm flare}$ [1000 K]} & $E_{G,f}$ [erg] & $A_{\rm flare}$ [cm$^2$]
%
}
\startdata
223941972 & A0IV &  4.90$\pm$0.02 &  8.84$\pm$0.83 & (33.4$\pm$4.9)$\times 10^{34}$ & (25.5$\pm$3.0)$\times 10^{19}$ \\
 &  &  &  &  &  \\
224008170 & A5IV &  1.06$\pm$0.01 &  6.44$\pm$0.66 & (92.6$\pm$6.6)$\times 10^{33}$ & (25.7$\pm$1.2)$\times 10^{20}$ \\
 &  &  &  &  &  \\
110685010 & F0II &  5.510$\pm$0.007 & 18.9$\pm$6.1 & (54.1$\pm$9.5)$\times 10^{32}$ & (57.6$\pm$4.4)$\times 10^{16}$ \\
 &  &  & 19.6$\pm$6.4 & (70.1$\pm$9.6)$\times 10^{32}$ & (16.7$\pm$1.9)$\times 10^{17}$ \\
 &  &  &  &  &  \\
102326330 & F5III &  4.141$\pm$0.001 &  8.2$\pm$1.4 & (72.0$\pm$2.4)$\times 10^{33}$ & (49.5$\pm$1.4)$\times 10^{19}$ \\
 &  &  &  9.1$\pm$1.7 & (88.5$\pm$2.5)$\times 10^{33}$ & (45.1$\pm$1.2)$\times 10^{19}$ \\
 &  &  &  &  &  \\
102332965 & F5III &  6.247$\pm$0.006 &  8.0$\pm$1.4 & (12.7$\pm$1.4)$\times 10^{33}$ & (20.1$\pm$2.1)$\times 10^{19}$ \\
 &  &  & 17.2$\pm$5.6 & (7.5$\pm$1.3)$\times 10^{33}$ & (30.7$\pm$3.4)$\times 10^{18}$ \\
 &  &  &  &  &  \\
102948867 & F5III &  8.79$\pm$0.01 &  9.3$\pm$1.8 & (5.0$\pm$1.4)$\times 10^{32}$ & (21.4$\pm$4.6)$\times 10^{18}$ \\
 &  &  & 11.2$\pm$2.5 & (6.0$\pm$1.5)$\times 10^{32}$ & (5.4$\pm$1.0)$\times 10^{18}$ \\
 &  &  & 11.2$\pm$2.5 & (7.9$\pm$1.9)$\times 10^{32}$ & (6.0$\pm$1.2)$\times 10^{18}$ \\
 &  &  & 16.1$\pm$4.6 & (11.3$\pm$1.1)$\times 10^{32}$ & (37.2$\pm$2.7)$\times 10^{17}$ \\
 &  &  & 16.6$\pm$4.9 & (18.7$\pm$1.3)$\times 10^{32}$ & (32.1$\pm$1.6)$\times 10^{17}$ \\
 &  &  & 20.3$\pm$6.1 & (43.7$\pm$8.5)$\times 10^{31}$ & (11.8$\pm$1.1)$\times 10^{17}$ \\
 &  &  & 33.8$\pm$9.5 & (7.4$\pm$2.4)$\times 10^{32}$ & (6.8$\pm$2.2)$\times 10^{17}$ \\
 &  &  & 37$\pm$10 & (17.2$\pm$5.7)$\times 10^{32}$ & (7.8$\pm$2.5)$\times 10^{17}$ \\
 &  &  &  &  &  \\
223940041 & F5III &  3.377$\pm$0.006 & 10.4$\pm$2.2 & (70.9$\pm$2.7)$\times 10^{33}$ & (32.7$\pm$1.0)$\times 10^{19}$ \\
 &  &  & 13.6$\pm$3.5 & (46.1$\pm$6.9)$\times 10^{32}$ & (38.3$\pm$3.0)$\times 10^{18}$ \\
 &  &  &  &  &  \\
223984608 & F5III &  6.20$\pm$0.02$^{t}$ & 10.0$\pm$2.1 & (63.8$\pm$2.3)$\times 10^{34}$ & (226.9$\pm$7.4)$\times 10^{19}$ \\
 &  &  & 18.8$\pm$7.1 & (17.1$\pm$3.1)$\times 10^{33}$ & (10.4$\pm$1.0)$\times 10^{19}$ \\
 &  &  & 19.8$\pm$8.1$^\ast$ & (15.7$\pm$2.2)$\times 10^{33}$ & (11.1$\pm$1.1)$\times 10^{19}$ \\
 &  &  &  &  &  \\
102925086 & F8I & 13$\pm$3 &  9.9$\pm$2.3 & (87.0$\pm$8.2)$\times 10^{32}$ & (109.6$\pm$6.3)$\times 10^{17}$ \\
 &  &  & 21.0$\pm$7.0 & (5.6$\pm$1.3)$\times 10^{33}$ & (13.5$\pm$2.4)$\times 10^{17}$ \\
 &  &  &  &  &  \\
110773079 & F8I &  5.29$\pm$0.05 &  6.19$\pm$0.95 & (12.9$\pm$2.5)$\times 10^{33}$ & (20.3$\pm$2.6)$\times 10^{19}$ \\
 &  &  &  6.27$\pm$0.97 & (32.8$\pm$8.2)$\times 10^{33}$ & (16.4$\pm$2.4)$\times 10^{19}$ \\
 &  &  &  9.1$\pm$1.9 & (7.8$\pm$2.4)$\times 10^{34}$ & (19.9$\pm$4.9)$\times 10^{19}$ \\
 &  &  & 12.5$\pm$3.3 & (37.3$\pm$6.9)$\times 10^{33}$ & (26.6$\pm$3.6)$\times 10^{18}$ \\
 &  &  &  &  &  \\
211607284 & F8IV &  2.013$\pm$0.001 &  7.7$\pm$1.1 & (26.2$\pm$4.8)$\times 10^{31}$ & (68.2$\pm$8.9)$\times 10^{17}$ \\
 &  &  & 14.0$\pm$3.5 & (20.2$\pm$6.3)$\times 10^{31}$ & (13.1$\pm$3.1)$\times 10^{17}$ \\
 &  &  & 14.4$\pm$3.7 & (38.7$\pm$5.8)$\times 10^{31}$ & (21.7$\pm$2.5)$\times 10^{17}$ \\
 &  &  & 22.8$\pm$8.4 & (5.4$\pm$1.5)$\times 10^{32}$ & (13.6$\pm$3.4)$\times 10^{17}$ \\
 &  &  & 24.5$\pm$9.2 & (4.1$\pm$1.0)$\times 10^{32}$ & ( 9.9$\pm$2.1)$\times 10^{17}$ \\
 &  &  &  &  &  \\
221628176 & F8IV &  3.43$\pm$0.03 &  6.71$\pm$0.88 & (10.0$\pm$3.7)$\times 10^{34}$ & (3.5$\pm$1.3)$\times 10^{21}$ \\
 &  &  &  &  &  \\
223959652 & F8IV &  3.636$\pm$0.004$^{t}$ &  6.19$\pm$0.70 & (79.2$\pm$4.9)$\times 10^{33}$ & (45.9$\pm$1.9)$\times 10^{20}$ \\
 &  &  &  6.56$\pm$0.80 & (18.5$\pm$1.3)$\times 10^{34}$ & (28.5$\pm$1.3)$\times 10^{20}$ \\
 &  &  &  6.62$\pm$0.81 & (22.1$\pm$1.5)$\times 10^{34}$ & (31.9$\pm$1.4)$\times 10^{20}$ \\
 &  &  & 12.8$\pm$2.9 & (30.1$\pm$7.8)$\times 10^{33}$ & (7.9$\pm$1.6)$\times 10^{20}$ \\
 &  &  & 21.2$\pm$8.0$^\ast$ & (21.3$\pm$3.6)$\times 10^{33}$ & (14.5$\pm$1.3)$\times 10^{19}$ \\
 &  &  & 22.7$\pm$9.3$^\ast$ & (18.3$\pm$4.4)$\times 10^{33}$ & (13.9$\pm$1.5)$\times 10^{19}$ \\
 &  &  &  &  &  \\
223970694 & F8IV &  1.478$\pm$0.002$^{t}$ &  7.8$\pm$1.1 & (22.7$\pm$1.4)$\times 10^{34}$ & (29.8$\pm$1.4)$\times 10^{20}$ \\
 &  &  & 15.4$\pm$4.1 & (11.8$\pm$1.5)$\times 10^{34}$ & (28.3$\pm$1.7)$\times 10^{19}$ \\
 &  &  &  &  &  \\
223983509 & G0III &  2.383$\pm$0.003$^{at}$ &  8.0$\pm$1.3 & (25.1$\pm$2.1)$\times 10^{34}$ & (168.9$\pm$8.5)$\times 10^{19}$ \\
 &  &  &  8.5$\pm$1.4 & (27.7$\pm$1.5)$\times 10^{34}$ & (24.6$\pm$1.0)$\times 10^{20}$ \\
 &  &  &  &  &  \\
102899501 & G0III &  1.6236$\pm$0.0004 &  8.3$\pm$1.3 & (5.5$\pm$1.5)$\times 10^{34}$ & (20.6$\pm$2.8)$\times 10^{19}$ \\
 &  &  &  &  &  \\
104228454 & G0III & 13.69$\pm$0.02 &  4.57$\pm$0.54 & (8.1$\pm$1.0)$\times 10^{35}$ & ( 97.0$\pm$7.7)$\times 10^{20}$ \\
 &  &  &  &  &  \\
106054338 & G0III &  0.529$\pm$0.001 &  6.60$\pm$0.83 & (149.4$\pm$8.4)$\times 10^{32}$ & (39.5$\pm$1.3)$\times 10^{19}$ \\
 &  &  &  7.10$\pm$0.95 & (50.8$\pm$4.4)$\times 10^{32}$ & (27.6$\pm$1.4)$\times 10^{19}$ \\
 &  &  &  8.0$\pm$1.2 & (48.5$\pm$4.5)$\times 10^{32}$ & (156.1$\pm$8.2)$\times 10^{18}$ \\
 &  &  &  8.6$\pm$1.4 & (151.1$\pm$7.9)$\times 10^{32}$ & (29.5$\pm$1.1)$\times 10^{19}$ \\
 &  &  &  &  &  \\
102631863 & G0III & 47$\pm$11 & 21.1$\pm$6.8 & (18.0$\pm$5.4)$\times 10^{35}$ & (8.6$\pm$2.4)$\times 10^{21}$ \\
 &  &  &  &  &  \\
102715243 & G0IV &  2.2460$\pm$0.0004 & 10.9$\pm$2.3 & (41.9$\pm$4.3)$\times 10^{33}$ & (80.8$\pm$6.8)$\times 10^{19}$ \\
 &  &  & 19.4$\pm$6.1 & (32.1$\pm$6.8)$\times 10^{33}$ & (13.4$\pm$1.9)$\times 10^{19}$ \\
 &  &  &  &  &  \\
102859855 & G0IV &  8.13$\pm$0.05 &  8.7$\pm$1.5 & (64.9$\pm$8.8)$\times 10^{33}$ & (67.4$\pm$6.6)$\times 10^{18}$ \\
 &  &  &  &  &  \\
110744989 & G0IV &  0.413$\pm$0.001 &  8.7$\pm$1.5 & (17.8$\pm$1.0)$\times 10^{33}$ & (46.3$\pm$1.8)$\times 10^{19}$ \\
 &  &  & 14.3$\pm$3.9 & (10.4$\pm$1.1)$\times 10^{33}$ & (58.3$\pm$3.7)$\times 10^{18}$ \\
 &  &  &  &  &  \\
223980621 & G0V &  3.1$\pm$0.7$^{t}$ &  8.5$\pm$1.3 & (11.3$\pm$2.6)$\times 10^{34}$ & (56.5$\pm$9.5)$\times 10^{19}$ \\
 &  &  &  9.1$\pm$1.5 & (25.8$\pm$3.0)$\times 10^{34}$ & (19.9$\pm$2.1)$\times 10^{20}$ \\
 &  &  & 24.3$\pm$9.5 & (28.5$\pm$6.5)$\times 10^{33}$ & (19.4$\pm$2.6)$\times 10^{19}$ \\
 &  &  &  &  &  \\
105085606 & G2I &  5.248$\pm$0.007 &  8.7$\pm$1.7 & (21.3$\pm$2.1)$\times 10^{32}$ & (37.2$\pm$2.3)$\times 10^{18}$ \\
 &  &  & 11.9$\pm$2.9$^\ast$ & (8.4$\pm$1.5)$\times 10^{32}$ & (70.3$\pm$6.3)$\times 10^{17}$ \\
 &  &  &  &  &  \\
500007157 & G2I &  4.34$\pm$0.02$^{t}$ &  5.99$\pm$0.92 & (14.3$\pm$4.6)$\times 10^{34}$ & (13.7$\pm$3.5)$\times 10^{20}$ \\
 &  &  &  &  &  \\
223993277 & G2IV &  1.175$\pm$0.003$^{t}$ &  4.23$\pm$0.41 & (86.0$\pm$7.6)$\times 10^{33}$ & (19.3$\pm$1.1)$\times 10^{21}$ \\
 &  &  &  5.65$\pm$0.64 & (42.6$\pm$1.7)$\times 10^{34}$ & (84.3$\pm$3.3)$\times 10^{20}$ \\
 &  &  &  6.02$\pm$0.73 & (52.4$\pm$5.5)$\times 10^{34}$ & (15.6$\pm$1.0)$\times 10^{21}$ \\
 &  &  &  6.08$\pm$0.73$^\ast$ & (13.9$\pm$1.0)$\times 10^{34}$ & (69.5$\pm$3.5)$\times 10^{20}$ \\
 &  &  & 14.5$\pm$3.7$^\ast$ & (24.6$\pm$5.0)$\times 10^{33}$ & (25.9$\pm$2.4)$\times 10^{19}$ \\
 &  &  & 22.4$\pm$7.2$^\ast$ & (21.0$\pm$3.7)$\times 10^{33}$ & (11.1$\pm$1.4)$\times 10^{19}$ \\
 &  &  &  &  &  \\
102854684 & G2IV &  1.097$\pm$0.003$^{b}$ &  6.15$\pm$0.79 & (48.6$\pm$1.7)$\times 10^{35}$ & (46.5$\pm$1.7)$\times 10^{21}$ \\
 &  &  &  &  &  \\
604183778 & G2V &  0.740$\pm$0.003 & 13.3$\pm$3.2 & (24.1$\pm$4.4)$\times 10^{33}$ & (41.8$\pm$4.5)$\times 10^{19}$ \\
 &  &  &  &  &  \\
102926046 & G2V &  2.368$\pm$0.004 &  6.77$\pm$0.84 & (164.2$\pm$7.8)$\times 10^{33}$ & (239.8$\pm$8.2)$\times 10^{19}$ \\
 &  &  &  &  &  \\
221649345 & G2V &  1.536$\pm$0.006 &  8.6$\pm$1.4 & (70.3$\pm$5.0)$\times 10^{33}$ & (75.8$\pm$3.1)$\times 10^{19}$ \\
 &  &  &  &  &  \\
223945488 & G2V &  3.142$\pm$0.004 &  5.90$\pm$0.75 & (24.6$\pm$2.8)$\times 10^{34}$ & (135.7$\pm$8.2)$\times 10^{19}$ \\
 &  &  &  &  &  \\
315258285 & G2V &  0.897$\pm$0.002 &  4.78$\pm$0.48 & (10.4$\pm$1.2)$\times 10^{33}$ & (119.0$\pm$7.4)$\times 10^{19}$ \\
 &  &  &  &  &  \\
102632674 & G5II &  6.07$\pm$0.02 &  5.12$\pm$0.80 & (13.5$\pm$3.3)$\times 10^{32}$ & (19.2$\pm$2.5)$\times 10^{19}$ \\
 &  &  &  6.3$\pm$1.1 & (37.9$\pm$2.4)$\times 10^{32}$ & (150.3$\pm$7.0)$\times 10^{18}$ \\
 &  &  &  6.5$\pm$1.2 & (38.3$\pm$2.5)$\times 10^{32}$ & (149.8$\pm$6.7)$\times 10^{18}$ \\
 &  &  &  6.6$\pm$1.2 & (12.3$\pm$1.5)$\times 10^{32}$ & (78.3$\pm$5.7)$\times 10^{18}$ \\
 &  &  &  6.7$\pm$1.2 & (16.9$\pm$5.2)$\times 10^{32}$ & (10.8$\pm$2.1)$\times 10^{19}$ \\
 &  &  &  7.0$\pm$1.3 & (34.9$\pm$2.9)$\times 10^{32}$ & (42.8$\pm$2.1)$\times 10^{18}$ \\
 &  &  &  7.2$\pm$1.4 & (29.2$\pm$1.8)$\times 10^{32}$ & (61.6$\pm$2.7)$\times 10^{18}$ \\
 &  &  &  7.2$\pm$1.4 & (10.5$\pm$1.1)$\times 10^{32}$ & (51.5$\pm$3.4)$\times 10^{18}$ \\
 &  &  &  7.2$\pm$1.4 & (45.7$\pm$2.4)$\times 10^{32}$ & (68.7$\pm$2.5)$\times 10^{18}$ \\
 &  &  &  7.7$\pm$1.5 & (22.7$\pm$1.8)$\times 10^{32}$ & (63.7$\pm$3.3)$\times 10^{18}$ \\
 &  &  &  7.7$\pm$1.5 & (20.0$\pm$2.2)$\times 10^{32}$ & (48.6$\pm$3.6)$\times 10^{18}$ \\
 &  &  &  8.4$\pm$1.8 & (12.9$\pm$1.3)$\times 10^{32}$ & (41.8$\pm$2.5)$\times 10^{18}$ \\
 &  &  &  9.3$\pm$2.1 & (17.0$\pm$3.0)$\times 10^{32}$ & (23.1$\pm$1.2)$\times 10^{18}$ \\
 &  &  &  &  &  \\
110680553 & G5II &  6.567$\pm$0.002 &  6.13$\pm$1.00 & (37.1$\pm$5.4)$\times 10^{33}$ & (89.6$\pm$7.0)$\times 10^{18}$ \\
 &  &  &  &  &  \\
104150155 & G5III &  6.867$\pm$0.002 &  4.89$\pm$0.74 & (61.0$\pm$9.3)$\times 10^{34}$ & (57.1$\pm$5.6)$\times 10^{20}$ \\
 &  &  &  5.15$\pm$0.80 & (18.0$\pm$3.0)$\times 10^{34}$ & (34.9$\pm$3.1)$\times 10^{20}$ \\
 &  &  &  &  &  \\
102615551 & G8III &  3.8771$\pm$0.0004$^{a\dagger}$ &  6.16$\pm$0.91 & (11.5$\pm$2.1)$\times 10^{32}$ & (8.1$\pm$1.2)$\times 10^{19}$ \\
 &  &  &  7.9$\pm$1.4 & (33.6$\pm$5.2)$\times 10^{32}$ & (37.1$\pm$5.2)$\times 10^{18}$ \\
 &  &  &  8.7$\pm$1.8 & (20.5$\pm$3.7)$\times 10^{32}$ & (40.7$\pm$6.0)$\times 10^{18}$ \\
 &  &  &  8.8$\pm$1.8 & (16.6$\pm$2.6)$\times 10^{32}$ & (39.9$\pm$5.6)$\times 10^{18}$ \\
 &  &  &  9.7$\pm$2.1$^\ast$ & (20.2$\pm$3.1)$\times 10^{32}$ & (26.6$\pm$3.8)$\times 10^{18}$ \\
 &  &  &  9.8$\pm$2.2$^\ast$ & (14.1$\pm$2.5)$\times 10^{32}$ & (35.1$\pm$5.7)$\times 10^{18}$ \\
 &  &  & 11.1$\pm$2.8$^\ast$ & (8.0$\pm$2.2)$\times 10^{32}$ & (15.2$\pm$3.0)$\times 10^{18}$ \\
 &  &  & 12.3$\pm$3.3$^\ast$ & (12.9$\pm$8.7)$\times 10^{32}$ & (2.3$\pm$1.1)$\times 10^{19}$ \\
 &  &  &  &  &  \\
110662866 & G8III &  2.959$\pm$0.001 &  5.45$\pm$0.70 & (19.6$\pm$4.2)$\times 10^{34}$ & (13.8$\pm$2.2)$\times 10^{20}$ \\
 &  &  &  &  &  \\
102590771 & G8V &  3.71$\pm$0.03 &  3.57$\pm$0.35 & (9.6$\pm$1.6)$\times 10^{34}$ & (8.7$\pm$1.2)$\times 10^{21}$ \\
 &  &  &  4.88$\pm$0.54 & (7.0$\pm$1.5)$\times 10^{34}$ & (12.3$\pm$2.0)$\times 10^{20}$ \\
 &  &  &  8.7$\pm$1.5 & (3.5$\pm$1.5)$\times 10^{34}$ & (12.8$\pm$5.0)$\times 10^{19}$ \\
 &  &  &  &  &  \\
223961132 & G8V &  3.829$\pm$0.005$^{t}$ &  4.59$\pm$0.58 & (9.5$\pm$3.0)$\times 10^{35}$ & (9.4$\pm$2.8)$\times 10^{22}$ \\
 &  &  &  4.90$\pm$0.50 & (60.4$\pm$8.6)$\times 10^{34}$ & (17.7$\pm$2.2)$\times 10^{21}$ \\
 &  &  &  5.75$\pm$0.65 & (10.0$\pm$1.3)$\times 10^{35}$ & (16.9$\pm$2.1)$\times 10^{21}$ \\
 &  &  &  6.75$\pm$0.89 & (44.7$\pm$5.8)$\times 10^{34}$ & (74.8$\pm$9.3)$\times 10^{20}$ \\
 &  &  &  7.9$\pm$1.4 & (21.3$\pm$3.1)$\times 10^{34}$ & (25.3$\pm$3.2)$\times 10^{20}$ \\
 &  &  &  9.8$\pm$2.0$^\ast$ & (4.9$\pm$1.2)$\times 10^{34}$ & (12.5$\pm$2.3)$\times 10^{20}$ \\
 &  &  & 11.3$\pm$2.6$^\ast$ & (4.2$\pm$1.3)$\times 10^{34}$ & (10.6$\pm$2.6)$\times 10^{20}$ \\
 &  &  & 17.2$\pm$6.2$^\ast$ & (45.9$\pm$8.2)$\times 10^{33}$ & (33.1$\pm$4.8)$\times 10^{19}$ \\
 &  &  &  &  &  \\
110681935 & K0III &  4.33$\pm$0.10 &  7.1$\pm$1.1 & (5.5$\pm$1.9)$\times 10^{34}$ & (4.2$\pm$1.2)$\times 10^{20}$ \\
 &  &  &  7.3$\pm$1.2 & (30.8$\pm$7.3)$\times 10^{33}$ & (11.1$\pm$1.5)$\times 10^{19}$ \\
 &  &  & 13.1$\pm$3.6 & (3.6$\pm$1.3)$\times 10^{34}$ & (4.0$\pm$1.3)$\times 10^{19}$ \\
 &  &  &  &  &  \\
110747714 & K0III &  4.775$\pm$0.006 &  5.71$\pm$0.75 & (18.5$\pm$5.5)$\times 10^{34}$ & (13.8$\pm$3.5)$\times 10^{20}$ \\
 &  &  &  6.51$\pm$0.94 & (7.9$\pm$1.1)$\times 10^{34}$ & (33.0$\pm$2.4)$\times 10^{19}$ \\
 &  &  &  &  &  \\
221647275 & K0III &  3.732$\pm$0.001 &  5.11$\pm$0.66 & (34.5$\pm$3.3)$\times 10^{32}$ & (52.6$\pm$3.6)$\times 10^{19}$ \\
 &  &  &  5.69$\pm$0.77 & (30.0$\pm$1.8)$\times 10^{32}$ & (21.9$\pm$1.1)$\times 10^{19}$ \\
 &  &  &  7.3$\pm$1.2 & (35.9$\pm$1.6)$\times 10^{32}$ & (72.6$\pm$2.4)$\times 10^{18}$ \\
 &  &  &  8.2$\pm$1.4 & (29.9$\pm$3.0)$\times 10^{32}$ & (59.4$\pm$3.0)$\times 10^{18}$ \\
 &  &  &  &  &  \\
102718810 & K0III &  1.86$\pm$0.03$^{b}$ &  4.93$\pm$0.61 & (42.1$\pm$4.4)$\times 10^{32}$ & (88.0$\pm$5.4)$\times 10^{19}$ \\
 &  &  &  5.86$\pm$0.79 & (31.4$\pm$1.5)$\times 10^{33}$ & (81.6$\pm$3.1)$\times 10^{19}$ \\
 &  &  &  6.9$\pm$1.1 & (26.8$\pm$1.2)$\times 10^{33}$ & (43.5$\pm$1.2)$\times 10^{19}$ \\
 &  &  &  7.5$\pm$1.2 & (60.4$\pm$4.1)$\times 10^{32}$ & (21.2$\pm$1.0)$\times 10^{19}$ \\
 &  &  &  8.0$\pm$1.3 & (40.6$\pm$7.1)$\times 10^{32}$ & (17.9$\pm$2.4)$\times 10^{19}$ \\
 &  &  &  8.3$\pm$1.4 & (28.9$\pm$4.2)$\times 10^{32}$ & (25.4$\pm$2.9)$\times 10^{19}$ \\
 &  &  &  8.4$\pm$1.5 & (157.6$\pm$9.5)$\times 10^{32}$ & (206.1$\pm$7.6)$\times 10^{18}$ \\
 &  &  &  8.7$\pm$1.5 & (44.4$\pm$2.0)$\times 10^{33}$ & (281.8$\pm$9.4)$\times 10^{18}$ \\
 &  &  &  8.9$\pm$1.6 & (101.1$\pm$7.1)$\times 10^{32}$ & (18.8$\pm$1.0)$\times 10^{19}$ \\
 &  &  &  9.1$\pm$1.7 & (43.6$\pm$7.0)$\times 10^{32}$ & (14.4$\pm$1.0)$\times 10^{19}$ \\
 &  &  &  9.5$\pm$1.8 & (154.9$\pm$9.3)$\times 10^{32}$ & (120.5$\pm$4.0)$\times 10^{18}$ \\
 &  &  & 10.0$\pm$2.0 & (109.1$\pm$6.8)$\times 10^{32}$ & (81.2$\pm$3.6)$\times 10^{18}$ \\
 &  &  & 10.6$\pm$2.3 & (109.4$\pm$8.8)$\times 10^{32}$ & (82.4$\pm$3.7)$\times 10^{18}$ \\
 &  &  & 10.9$\pm$2.4 & (76.8$\pm$7.0)$\times 10^{32}$ & (107.9$\pm$6.2)$\times 10^{18}$ \\
 &  &  & 14.6$\pm$3.9 & (5.3$\pm$2.2)$\times 10^{33}$ & (34.6$\pm$5.4)$\times 10^{18}$ \\
 &  &  &  &  &  \\
223968260 & K0V &  1.645$\pm$0.001 &  5.26$\pm$0.70 & (11.7$\pm$3.5)$\times 10^{34}$ & (18.3$\pm$3.1)$\times 10^{20}$ \\
 &  &  &  &  &  \\
110660270 & K0V &  1.969$\pm$0.001 &  5.77$\pm$0.74 & (32.1$\pm$2.7)$\times 10^{34}$ & (110.9$\pm$8.3)$\times 10^{20}$ \\
 &  &  &  &  &  \\
110757572 & K0V &  6.676$\pm$0.002 &  4.60$\pm$0.51 & (223.2$\pm$9.1)$\times 10^{32}$ & (35.7$\pm$1.2)$\times 10^{20}$ \\
 &  &  &  4.70$\pm$0.53 & (24.1$\pm$1.4)$\times 10^{33}$ & (35.2$\pm$1.3)$\times 10^{20}$ \\
 &  &  &  4.77$\pm$0.54 & (101.5$\pm$7.5)$\times 10^{32}$ & (167.2$\pm$7.5)$\times 10^{19}$ \\
 &  &  &  5.12$\pm$0.60 & (114.8$\pm$8.6)$\times 10^{32}$ & (152.9$\pm$6.9)$\times 10^{19}$ \\
 &  &  &  5.42$\pm$0.65 & (36.1$\pm$2.2)$\times 10^{33}$ & (188.3$\pm$7.1)$\times 10^{19}$ \\
 &  &  &  5.87$\pm$0.75 & (27.0$\pm$2.2)$\times 10^{33}$ & (73.4$\pm$3.8)$\times 10^{19}$ \\
 &  &  &  6.86$\pm$0.99 & (31.3$\pm$1.6)$\times 10^{33}$ & (45.3$\pm$1.6)$\times 10^{19}$ \\
 &  &  &  7.0$\pm$1.0 & (21.4$\pm$1.3)$\times 10^{33}$ & (82.6$\pm$3.9)$\times 10^{19}$ \\
 &  &  &  7.9$\pm$1.3 & (17.1$\pm$1.7)$\times 10^{33}$ & (28.8$\pm$1.4)$\times 10^{19}$ \\
 &  &  &  &  &  \\
211607099 & K0V &  0.974$\pm$0.001 & 10.3$\pm$2.3 & (4.8$\pm$1.1)$\times 10^{34}$ & (7.4$\pm$1.0)$\times 10^{19}$ \\
 &  &  &  &  &  \\
101282630 & K1III & 13.866$\pm$0.004 &  4.15$\pm$0.48 & (95.8$\pm$9.5)$\times 10^{34}$ & (27.2$\pm$1.9)$\times 10^{21}$ \\
 &  &  &  4.41$\pm$0.52 & (137.2$\pm$9.2)$\times 10^{34}$ & (41.0$\pm$2.5)$\times 10^{21}$ \\
 &  &  &  &  &  \\
104939145 & K1III & 26$\pm$6 &  5.96$\pm$0.77 & (28.6$\pm$4.8)$\times 10^{34}$ & (52.9$\pm$8.1)$\times 10^{21}$ \\
 &  &  &  &  &  \\
110661118 & K1III &  4.024$\pm$0.002 &  5.02$\pm$0.59 & (5.7$\pm$1.9)$\times 10^{34}$ & (8.3$\pm$1.3)$\times 10^{20}$ \\
 &  &  &  &  &  \\
223988965 & K1III &  3.4$\pm$0.5$^{t}$ &  5.18$\pm$0.73 & (10.9$\pm$6.8)$\times 10^{35}$ & (7.0$\pm$4.3)$\times 10^{21}$ \\
 &  &  &  &  &  \\
101309841 & K2III &  8.587$\pm$0.007 &  4.42$\pm$0.62 & (11.7$\pm$1.3)$\times 10^{35}$ & (76.1$\pm$4.9)$\times 10^{20}$ \\
 &  &  &  4.87$\pm$0.73 & (94.0$\pm$8.8)$\times 10^{34}$ & (44.3$\pm$2.4)$\times 10^{20}$ \\
 &  &  & 12.6$\pm$3.6 & (45.9$\pm$3.7)$\times 10^{34}$ & (16.0$\pm$1.0)$\times 10^{19}$ \\
 &  &  &  &  &  \\
105271299 & K2III & 10.48$\pm$0.03 &  3.93$\pm$0.52 & (14.1$\pm$2.5)$\times 10^{34}$ & (71.9$\pm$7.2)$\times 10^{20}$ \\
 &  &  &  5.07$\pm$0.75 & (20.5$\pm$6.8)$\times 10^{34}$ & (4.9$\pm$1.4)$\times 10^{21}$ \\
 &  &  &  &  &  \\
110669882 & K2III &  8.052$\pm$0.003 &  4.58$\pm$0.66 & (102.0$\pm$9.3)$\times 10^{34}$ & (15.3$\pm$1.0)$\times 10^{21}$ \\
 &  &  &  4.68$\pm$0.68 & (78.8$\pm$7.2)$\times 10^{34}$ & (110.9$\pm$6.9)$\times 10^{20}$ \\
 &  &  &  6.0$\pm$1.0 & (110.0$\pm$5.7)$\times 10^{34}$ & (79.2$\pm$3.4)$\times 10^{20}$ \\
 &  &  &  &  &  \\
110741064 & K2III & --- &  6.7$\pm$1.2 & (40.0$\pm$3.6)$\times 10^{32}$ & (36.8$\pm$2.1)$\times 10^{19}$ \\
 &  &  &  &  &  \\
110752597 & K2III &  4.216$\pm$0.002 &  4.05$\pm$0.57 & (42.5$\pm$8.7)$\times 10^{33}$ & (11.5$\pm$1.1)$\times 10^{20}$ \\
 &  &  &  &  &  \\
105132247 & K2V &  4.75$\pm$0.05 &  7.1$\pm$1.3 & (4.9$\pm$1.3)$\times 10^{34}$ & (39.0$\pm$4.0)$\times 10^{20}$ \\
 &  &  &  &  &  \\
211644521 & K2V &  4.25$\pm$0.02 &  5.43$\pm$0.84 & (10.9$\pm$1.1)$\times 10^{33}$ & ( 96.9$\pm$6.3)$\times 10^{19}$ \\
 &  &  &  &  &  \\
105628093 & K2V &  0.5073$\pm$0.0008 &  4.38$\pm$0.59 & (9.3$\pm$1.5)$\times 10^{33}$ & (24.6$\pm$3.3)$\times 10^{20}$ \\
 &  &  &  &  &  \\
300002990 & K2V &  6.56$\pm$0.03 &  4.63$\pm$0.66 & (49.3$\pm$8.2)$\times 10^{33}$ & (22.5$\pm$2.6)$\times 10^{20}$ \\
 &  &  &  4.70$\pm$0.67 & (12.8$\pm$1.2)$\times 10^{34}$ & (119.8$\pm$6.8)$\times 10^{19}$ \\
 &  &  &  5.23$\pm$0.79 & (5.4$\pm$1.1)$\times 10^{34}$ & (91.5$\pm$7.9)$\times 10^{19}$ \\
 &  &  &  6.7$\pm$1.2 & (9.4$\pm$1.5)$\times 10^{34}$ & (36.5$\pm$3.7)$\times 10^{19}$ \\
 &  &  &  8.0$\pm$1.6 & (14.6$\pm$1.2)$\times 10^{34}$ & (174.9$\pm$9.1)$\times 10^{18}$ \\
 &  &  &  &  &  \\
500007248 & K3I &  3.78$\pm$0.02$^{t}$ &  5.57$\pm$0.85 & (32.4$\pm$7.6)$\times 10^{33}$ & (51.8$\pm$8.0)$\times 10^{19}$ \\
 &  &  &  &  &  \\
102646977 & K3II & 20.8$\pm$0.3 &  9.5$\pm$2.1 & (4.4$\pm$1.0)$\times 10^{35}$ & (12.5$\pm$1.9)$\times 10^{21}$ \\
 &  &  & 10.5$\pm$2.5 & (27.5$\pm$5.4)$\times 10^{34}$ & (9.5$\pm$1.5)$\times 10^{21}$ \\
 &  &  & 10.8$\pm$2.7 & (36.4$\pm$7.1)$\times 10^{34}$ & (7.4$\pm$1.2)$\times 10^{21}$ \\
 &  &  &  &  &  \\
500007323 & K3II &  6$\pm$1 &  7.2$\pm$1.4 & (22.0$\pm$2.4)$\times 10^{34}$ & (15.1$\pm$1.4)$\times 10^{19}$ \\
 &  &  &  &  &  \\
102379742 & K3III &  0.4231$\pm$0.0008 &  3.90$\pm$0.51 & (1.7$\pm$1.1)$\times 10^{32}$ & (5.5$\pm$3.8)$\times 10^{19}$ \\
 &  &  &  4.27$\pm$0.57 & (5.2$\pm$3.6)$\times 10^{33}$ & (11.8$\pm$8.1)$\times 10^{20}$ \\
 &  &  &  4.96$\pm$0.71 & (2.2$\pm$1.6)$\times 10^{33}$ & (5.2$\pm$3.6)$\times 10^{20}$ \\
 &  &  &  5.11$\pm$0.74 & (2.1$\pm$1.4)$\times 10^{34}$ & (12.5$\pm$8.6)$\times 10^{20}$ \\
 &  &  &  5.18$\pm$0.76 & (10.8$\pm$7.4)$\times 10^{33}$ & (8.8$\pm$6.0)$\times 10^{20}$ \\
 &  &  &  5.33$\pm$0.79 & (8.8$\pm$6.4)$\times 10^{33}$ & (8.0$\pm$5.6)$\times 10^{20}$ \\
 &  &  &  5.72$\pm$0.88 & (5.2$\pm$3.6)$\times 10^{33}$ & (3.2$\pm$2.2)$\times 10^{20}$ \\
 &  &  &  5.76$\pm$0.89 & (7.2$\pm$5.1)$\times 10^{33}$ & (6.1$\pm$4.3)$\times 10^{20}$ \\
 &  &  &  5.89$\pm$0.93 & (8.0$\pm$5.5)$\times 10^{33}$ & (5.9$\pm$4.0)$\times 10^{20}$ \\
 &  &  &  6.2$\pm$1.0 & (6.6$\pm$4.5)$\times 10^{33}$ & (3.3$\pm$2.3)$\times 10^{20}$ \\
 &  &  &  6.4$\pm$1.1 & (2.4$\pm$1.7)$\times 10^{33}$ & (2.5$\pm$1.7)$\times 10^{20}$ \\
 &  &  &  6.6$\pm$1.1 & (9.5$\pm$6.6)$\times 10^{33}$ & (2.2$\pm$1.5)$\times 10^{20}$ \\
 &  &  &  6.6$\pm$1.1 & (12.8$\pm$8.8)$\times 10^{33}$ & (3.3$\pm$2.2)$\times 10^{20}$ \\
 &  &  &  7.1$\pm$1.3 & (4.1$\pm$3.1)$\times 10^{33}$ & (12.9$\pm$9.0)$\times 10^{19}$ \\
 &  &  &  7.3$\pm$1.3 & (12.5$\pm$8.9)$\times 10^{32}$ & (1.7$\pm$1.2)$\times 10^{20}$ \\
 &  &  &  &  &  \\
102646279 & K3III & 10.52$\pm$0.04 & 10.2$\pm$2.6 & (15.5$\pm$1.7)$\times 10^{35}$ & (20.0$\pm$1.6)$\times 10^{21}$ \\
 &  &  & 12.1$\pm$3.6$^\ast$ & (9.3$\pm$1.2)$\times 10^{35}$ & (12.5$\pm$1.1)$\times 10^{21}$ \\
 &  &  &  &  &  \\
102691871 & K3III &  7.138$\pm$0.004 &  5.9$\pm$1.0 & (23.1$\pm$2.0)$\times 10^{35}$ & (91.4$\pm$6.3)$\times 10^{20}$ \\
 &  &  &  &  &  \\
101562259 & K5III &  7.4867$\pm$0.0008 &  4.43$\pm$0.77 & (133.7$\pm$9.5)$\times 10^{33}$ & (94.8$\pm$5.0)$\times 10^{20}$ \\
 &  &  &  4.75$\pm$0.85 & (42.0$\pm$3.2)$\times 10^{33}$ & (36.9$\pm$1.9)$\times 10^{20}$ \\
 &  &  &  4.84$\pm$0.87 & (37.4$\pm$1.3)$\times 10^{34}$ & (151.2$\pm$5.6)$\times 10^{20}$ \\
 &  &  &  5.6$\pm$1.1 & (54.5$\pm$4.9)$\times 10^{33}$ & (21.1$\pm$1.1)$\times 10^{20}$ \\
 &  &  &  6.5$\pm$1.4 & (14.7$\pm$3.4)$\times 10^{33}$ & (13.7$\pm$1.5)$\times 10^{20}$ \\
 &  &  &  &  &  \\
102602133 & K5III & 21.3$\pm$0.1 &  3.70$\pm$0.62 & (93.1$\pm$7.1)$\times 10^{35}$ & (17.5$\pm$1.1)$\times 10^{22}$ \\
 &  &  &  5.5$\pm$1.2$^\ast$ & (24.3$\pm$4.4)$\times 10^{34}$ & (11.0$\pm$1.7)$\times 10^{22}$ \\
 &  &  &  5.8$\pm$1.3$^\ast$ & (36.0$\pm$5.2)$\times 10^{34}$ & (55.7$\pm$6.5)$\times 10^{21}$ \\
 &  &  &  6.5$\pm$1.7$^\ast$ & (44.7$\pm$7.2)$\times 10^{34}$ & (35.7$\pm$4.4)$\times 10^{21}$ \\
 &  &  &  &  &  \\
106048015 & K5III & 15.430$\pm$0.009 &  3.19$\pm$0.61 & (22.5$\pm$3.9)$\times 10^{34}$ & (29.8$\pm$2.5)$\times 10^{21}$ \\
 &  &  &  3.35$\pm$0.65 & (15.9$\pm$5.6)$\times 10^{34}$ & (4.0$\pm$1.2)$\times 10^{22}$ \\
 &  &  &  3.41$\pm$0.67 & (13.1$\pm$2.8)$\times 10^{34}$ & (16.5$\pm$2.3)$\times 10^{21}$ \\
 &  &  &  3.53$\pm$0.70 & (10.2$\pm$1.8)$\times 10^{34}$ & (17.8$\pm$1.9)$\times 10^{21}$ \\
 &  &  &  3.77$\pm$0.77 & (20.3$\pm$2.7)$\times 10^{34}$ & (17.2$\pm$2.0)$\times 10^{21}$ \\
 &  &  &  4.20$\pm$0.90 & (24.1$\pm$2.9)$\times 10^{34}$ & (66.6$\pm$6.4)$\times 10^{20}$ \\
 &  &  &  4.33$\pm$0.95 & (34.3$\pm$4.9)$\times 10^{34}$ & (84.3$\pm$7.8)$\times 10^{20}$ \\
 &  &  &  4.6$\pm$1.0 & (15.1$\pm$2.4)$\times 10^{34}$ & (27.2$\pm$2.9)$\times 10^{20}$ \\
 &  &  &  5.0$\pm$1.2 & (65.7$\pm$8.8)$\times 10^{34}$ & (42.1$\pm$4.2)$\times 10^{20}$ \\
 &  &  &  &  &  \\
  \enddata
\end{deluxetable*}
\end{center}
}


\section{Calibration}

The accuracy of Gaia photometry is better than any other major catalog currently available,
even more with the improvements
from Gaia Early Data Release 3 - Gaia EDR3 \citep{riello2021}.
We used the fact that the GAIA G band is very similar to the CoRoT White channel \citep[see, for example, Fig.1 in][]{nemec2020} 
to calibrate the analyzed flare flux.

\begin{figure}
	\includegraphics[width=\columnwidth]{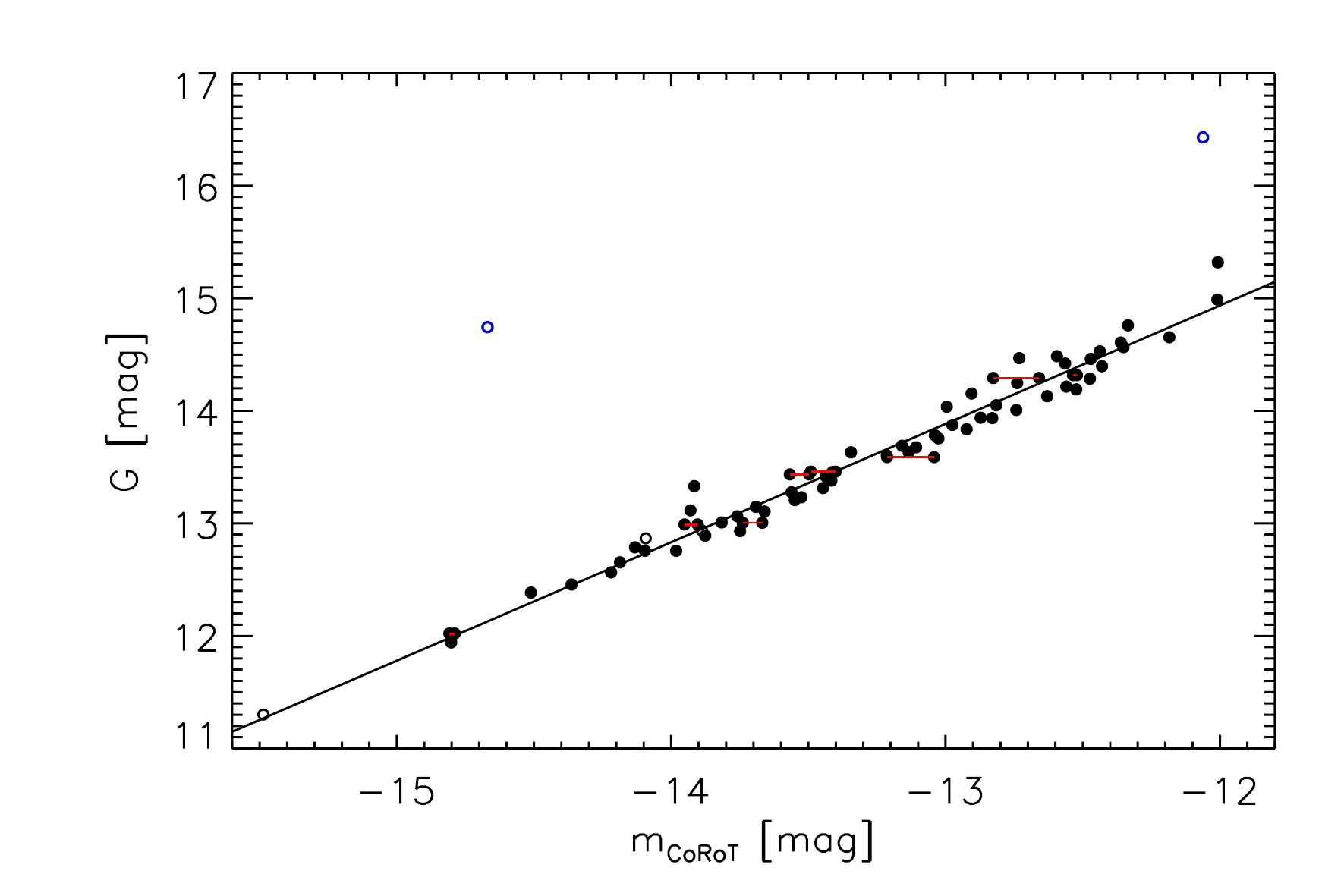}
    \caption{Observed GAIA G magnitude as a function of CoRoT White channel magnitude for all analyzed stars. 
    The three empty black circles are more than 1 arcsec from the nearest star in the GAIA EDR3 catalog.
    One of the three has $m_{\rm CoRoT}$ = -13.9 mag and is not easily visible in the figure.
    The two empty blue circles are more than 4$\sigma$ away from the fitted straight line and were also removed from the fitting.
    The small red horizontal lines link the CoRoT magnitudes of stars that have been observed twice.
    }
    \label{fig:gaia}
\end{figure}

\begin{figure}
	\includegraphics[width=\columnwidth]{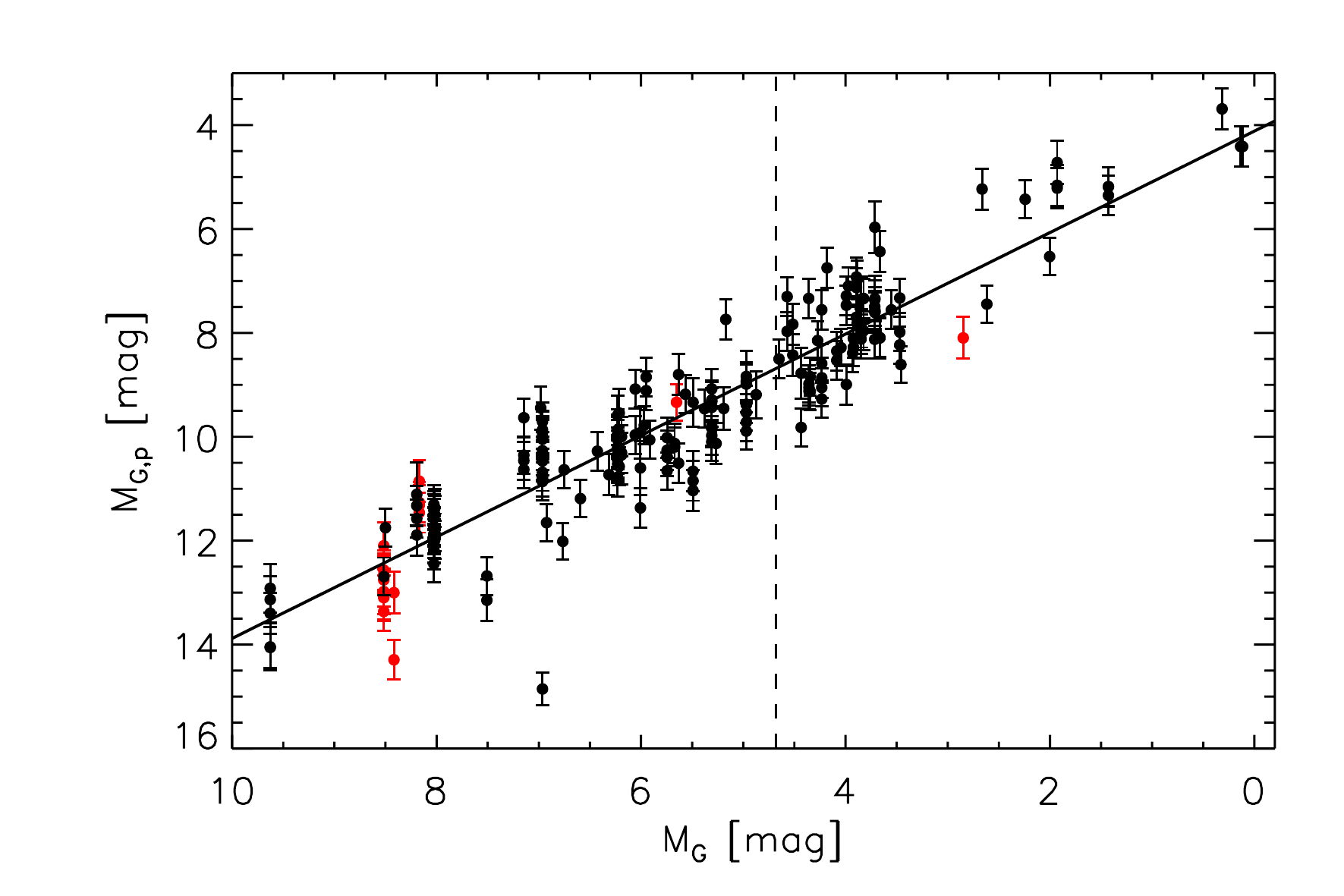}
    \caption{
    Absolute magnitude of the flare peak as a function of absolute stellar magnitude.
    Red dots correspond to flares in stars with an unreliable match to a star in the Gaia EDR3 catalog (i.e., empty circles in Fig.~\ref{fig:gaia}) and were not used in the fitting (full line).
    The vertical line shows the solar absolute magnitude estimated for the G band, $M_{G,\odot}$=4.68 mag \citep{andrae2018}.
    }
    \label{fig:abs_mag}
\end{figure}

Firstly, we searched the Gaia EDR3 for stars 
at the same coordinates ($\alpha$, $\delta$) as ours. 
We compared their coordinates differences 
and iteratively removed outliers. 
We ended up with a standard deviation of 0.15 arcsec in
$\alpha$ and $\delta$.
We selected only stars of Gaia that differ by less than 0.6 arcsec (4$\sigma$) in $\alpha$ and $\delta$
and its angular distance 
is less than 0.7 arcsec from the CoRoT
coordinates, we found 66 out of 69 stars.
The stars observed by Gaia that are closer to
the three missing ones
(CoRoT 102948867, 110685010 and 221647275)
are 1.3, 2.1, and 2.7 arcsec away, respectively.
Table \ref{tab:gaiaid} shows the corresponding Gaia ID EDR3.
In Figure~\ref{fig:gaia}, we plotted GAIA G-band mean apparent magnitude (phot$\_$g$\_$mean$\_$mag),
which is computed from the G-band mean flux applying the magnitude zero-point in the Vega scale, in the VEGAMAG  photometric system
\citep[equation 1 in][]{andrae2018}
versus
the CoRoT magnitude in the White channel given by: -2.5 log$_{10}$ F$_{\rm white}$.
The CoRoT flux is in electrons per 32\,s \citep{chaintreuil2016}.
%
After fitting a straight line to 66 stars and removing outliers from the residuals 
larger than 4$\sigma$ (empty blue circles), 
we fitted again a straight line through the selected full circles (corresponding to 64 stars) and obtained:
\begin{equation}
G = c_0 + c_1 \cdot m_{\rm CoRoT}
\label{eq:G}
\end{equation}
where $c_0$ = (27.6$\pm$0.3) mag and $c_1$ = 1.05$\pm$0.02.
The two discarded outliers, CoRoT 500007248 and 500007323,
are fainter than expected in the White channel, which
may be due to an observational artifact or misidentification due to field contamination. Each of these stars contributes with only one flare in our analysis.
There are eight stars in our set that have been observed twice (in 2008 and 2012).
Their two CoRoT magnitudes are connected with a red horizontal line in Figure~\ref{fig:gaia}.
The flare peak flux, $a_0$, observed in the White channel (Eq.~\ref{eq:flarefitting}) can be converted to 
GAIA G-band magnitude using Eq.~\ref{eq:G}:
\begin{equation}
   G_p = c_0 - 2.5 c_1 \log_{10}(a_0) 
   \label{eq:Gp}
\end{equation}
The uncertainty in estimating the G-band flare peak
was obtained using standard error propagation.
There is no perceptible dependence of the G-band flare peak, $G_p$, on the stellar magnitude, $G$.
Using Eq.~\ref{eq:G}, 
the average apparent stellar magnitude of our 69 stars and its standard deviation are equal to
G = (13.6$\pm$0.8) mag.
However, their distances, $d$, range from 37\,pc to 6.2\,kpc.
These stars are all in the galactic disk,
as the CoRoT observations were in an area toward the Galactic center and another at the Galactic anticenter.
 We calculated the absolute stellar magnitude $M_G = G - 5 \log_{10} d + 5$
 and similarly the absolute magnitude of the flare peak, $M_{G,p}$, without taking the extinction into account.
The flare peak amplitudes are in the range of $4\,<\,M_{G,p}\,<\,14$ mag. 
 There is a good correlation between them (Figure~\ref{fig:abs_mag}). 
 Fitting a straight line, we obtained:
 \begin{equation}
 M_{G,p} = (4.12\pm0.09) + (0.98\pm0.01) M_G
 \end{equation}
 For a linear coefficient equal to 1, we have that
the peak flare flux is $\sim$2.5\% of the stellar flux
in the Gaia G band.

\begin{table*}
\begin{center}
 \caption{
EDR3 Gaia identification used for each analyzed CoRoT star.
The EDR3 Gaia observation corresponding to CoRoT 102379742
does not have an estimated parallax, and has been replaced by Gaia DR2 (labeled `R2').
The three EDR3 Gaia observations that are 
more than one arcsec away from the CoRoT target are marked with a $d$ and
the two that are fainter than expected are indicated by an $F$.
}
\label{tab:gaiaid}
 \begin{tabular}{llRlRlRlRl}
 CoRoT ID & Gaia ID & CoRoT ID & Gaia ID & CoRoT ID & Gaia ID \\
\hline
101282630 & 4263419769100038016 & 104228454 & 4479364776186624640 & 211644521 & 4269234708175942784 \\
101309841 & 4287737908286241408 & 104939145 & 4284794275144637824 & 221628176 & 3323880005037073280 \\
101562259 & 4287508488312311168 & 105085606 & 4283813132812345856 & 221647275 & 3324510948616275200$^d$ \\
102326330 & 3317450576432937984 & 105132247 & 4284031866916578688 & 221649345 & 3323856709134836480 \\
102332965 & 3318597019166336896 & 105271299 & 4286729793567044352 & 223940041 & 3326099953372611968 \\
102379742 & 3317374843273655040$^{R2}$ & 105628093 & 4285482809953521920 & 223941972 & 3326973098749988736 \\
102590771 & 3107330708913023488 & 106048015 & 4285502772956212480 & 223945488 & 3133881406461641088 \\
102602133 & 3119470622952565504 & 106054338 & 4285618908854461184 & 223959652 & 3326704852270253056 \\
102615551 & 3119353524964187520 & 110660270 & 3101935229255761408 & 223961132 & 3326895724912475648 \\
102631863 & 3107228694850611968 & 110661118 & 3102183104708835328 & 223968260 & 3133861271655267072 \\
102632674 & 3125806455627199104 & 110662866 & 3101991922824716416 & 223970694 & 3326703512240456064 \\
102646279 & 3107323179828911104 & 110669882 & 3102084560979475840 & 223980621 & 3326929427518677504 \\
102646977 & 3107338577293220480 & 110680553 & 3102338273286874240 & 223983509 & 3326931699558507648 \\
102691871 & 3119514878295078144 & 110681935 & 3101134647354726272 & 223984608 & 3326685443313414144 \\
102715243 & 3107267791930670976 & 110685010 & 3101901526654756864$^d$ & 223988965 & 3326697984617869824 \\
102718810 & 3107362079353715584 & 110741064 & 3102360362299563520 & 223993277 & 3326722276952860672 \\
102854684 & 3106096674609281920 & 110744989 & 3101876478405414016 & 224008170 & 3326640844372957824 \\
102859855 & 3106107330427498752 & 110747714 & 3101862833286509568 & 300002990 & 3102196814244084352 \\
102899501 & 3106093827050203520 & 110752597 & 3102390912405546880 & 315258285 & 3120331437476706688 \\
102925086 & 3106193569071435264 & 110757572 & 3101907569665841536 & 500007157 & 3326739903498311808 \\
102926046 & 3106292422038384384 & 110773079 & 3102022880948173056 & 500007248 & 3326739869138571776$^F$ \\
102948867 & 3105880590514179456$^d$ & 211607099 & 4280865410836745984 & 500007323 & 3326895342663197056$^F$ \\
104150155 & 4478967135257757056 & 211607284 & 4268382891546982528 & 604183778 & 3322004650516148992 \\
 \end{tabular}
 \end{center}
\end{table*}

\begin{figure*}
	\includegraphics[width=\textwidth]
	{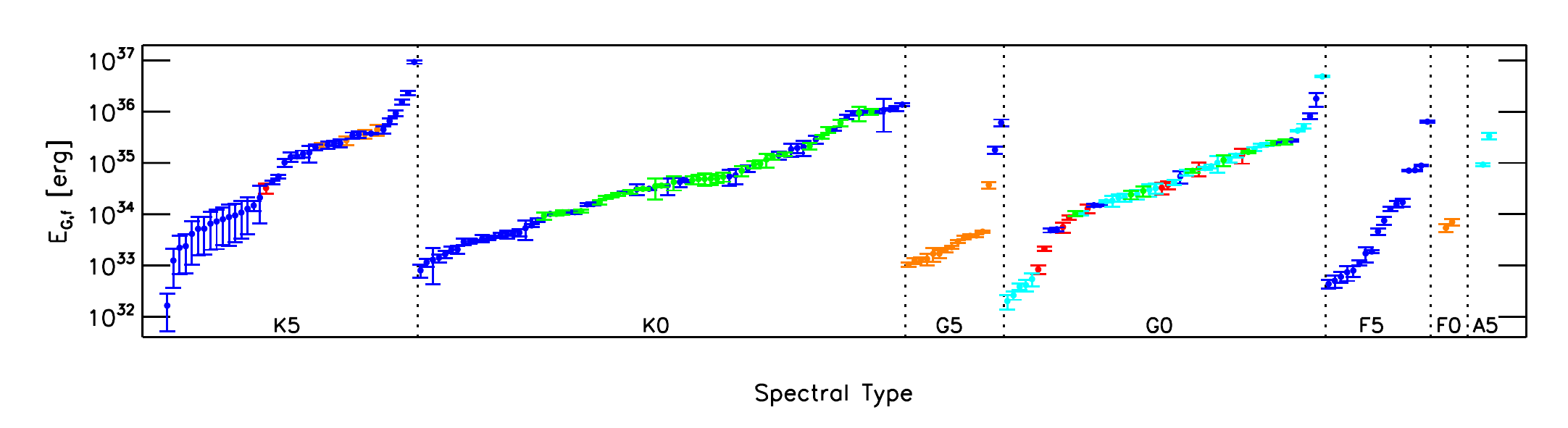}
	\includegraphics[width=\textwidth]
	{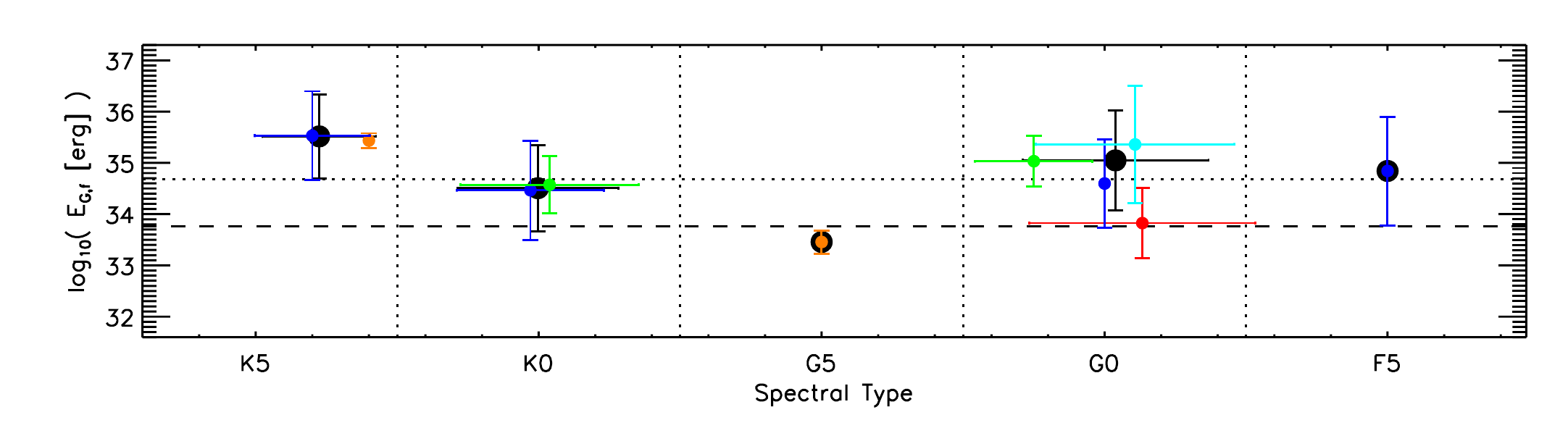}
    \caption{Top panel. Gaia flare energy as function of spectral type, going from late to early type stars,
in intervals of half a spectral class. 
The spectral type at the center of each interval is indicated at the bottom of the plot.
The flare energies are in ascending order inside each interval.
The different colors represent the different luminosity classes, as in Fig.~\ref{fig:tmean},
where classes I to V are in red, orange, blue, cyan and, green respectively.
Bottom panel. Weighted average of the flare energy
as a function of the average spectral subclass corresponding to the flares within each interval of half a spectral class shown in the top panel.  
The averages for a given luminosity class, when there are 4 or more flares, are shown in different colors (as in the top panel). The averages including all luminosity classes are shown in black.
%
The vertical error bars correspond to the standard deviation of the flare energy weighted average
and
the horizontal error bars to the standard deviation of the average of the spectral subclass.
The absence of a horizontal bar means that only one spectral type was used in the energy averaged.
       }
    \label{fig:emean}
\end{figure*}

The Gaia flare energy, $E_{G,f}$, was calculated by multiplying the flare equivalent duration ED (Eq.~\ref{eq:ed}) observed in the White channel, measured in units of time, by the stellar luminosity in the Gaia bandpass filter:
\begin{equation}
    E_{G,f} = {\rm ED}_{\rm White} \cdot L_G 
    \label{eq:egf}
\end{equation}
As the flare equivalent duration varies with the observed bandpass filter by two or more orders of magnitude \citep{hawley2014},
we did not attempt to estimate the bolometric energy.
To calculate the stellar luminosity in Gaia's G-band, $L_G$,
we first converted mean flux $f_G$ (phot$\_$g$\_$mean$\_$flux [$e^- s^{-1}$]) 
into mean energy [W m$^{-2}$ nm$^{-1}$] 
by multiplying it by $c_\lambda$ = 1.346109 $\times 10^{-21}$
as described in the Gaia photometric calibration 
(Montegriffo, 2021\footnote{Gaia Early Data Release 3, Documentation release 1.1, Chapter 5. Photometric Data, subsection 5.4.1. Calibration
(https://gea.esac.esa.int/archive/documentation/GEDR3/index.html)}). 
Then we multiplied the stellar energy by 
4$\pi d^2$ and by Gaia filter FWHM, $\Gamma_G$,
to obtain $L_G$ \citep[as in][]{hawley2014}:
\begin{equation}
L_G = f_G \cdot c_\lambda \cdot 4\pi d^2 \cdot \Gamma_G
\label{eq:lg}
\end{equation}
where $\Gamma_G$ = 454.82\,nm \citep[Table 3 in][]{riello2021}.
The observed flare energies are in the range $10^{32}\,<\, E_{G,f}\,<\,10^{37}$ erg,
and are given in Table~\ref{tab:temperature}.
The energy of sixteen flares observed at five stars, for which we did not find a good match with Gaia observations (labeled in Table~\ref{tab:gaiaid}), was not used in our analysis, and 
their values in the table should be used with caution.
Fitting a straight line to the logarithm of the flare energy as a function of the logarithm of the stellar luminosity, we get: 
 \begin{equation}
E_{G,f} = (4\pm1) \times 10^{-6} \, L_G^{(1.218\pm0.005)}
\label{eq:eg_lg}
 \end{equation}
where $E_{G,f}$ and $L_G$ are in erg and erg s$^{-1}$ respectively.
Flare energy exponentially increases with stellar luminosity.
%
As in Fig.~\ref{fig:tmean} (2nd row),
the top panel in Figure~\ref{fig:emean} shows the flare energy
in intervals of half-spectral type of its star, where the energies are in ascending order inside each range. 
%
The weighted average of the flare energy over each interval is in the bottom panel. 
The dashed (dotted) line indicates the weighted (unweighted) average of the mean energy including all stellar luminosity classes (black circles).
There is some weak indication of a possible increase of flare energy for late-type stars.

\citet{howard2020} reported a weak correlation between the flare temperature, integrated over the entire flare, and flare energy in the g' band.
We do not see any indication of a dependence between our estimation of flare energy in the Gaia G band and flare temperature.
Furthermore, we do not see a variation of the flare energy with the stellar rotation
(Fig.~\ref{fig:rotation}).
We looked for a complex dependence 
using the theoretical estimate by \citet{namekata2017} of 
the bolometric flare energy:
\begin{equation}
E_{\rm bol} = c \, d^3 B^5
\label{eq:namekata}
\end{equation}
where 
$c = 5.8 \times 10^{19}$ erg min$^{-3}$ G$^{-5}$.
In addition to the implication that a flare with a
longer duration ($d$) will have higher energy, they also
deduced, considering the dependence of the Alfvén velocity on the magnetic field, that a stronger local magnetic field $B$
will imply a flare with larger energy.
It is expected that greater amounts of magnetic energy stored, for example, at large spots,
lead to more energetic flares
\citep[e.g.][]{candelaresi2014}.
The coefficient $c$
was estimated from the average values of their solar flare observations, 
which are: $d$ = 3.5 minutes, $E_{bol}$ = 1.5 $\times 10^{30}$\,erg, and $B$=  57\,G.

\begin{figure}
	\includegraphics[scale=0.5]
	{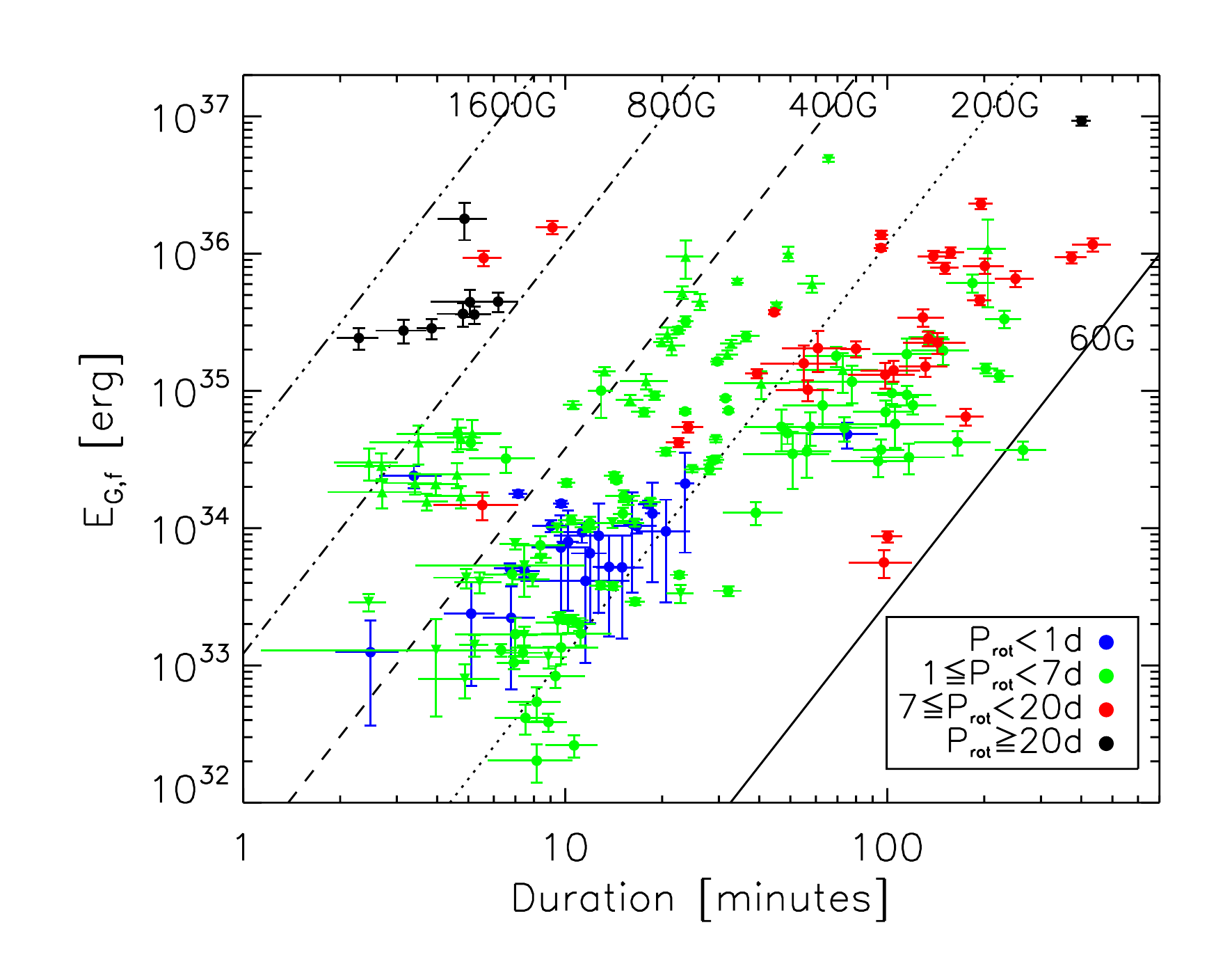}\\
	\includegraphics[scale=0.5]
	{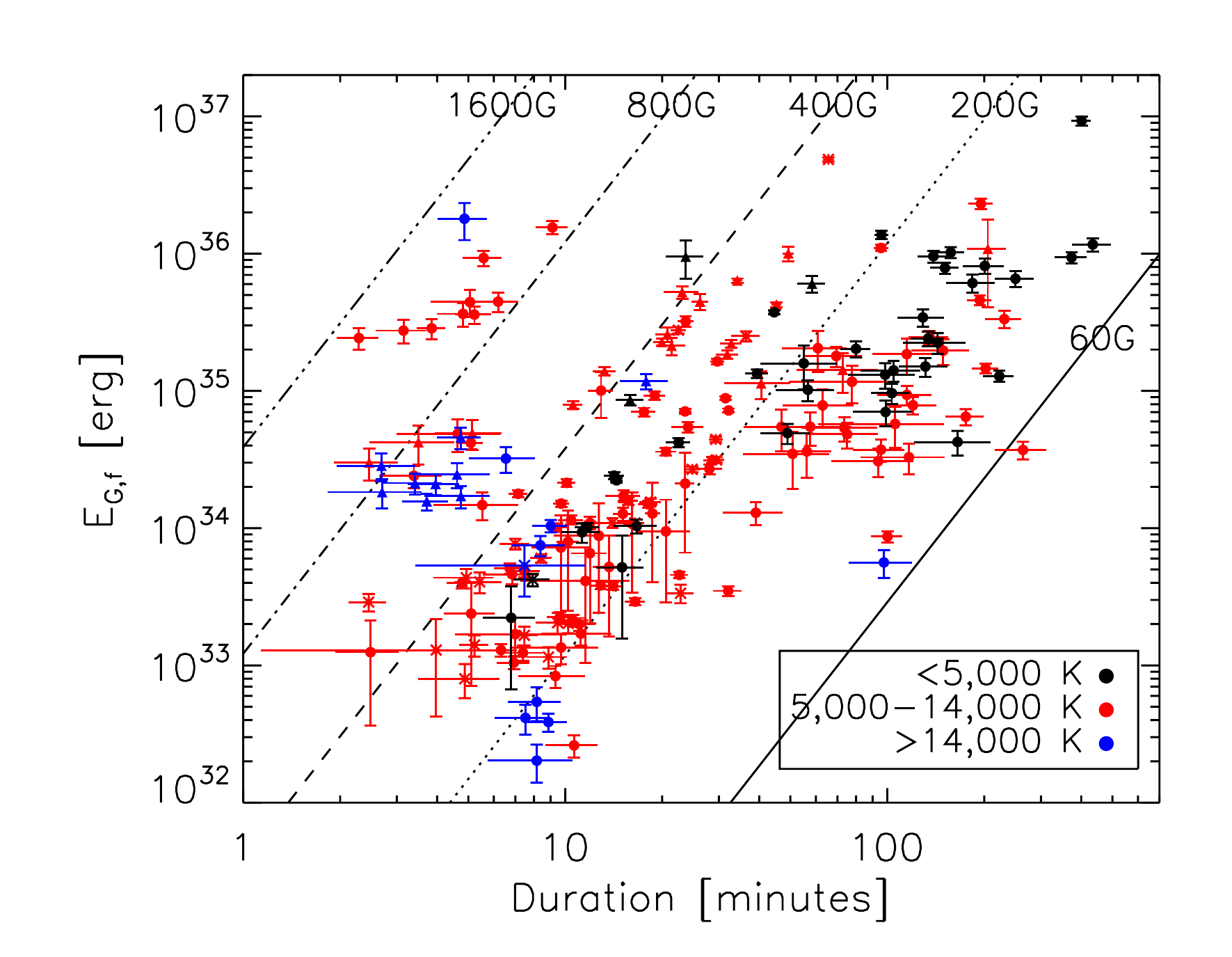}\\
    \caption{Gaia flare energy $E_{G,f}$ versus flare duration $d_{r+d}$.
    The black lines correspond to: $E_{G,f} = c' d_{r+d}^3 B^5$ 
    for different B values (see text).
    The different colors stand 
    for different intervals of stellar rotation in the top panel and
    for different intervals of flare temperature in the bottom panel.
    Eclipsing binaries are shown by stars and T Tauris by triangles.
    }
    \label{fig:energy_duration}
\end{figure}

Figure~\ref{fig:energy_duration} shows the flare energy in the Gaia band versus the flare duration given by the e-folding rise time plus the e-folding decay time observed by CoRoT.
In the top panel, the different colors correspond to four different stellar rotation ranges and, in the bottom panel, they correspond to three different flare temperature intervals.
Similarly to the relation given by Eq.~\ref{eq:namekata}, the panels show, for various values of $B$, the following relationship:
$E_{G,f} = c' d_{r+d}^3 B^5$
where
$c'$ = 3.7 $\times 10^{18}$ erg min$^{-3}$ G$^{-5}$.
To obtain $c'$, we arbitrarily increased the flare duration value by 25\% to account for the flare rising phase (not included before) and divided their bolometric energy estimate by 8 to adjust for the flare energy in the Gaia G band.
This factor was roughly estimated using the relation in Eq.~\ref{eq:egf},
where we assumed that the flare duration in the Gaia passband is the same as 
that at all wavelengths, we used the GAIA DR2 estimate of stellar luminosity (`lum$\_$val'),
and we extrapolated to the low value of the average solar flare energy.
\citet{namekata2017} also used the duration of the flare observed in visible light (more specifically, in the SDO/HMI's narrow passband around the Fe I 6173.3\AA\, line) to estimate the flare bolometric energy.
%
%
Since the $c'$ coefficient was not precisely determined, all straight lines in Fig.~\ref{fig:energy_duration} might be horizontally displaced by a small amount.
We can see in the upper panel that
flares observed in fast rotators, $P_{\rm rot}<$1\,d (blue circles),
are associated with magnetic fields slightly larger than $\sim$200\,G, while slower rotators, $7\leq P_{\rm rot}<$20\,d (red circles), are mostly associated with B slightly smaller than $\sim$200G.
Interestingly, flares in stars with longer rotation periods, $P_{\rm rot}>$20\,d (black circles), unexpectedly correspond to large magnetic fields ($B\gtrsim$800\,G).
These nine flares were observed on 4 stars (CoRoT 102602133, 102631863, 104939145, and 102646977), none of which were identified as an eclipsing binary star or a T Tauri star.
Four of these flares were observed on CoRoT 102602133, but only one  corresponds to a lower magnetic field as expected (B$\lesssim$200\,G). This flare also differs from the others for having a much longer duration and a low temperature:
$T_{\rm flare}$ = (3700$\pm$600)\,K.
One of the flares on a star with 
$P_{\rm rot}>$20\,d (black circles)
has a very high temperature: $T_{\rm flare}$=(21100$\pm$6800)\,K
(the one with $E_{G,f}$ = 1.8 $\times 10^{36}$\,erg in the Figure)
in contrast to the low temperature just mentioned.
The bottom panel shows the variation with the flare temperature, where we see no clear interrelationship.
We note, however, that
flares with $T_{\rm flare}<$5,000\,K (black symbols)
correspond to $B \lesssim$ 400\,G.
In addition, most high-temperature flares ($\gtrsim$ 14,000\,K) are short-lived (less than 10 minutes).
For 40\% of flares that last less than 10 minutes,  
the ascending phase is longer than the descending phase (i.e., $a_2 < a_3$ in Equation~\ref{eq:flarefitting}).
These flares correspond to 12\% of the analyzed flares, and all but two have a duration smaller than 10 minutes.
%

\begin{figure}
    \includegraphics[width=\columnwidth]
	{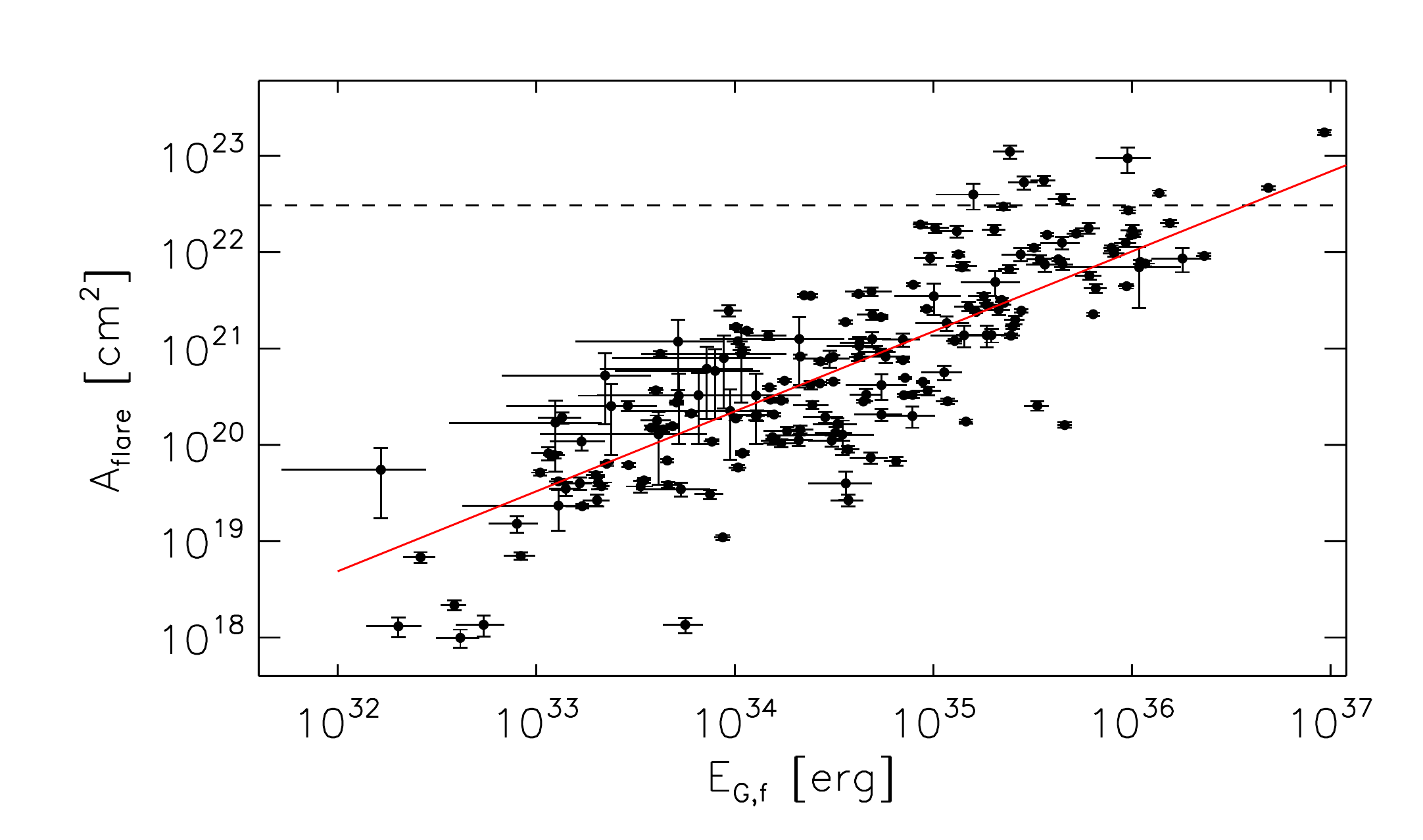}\\
	\includegraphics[width=\columnwidth]
	{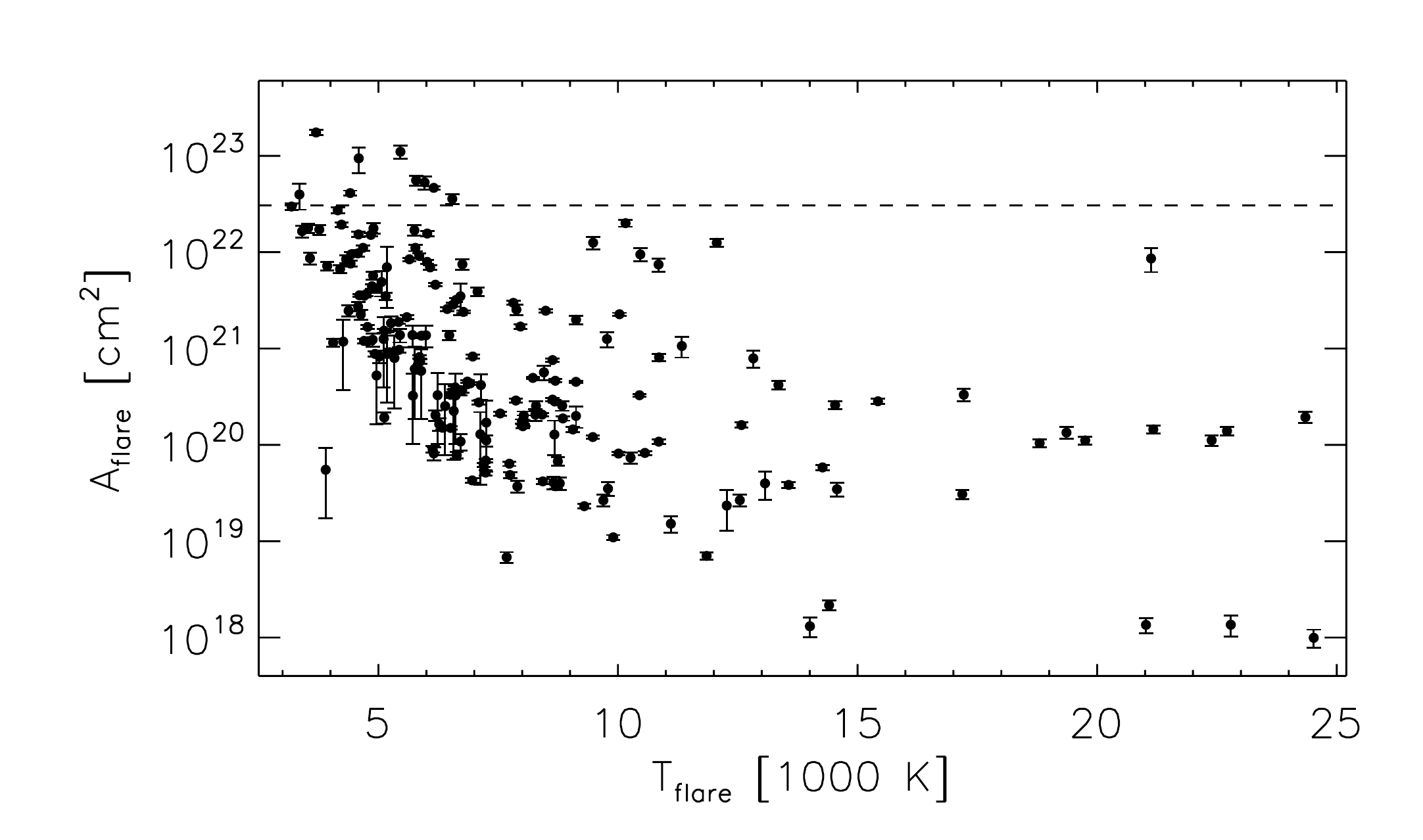}\\
	\includegraphics[width=\columnwidth]
	{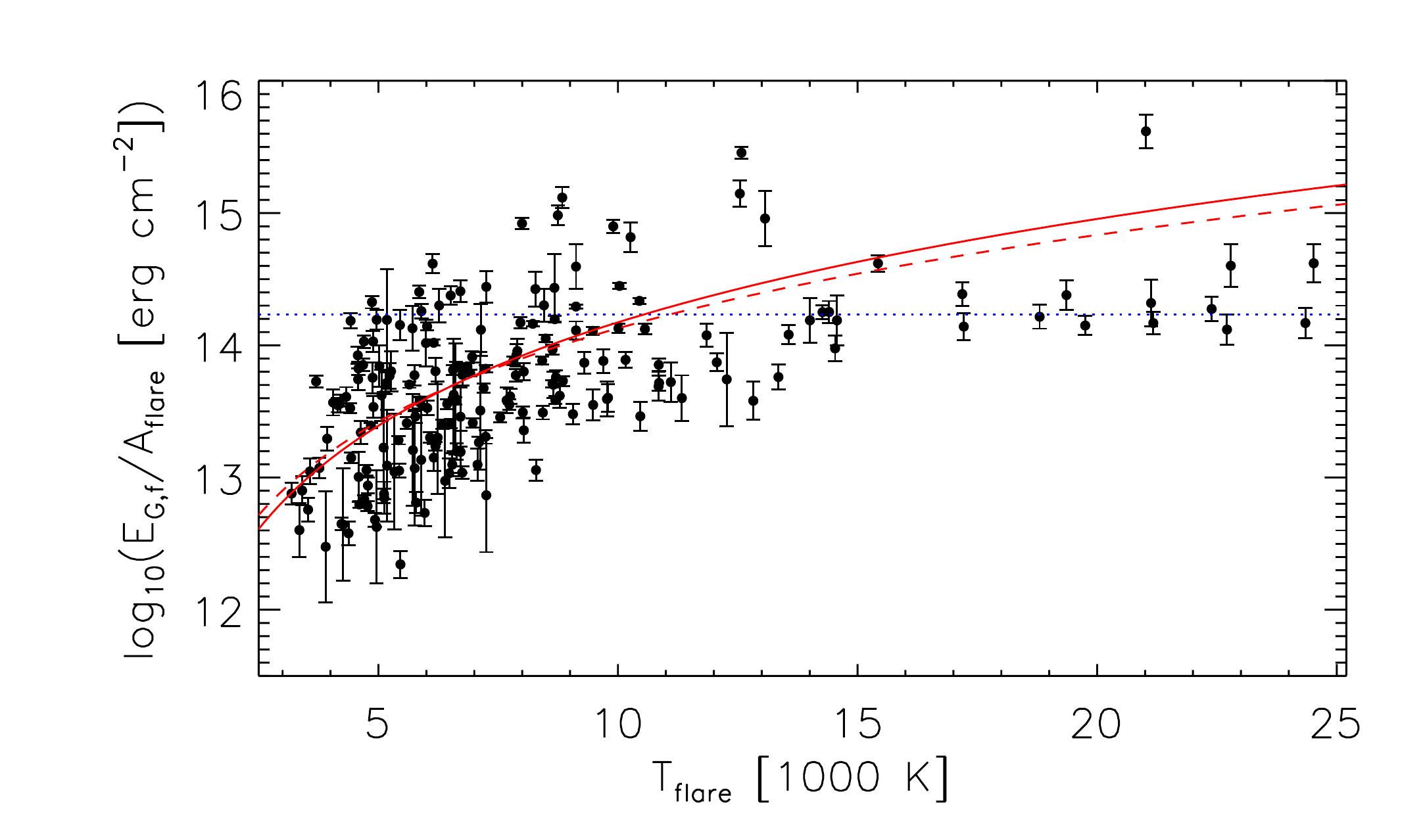}\\
	\caption{Flare area versus flare energy (top) and flare temperature (middle).
	The horizontal dashed line corresponds to the area of one solar hemisphere.
	The red line in the top panel shows the linear fit to the data.
	The bottom panel shows the flare energy divided by flare area, in cgs units, versus flare temperature.
	The red dashed line shows the fit to all points.
	The full red line shows a fit to flares with $T_{\rm flare} <$ 10,000\,K.
	The blue dotted horizontal line is the weighted mean for flares with $T_{\rm flare} >$ 10,000\,K.
    The flare temperature uncertainties were omitted in the middle and bottom panel for a better visualization.
	}
	    \label{fig:area_temp}
\end{figure}

We estimated the area of the flares, $A_{\rm flare}$, observed by the CoRoT White channel 
using Eq.~\ref{eq:ed_area} in units of energy, 
where the stellar luminosity, $L_G$, is given by Eq.~\ref{eq:lg}.
The flare area varies by 5 orders of magnitude, from $10^{18}$ to $10^{23}$ cm$^2$, i.e., 30\,$\mu$sh to 3\,sh (solar hemisphere), and is displayed in Table~\ref{tab:temperature}.
The area uncertainty was estimated by propagating the errors of ED, duration, $L_G$, and flare surface brightness.
Figure~\ref{fig:area_temp} (top panel) shows the variation of flare area with flare energy. 
The flare area increases (almost linearly) with flare energy:
$\log_{10}A_{\rm flare} = -(7.87\pm0.09) + (0.830\pm0.002) \log_{10}E_{G,f}$,
in cgs units (shown as red line in the top panel),
This is expected (Eq.~\ref{eq:ed_area} and~\ref{eq:egf}),
as well as that the flare area is proportional to $L_G^{(1.144\pm0.003)}$.
%
The stellar luminosity is proportional to its surface area, 
for a given effective temperature, therefore, in principle, 
the area of the flare is increasing with the area of the star.
However, we do not observe any clear dependence of flare area 
with spectral type or luminosity class in our data.
%
Figure~\ref{fig:area_temp} (middle panel) shows the variation of flare area with flare temperature. 
For $T_{\rm flare} \lesssim$ 7,500\,K, 
the flare area decreases, in most cases, as temperature increases.
Analyzing the flare energy-to-area ratio versus temperature (Fig.~\ref{fig:area_temp} bottom panel),
we did a log-log fit and obtained that the ratio, in cgs units, 
is proportional to T$_{\rm flare}^{(2.35\pm0.03)}$ (red dashed line). 
%
It seems that there is a change in behavior where, for flares with T$_{\rm flare} >$ 10,000\,K, 
the ratio seems to be constant, its weighted average is equal to 
E$_{G,f}$/A$_{\rm flare} = 1.7 \times 10^{14}$\,erg cm$^{-2}$
(horizontal dotted blue line). 
Fitting again now only for $T_{\rm flare} <$ 10,000\,K, we obtain:
E$_{G,f}$/$A_{\rm flare} \propto$ T$_{\rm flare}^{(2.60\pm0.04)}$ (red full line).
%
As the temperature percentage uncertainties increase with temperature 
and the number of flares decreases, 
this possible saturation at T$_{\rm flare} \gtrsim$ 10,000\,K should be regarded with caution.

\section{Conclusions}

Assuming that the spectrum of white-light flares can be described by blackbody radiation, we determined the effective flare temperature for more than 200 flares observed by CoRoT in 69 F, G, and K-type stars.
The temperatures were obtained using the flare equivalent duration and stellar flux observed in the Blue and Red channels.
The wavelength limits for the color channels were estimated using stellar spectra from the Pickles library 
for the spectral type and luminosity class of the star given by the Exodat Database.
We took the uncertainties in the stellar classification into account and carefully propagated them in our estimate of the flare temperature.

The estimated flare temperatures vary from 3,000\,K up to 30,000\,K.
Its uncertainties steadily increase from 10\% at 3,000 K to about 40\% at 25,000 K,
as the blackbody radiation maximum moves toward UV wavelengths and out of the CoRoT Response Function.
The expected value of the temperature distribution of the 209 analyzed flares is equal to 6,400\,K with a standard deviation of 2,800\,K.
These flares have been observed in A5$-$K5 stars.
The mean spectral type, weighted by the number of flares in each spectral subclass, is equal to G6.
The narrowest range that includes 75\% of the temperature distribution 
occurs between 3400\,K and 8100\,K.
Only 24 stars (35\%) of the analyzed stars have 3 or more flares with an estimated flare temperature. 
In most of them, the temperature variation of the flares observed on the same star is within the individual uncertainties, except for 10 stars.
The largest variation is for CoRoT 223993277 (shown in Fig.~\ref{fig:many}) which has a standard deviation equal to 70\% of its mean temperature.
 
It is often assumed that a stellar flare 
is radiated by a blackbody with a temperature of 9,000\,K or 10,000\,K (see Introduction).
The estimated temperature probability density function 
(shown as a blue line in Fig.~\ref{fig:histo})
gives a 6\% probability that the temperature is within 9,000$\pm$500\,K
and even lower around 10,000\,K.
The probability increases to 21\% at 5,300$\pm$500\,K (i.e., at the mode).
If we disregard the temperature uncertainties that give less weight to high temperatures, 
the PDF (shown as a red full line in Fig.~\ref{fig:histo}) 
is more spread and the probability increases to 8\% at 9,000$\pm$500\,K, while it decreases to 15\% at its mode (5,800+$\pm$500\,K).
In conclusion, our results do not favor a 9,000\,K or higher as a representative 
value for the flare temperature.
However, they agree with estimates of solar white-flare, around 6,000 K.

M-type stars are known to have many energetic flares \citep[e.g.][]{maehara2021}.
Nonetheless,
our analysis suggests that the mean flare temperature 
decreases in stars 
from early to late type
by (1,800$\pm$1,400)\,K per spectral type.
The average temperature seems to
decrease from $\sim$6,000\,K for flares in G6-type stars, the average spectral type of the analyzed stars, to less than 4,000\,K in M stars.
Furthermore, we found no indication that an energetic flare (observed by the CoRoT White channel)
has a high temperature.
%
The extrapolated low-temperature value for M stars 
does not agree with 
other estimations using observations in the visible.
Flares on the M dwarf AD Leo
were estimated to have a black-body temperature of $\sim$9,000\,K by \citet{hawley2003}
and
$14,000^{+17,000}_{-8,000}$\,K 
by \citet{namekata2020}.
%
While
the values obtained by
\citet{howard2020}  
for 40+ flares observed on K5-M5 dwarfs
have a median of 7,500\,K
and a large standard deviation, equal to 6600\,K,
%
with values ranging from 4,400\,K to 50,000\,K
(given by `Total T$_{\rm eff}$'  
in their Table 1).

We calibrated the amplitude and energy of the analyzed flares observed in the White CoRoT channel to the Gaia photometric system.
The absolute magnitude in the Gaia's G band of the analyzed stars varies from 0 to 10 mag while the flare peaks, from 4 to 14 mag.
The flare luminosity during its maximum is, on average, $\sim$2.5\% of the stellar luminosity in the G-band ($L_G$). 
The G-band flare energy was calculated by multiplying $L_G$ by the equivalent duration of the flare observed in the CoRoT White channel.
The Gaia flare energy varies between 10$^{32}$ and 10$^{37}$ erg
and is roughly proportional to $L_G^{1.22}$.    

There is no simple or clear dependence between flare temperature, duration, energy, stellar rotation period, and local magnetic field.
The latter was included in our comparison using the relation given by \citet{namekata2017} where 
the flare energy is proportional to the cube of its duration and the fifth power of the local magnetic field. 
We arbitrarily divided the stars into four rotation period intervals.
For our flare sample, we observed that
flares in stars with $P_{\rm rot}< $1\,d appear to have short duration ($\lesssim$ 20') and low energy ($\lesssim 10^{34}$ erg) with $B\simeq$ 200$-$400\,G.
While flares in stars with $P_{\rm rot}$=7$-$20\,d have the opposite behavior, that is, 
long duration ($\gtrsim$ 20') and high energy ($\gtrsim 10^{34}$ erg) with $B\lesssim$ 200\,G.
Flares in stars with intermediate rotation period, $P_{\rm rot}$ = 1$-$7\,d,
have a wide range of duration, energy, and magnetic field.
Unexpectedly, stars with a long rotation period, $P_{\rm rot}>$ 20\,d,
are associated with
larger magnetic fields ($B\gtrsim$800 \,G), with
shorter duration ($\lesssim$10') and higher energies ($\gtrsim 10^{35}$ erg).
We do not notice any correlation with flare temperature, except for some evidence 
that low-temperature flares ($T_{\rm flare} <$5,000\,K) do not have strong magnetic fields
($B\lesssim$ 400\,G).

We estimated the flare area based on our flare temperature estimate and by calibrating the CoRoT observations to the Gaia photometric system.
Observed in the Gaia G band, the flare area increases with flare energy as $E_{G,f}^{0.83}$ in cgs units,
varying from 3$\times 10^{-5}$ sh to 3\,sh (solar hemisphere).
Assuming the flare ribbon area is similar to the spot area \citep[e.g.][]{toriumi2017},
these values range from 2 orders of magnitude smaller to three times larger than spot areas reported by others.
\citet{okamoto2021} estimated spot areas between 1.6$\times10^{-3}$ to 13$\times10^{-3}$ sh for Sun-like stars.
While \citet{herbst2021} found, for K-M stars, spot areas ranging from 5$\times10^{-3}$ to 1 sh.
Although it increases almost linearly with Gaia stellar luminosity, we did not find a variation with spectral type or luminosity class.
A careful stellar classification using spectroscopic observations would help clarify this.
The energy output per area, given by $E_{G,f}/A_{\rm flare}$, is proportional to 
$T_{\rm flare}^{2.6}$ and it seems to saturate around 
10,000\,K at 1.7 $\times 10^{14}$\,erg s$^{-1}$ cm$^{-2}$.

The lack of high-cadence, long-duration observations of stellar spectra or
high-precision photometry in multiple (more than one) passband filters obtained simultaneously,
has hampered the determination of the temperature of the flares.
Hence, their quantity is small.
This work significantly increases
the number of temperature determinations.
Furthermore, it shows the importance of multi-filter space missions like Plato \citep{rauer2014,rauer2016}.
To the best of our knowledge, this is the first estimation of WLF temperature for G-dwarfs other than the Sun using multicolor photometry.
While there is still some doubt that the observed flaring G dwarfs are indeed 
Sun-like stars \citep{cliver2022}, the possibility
remains that the Sun is capable of producing flares as energetic as those observed in other stars, dubbed superflares.
Furthermore, the study of superflares in stars of the same spectral class as our Sun is relevant not only for the possibility that this happens on our Sun but also in the search for life on exoplanets \citep{maehara2012,namekata2021}.
Superflares may produce coronal mass ejections (CMEs) much larger than the largest observed on the Sun and affect the environment, habitability, and development of life on nearby exoplanets \citep{airapetian2020}.


\begin{acknowledgments}
M.C.R.S is supported by NASA Contract NAS5-02139 (HMI) to Stanford University.
MCF and BPLF are grateful for the support received from CNPq, Conselho Nacional de Desenvolvimento Científico  e Tecnológico - Brazil (Doctoral Scholarship).

This work 
used data from the CoRoT public archive
(http://idoc-corot.ias.u-psud.fr/) 
and
employed the ExoDat Database, operated at LAM-OAMP, Marseille, France, on behalf of the CoRoT/Exoplanet program.

This work has made use of results
from the European Space 
Agency (ESA) mission Gaia (https://www.cosmos.esa.int/
gaia), processed by the Gaia Data Processing and Analysis Consortium (DPAC).
Funding for the DPAC has been provided by national institutions, in particular the institutions participating in the Gaia Multilateral Agreement.

\end{acknowledgments}


%

\facilities{CoRoT, Gaia}


\software{IDL,
matplotlib \citep{Hunter:2007},
astropy \citep{astropy2013,astropy2018},
LMFIT \citep{lmfit2014},
scipy \citep{scipy2020}
}




\bibliography{main}{}

\begin{thebibliography}{}
\expandafter\ifx\csname natexlab\endcsname\relax\def\natexlab#1{#1}\fi
\providecommand{\url}[1]{\href{#1}{#1}}
\providecommand{\dodoi}[1]{doi:~\href{http://doi.org/#1}{\nolinkurl{#1}}}
\providecommand{\doeprint}[1]{\href{http://ascl.net/#1}{\nolinkurl{http://ascl.net/#1}}}
\providecommand{\doarXiv}[1]{\href{https://arxiv.org/abs/#1}{\nolinkurl{https://arxiv.org/abs/#1}}}

\bibitem[{{Airapetian} {et~al.}(2020){Airapetian}, {Barnes}, {Cohen},
  {Collinson}, {Danchi}, {Dong}, {Del Genio}, {France}, {Garcia-Sage},
  {Glocer}, {Gopalswamy}, {Grenfell}, {Gronoff}, {G{\"u}del}, {Herbst},
  {Henning}, {Jackman}, {Jin}, {Johnstone}, {Kaltenegger}, {Kay}, {Kobayashi},
  {Kuang}, {Li}, {Lynch}, {L{\"u}ftinger}, {Luhmann}, {Maehara}, {Mlynczak},
  {Notsu}, {Osten}, {Ramirez}, {Rugheimer}, {Scheucher}, {Schlieder},
  {Shibata}, {Sousa-Silva}, {Stamenkovi{\'c}}, {Strangeway}, {Usmanov},
  {Vergados}, {Verkhoglyadova}, {Vidotto}, {Voytek}, {Way}, {Zank}, \&
  {Yamashiki}}]{airapetian2020}
{Airapetian}, V.~S., {Barnes}, R., {Cohen}, O., {et~al.} 2020, International
  Journal of Astrobiology, 19, 136, \dodoi{10.1017/S1473550419000132}

\bibitem[{{Andrae} {et~al.}(2018){Andrae}, {Fouesneau}, {Creevey}, {Ordenovic},
  {Mary}, {Burlacu}, {Chaoul}, {Jean-Antoine-Piccolo}, {Kordopatis}, {Korn},
  {Lebreton}, {Panem}, {Pichon}, {Th{\'e}venin}, {Walmsley}, \&
  {Bailer-Jones}}]{andrae2018}
{Andrae}, R., {Fouesneau}, M., {Creevey}, O., {et~al.} 2018, \aap, 616, A8,
  \dodoi{10.1051/0004-6361/201732516}

\bibitem[{{Astropy Collaboration} {et~al.}(2013){Astropy Collaboration},
  {Robitaille}, {Tollerud}, {Greenfield}, {Droettboom}, {Bray}, {Aldcroft},
  {Davis}, {Ginsburg}, {Price-Whelan}, {Kerzendorf}, {Conley}, {Crighton},
  {Barbary}, {Muna}, {Ferguson}, {Grollier}, {Parikh}, {Nair}, {Unther},
  {Deil}, {Woillez}, {Conseil}, {Kramer}, {Turner}, {Singer}, {Fox}, {Weaver},
  {Zabalza}, {Edwards}, {Azalee Bostroem}, {Burke}, {Casey}, {Crawford},
  {Dencheva}, {Ely}, {Jenness}, {Labrie}, {Lim}, {Pierfederici}, {Pontzen},
  {Ptak}, {Refsdal}, {Servillat}, \& {Streicher}}]{astropy2013}
{Astropy Collaboration}, {Robitaille}, T.~P., {Tollerud}, E.~J., {et~al.} 2013,
  \aap, 558, A33, \dodoi{10.1051/0004-6361/201322068}

\bibitem[{{Astropy Collaboration} {et~al.}(2018){Astropy Collaboration},
  {Price-Whelan}, {Sip{\H{o}}cz}, {G{\"u}nther}, {Lim}, {Crawford}, {Conseil},
  {Shupe}, {Craig}, {Dencheva}, {Ginsburg}, {VanderPlas}, {Bradley},
  {P{\'e}rez-Su{\'a}rez}, {de Val-Borro}, {Aldcroft}, {Cruz}, {Robitaille},
  {Tollerud}, {Ardelean}, {Babej}, {Bach}, {Bachetti}, {Bakanov}, {Bamford},
  {Barentsen}, {Barmby}, {Baumbach}, {Berry}, {Biscani}, {Boquien}, {Bostroem},
  {Bouma}, {Brammer}, {Bray}, {Breytenbach}, {Buddelmeijer}, {Burke},
  {Calderone}, {Cano Rodr{\'\i}guez}, {Cara}, {Cardoso}, {Cheedella}, {Copin},
  {Corrales}, {Crichton}, {D'Avella}, {Deil}, {Depagne}, {Dietrich}, {Donath},
  {Droettboom}, {Earl}, {Erben}, {Fabbro}, {Ferreira}, {Finethy}, {Fox},
  {Garrison}, {Gibbons}, {Goldstein}, {Gommers}, {Greco}, {Greenfield},
  {Groener}, {Grollier}, {Hagen}, {Hirst}, {Homeier}, {Horton}, {Hosseinzadeh},
  {Hu}, {Hunkeler}, {Ivezi{\'c}}, {Jain}, {Jenness}, {Kanarek}, {Kendrew},
  {Kern}, {Kerzendorf}, {Khvalko}, {King}, {Kirkby}, {Kulkarni}, {Kumar},
  {Lee}, {Lenz}, {Littlefair}, {Ma}, {Macleod}, {Mastropietro}, {McCully},
  {Montagnac}, {Morris}, {Mueller}, {Mumford}, {Muna}, {Murphy}, {Nelson},
  {Nguyen}, {Ninan}, {N{\"o}the}, {Ogaz}, {Oh}, {Parejko}, {Parley}, {Pascual},
  {Patil}, {Patil}, {Plunkett}, {Prochaska}, {Rastogi}, {Reddy Janga},
  {Sabater}, {Sakurikar}, {Seifert}, {Sherbert}, {Sherwood-Taylor}, {Shih},
  {Sick}, {Silbiger}, {Singanamalla}, {Singer}, {Sladen}, {Sooley},
  {Sornarajah}, {Streicher}, {Teuben}, {Thomas}, {Tremblay}, {Turner},
  {Terr{\'o}n}, {van Kerkwijk}, {de la Vega}, {Watkins}, {Weaver}, {Whitmore},
  {Woillez}, {Zabalza}, \& {Astropy Contributors}}]{astropy2018}
{Astropy Collaboration}, {Price-Whelan}, A.~M., {Sip{\H{o}}cz}, B.~M., {et~al.}
  2018, \aj, 156, 123, \dodoi{10.3847/1538-3881/aabc4f}

\bibitem[{{Auvergne} {et~al.}(2009){Auvergne}, {Bodin}, {Boisnard}, {Buey},
  {Chaintreuil}, {Epstein}, {Jouret}, {Lam-Trong}, {Levacher}, {Magnan},
  {Perez}, {Plasson}, {Plesseria}, {Peter}, {Steller}, {Tiph{\`e}ne}, {Baglin},
  {Agogu{\'e}}, {Appourchaux}, {Barbet}, {Beaufort}, {Bellenger}, {Berlin},
  {Bernardi}, {Blouin}, {Boumier}, {Bonneau}, {Briet}, {Butler}, {Cautain},
  {Chiavassa}, {Costes}, {Cuvilho}, {Cunha-Parro}, {de Oliveira Fialho},
  {Decaudin}, {Defise}, {Djalal}, {Docclo}, {Drummond}, {Dupuis}, {Exil},
  {Faur{\'e}}, {Gaboriaud}, {Gamet}, {Gavalda}, {Grolleau}, {Gueguen},
  {Guivarc'h}, {Guterman}, {Hasiba}, {Huntzinger}, {Hustaix}, {Imbert},
  {Jeanville}, {Johlander}, {Jorda}, {Journoud}, {Karioty}, {Kerjean},
  {Lafond}, {Lapeyrere}, {Landiech}, {Larqu{\'e}}, {Laudet}, {Le Merrer},
  {Leporati}, {Leruyet}, {Levieuge}, {Llebaria}, {Martin}, {Mazy}, {Mesnager},
  {Michel}, {Moalic}, {Monjoin}, {Naudet}, {Neukirchner}, {Nguyen-Kim},
  {Ollivier}, {Orcesi}, {Ottacher}, {Oulali}, {Parisot}, {Perruchot},
  {Piacentino}, {Pinheiro da Silva}, {Platzer}, {Pontet}, {Pradines},
  {Quentin}, {Rohbeck}, {Rolland}, {Rollenhagen}, {Romagnan}, {Russ}, {Samadi},
  {Schmidt}, {Schwartz}, {Sebbag}, {Smit}, {Sunter}, {Tello}, {Toulouse},
  {Ulmer}, {Vandermarcq}, {Vergnault}, {Wallner}, {Waultier}, \&
  {Zanatta}}]{auvergne2009}
{Auvergne}, M., {Bodin}, P., {Boisnard}, L., {et~al.} 2009, \aap, 506, 411,
  \dodoi{10.1051/0004-6361/200810860}

\bibitem[{{Baglin} \& {CoRot Team}(2016)}]{baglin2016}
{Baglin}, A., \& {CoRot Team}. 2016, {I.1 The general framework} (EDP
  sciences), 5, \dodoi{10.1051/978-2-7598-1876-1.c011}

\bibitem[{{Bevington} \& {Robinson}(2003)}]{bevington2003}
{Bevington}, P.~R., \& {Robinson}, D.~K. 2003, {Data reduction and error
  analysis for the physical sciences} (McGraw-Hill)

\bibitem[{{Bord{\'e}} {et~al.}(2010){Bord{\'e}}, {Bouchy}, {Deleuil},
  {Cabrera}, {Jorda}, {Lovis}, {Csizmadia}, {Aigrain}, {Almenara}, {Alonso},
  {Auvergne}, {Baglin}, {Barge}, {Benz}, {Bonomo}, {Bruntt}, {Carone},
  {Carpano}, {Deeg}, {Dvorak}, {Erikson}, {Ferraz-Mello}, {Fridlund},
  {Gandolfi}, {Gazzano}, {Gillon}, {Guenther}, {Guillot}, {Guterman}, {Hatzes},
  {Havel}, {H{\'e}brard}, {Lammer}, {L{\'e}ger}, {Mayor}, {Mazeh}, {Moutou},
  {P{\"a}tzold}, {Pepe}, {Ollivier}, {Queloz}, {Rauer}, {Rouan}, {Samuel},
  {Santerne}, {Schneider}, {Tingley}, {Udry}, {Weingrill}, \&
  {Wuchterl}}]{borde2010}
{Bord{\'e}}, P., {Bouchy}, F., {Deleuil}, M., {et~al.} 2010, \aap, 520, A66,
  \dodoi{10.1051/0004-6361/201014775}

\bibitem[{{Borsa} \& {Poretti}(2013)}]{borsa2013}
{Borsa}, F., \& {Poretti}, E. 2013, \mnras, 428, 891,
  \dodoi{10.1093/mnras/sts087}

\bibitem[{{Candelaresi} {et~al.}(2014){Candelaresi}, {Hillier}, {Maehara},
  {Brandenburg}, \& {Shibata}}]{candelaresi2014}
{Candelaresi}, S., {Hillier}, A., {Maehara}, H., {Brandenburg}, A., \&
  {Shibata}, K. 2014, \apj, 792, 67, \dodoi{10.1088/0004-637X/792/1/67}

\bibitem[{{Carone} {et~al.}(2012){Carone}, {Gandolfi}, {Cabrera}, {Hatzes},
  {Deeg}, {Csizmadia}, {P{\"a}tzold}, {Weingrill}, {Aigrain}, {Alonso},
  {Alapini}, {Almenara}, {Auvergne}, {Baglin}, {Barge}, {Bonomo}, {Bord{\'e}},
  {Bouchy}, {Bruntt}, {Carpano}, {Cochran}, {Deleuil}, {D{\'\i}az}, {Dreizler},
  {Dvorak}, {Eisl{\"o}ffel}, {Eigm{\"u}ller}, {Endl}, {Erikson},
  {Ferraz-Mello}, {Fridlund}, {Gazzano}, {Gibson}, {Gillon}, {Gondoin},
  {Grziwa}, {G{\"u}nther}, {Guillot}, {Hartmann}, {Havel}, {H{\'e}brard},
  {Jorda}, {Kabath}, {L{\'e}ger}, {Llebaria}, {Lammer}, {Lovis}, {MacQueen},
  {Mayor}, {Mazeh}, {Moutou}, {Nortmann}, {Ofir}, {Ollivier}, {Parviainen},
  {Pepe}, {Pont}, {Queloz}, {Rabus}, {Rauer}, {R{\'e}gulo}, {Renner}, {de La
  Reza}, {Rouan}, {Santerne}, {Samuel}, {Schneider}, {Shporer}, {Stecklum},
  {Tal-Or}, {Tingley}, {Udry}, \& {Wuchterl}}]{carone2012}
{Carone}, L., {Gandolfi}, D., {Cabrera}, J., {et~al.} 2012, \aap, 538, A112,
  \dodoi{10.1051/0004-6361/201116968}

\bibitem[{{Chaintreuil} {et~al.}(2016){Chaintreuil}, {Deru}, {Baudin},
  {Ferrigno}, {Grolleau}, {Romagnan}, \& {CoRot Team}}]{chaintreuil2016}
{Chaintreuil}, S., {Deru}, A., {Baudin}, F., {et~al.} 2016, {II.4 The ``ready
  to use'' CoRoT data} (EDP sciences), 61,
  \dodoi{10.1051/978-2-7598-1876-1.c024}

\bibitem[{{Cliver} {et~al.}(2022){Cliver}, {Schrijver}, {Shibata}, \&
  {Usoskin}}]{cliver2022}
{Cliver}, E.~W., {Schrijver}, C.~J., {Shibata}, K., \& {Usoskin}, I.~G. 2022,
  Living Reviews in Solar Physics, 19, 2, \dodoi{10.1007/s41116-022-00033-8}

\bibitem[{{Damiani} {et~al.}(2016){Damiani}, {Meunier}, {Moutou}, {Deleuil},
  {Ysard}, {Baudin}, \& {Deeg}}]{damiani2016A&A}
{Damiani}, C., {Meunier}, J.~C., {Moutou}, C., {et~al.} 2016, \aap, 595, A95,
  \dodoi{10.1051/0004-6361/201628627}

\bibitem[{{Davenport} {et~al.}(2020){Davenport}, {Mendoza}, \&
  {Hawley}}]{davenport2020}
{Davenport}, J. R.~A., {Mendoza}, G.~T., \& {Hawley}, S.~L. 2020, \aj, 160, 36,
  \dodoi{10.3847/1538-3881/ab9536}

\bibitem[{{Davenport} {et~al.}(2014){Davenport}, {Hawley}, {Hebb},
  {Wisniewski}, {Kowalski}, {Johnson}, {Malatesta}, {Peraza}, {Keil},
  {Silverberg}, {Jansen}, {Scheffler}, {Berdis}, {Larsen}, \&
  {Hilton}}]{davenport2014}
{Davenport}, J. R.~A., {Hawley}, S.~L., {Hebb}, L., {et~al.} 2014, \apj, 797,
  122, \dodoi{10.1088/0004-637X/797/2/122}

\bibitem[{{Deleuil} {et~al.}(2009){Deleuil}, {Meunier}, {Moutou}, {Surace},
  {Deeg}, {Barbieri}, {Debosscher}, {Almenara}, {Agneray}, {Granet},
  {Guterman}, \& {Hodgkin}}]{deleuil2009}
{Deleuil}, M., {Meunier}, J.~C., {Moutou}, C., {et~al.} 2009, \aj, 138, 649,
  \dodoi{10.1088/0004-6256/138/2/649}

\bibitem[{{Deleuil} {et~al.}(2018){Deleuil}, {Aigrain}, {Moutou}, {Cabrera},
  {Bouchy}, {Deeg}, {Almenara}, {H{\'e}brard}, {Santerne}, {Alonso}, {Bonomo},
  {Bord{\'e}}, {Csizmadia}, {D{\`\i}az}, {Erikson}, {Fridlund}, {Gandolfi},
  {Guenther}, {Guillot}, {Guterman}, {Grziwa}, {Hatzes}, {L{\'e}ger}, {Mazeh},
  {Ofir}, {Ollivier}, {P{\"a}tzold}, {Parviainen}, {Rauer}, {Rouan},
  {Schneider}, {Titz-Weider}, {Tingley}, \& {Weingrill}}]{deleuil2018}
{Deleuil}, M., {Aigrain}, S., {Moutou}, C., {et~al.} 2018, \aap, 619, A97,
  \dodoi{10.1051/0004-6361/201731068}

\bibitem[{{Donati} \& {Landstreet}(2009)}]{donati2009}
{Donati}, J.~F., \& {Landstreet}, J.~D. 2009, \araa, 47, 333,
  \dodoi{10.1146/annurev-astro-082708-101833}

\bibitem[{{Drabent}(2012)}]{drabent2012}
{Drabent}, A. 2012, PhD thesis, Friedrich-Schiller-Universit\"at Jena

\bibitem[{{Foreman-Mackey} {et~al.}(2013){Foreman-Mackey}, {Hogg}, {Lang}, \&
  {Goodman}}]{foreman2013}
{Foreman-Mackey}, D., {Hogg}, D.~W., {Lang}, D., \& {Goodman}, J. 2013, \pasp,
  125, 306, \dodoi{10.1086/670067}

\bibitem[{{Gershberg}(1972)}]{gershberg1972}
{Gershberg}, R.~E. 1972, \apss, 19, 75, \dodoi{10.1007/BF00643168}

\bibitem[{{G{\"u}nther} {et~al.}(2020){G{\"u}nther}, {Zhan}, {Seager},
  {Rimmer}, {Ranjan}, {Stassun}, {Oelkers}, {Daylan}, {Newton}, {Kristiansen},
  {Olah}, {Gillen}, {Rappaport}, {Ricker}, {Vanderspek}, {Latham}, {Winn},
  {Jenkins}, {Glidden}, {Fausnaugh}, {Levine}, {Dittmann}, {Quinn},
  {Krishnamurthy}, \& {Ting}}]{gunther2020}
{G{\"u}nther}, M.~N., {Zhan}, Z., {Seager}, S., {et~al.} 2020, \aj, 159, 60,
  \dodoi{10.3847/1538-3881/ab5d3a}

\bibitem[{{Hawley} {et~al.}(2014){Hawley}, {Davenport}, {Kowalski},
  {Wisniewski}, {Hebb}, {Deitrick}, \& {Hilton}}]{hawley2014}
{Hawley}, S.~L., {Davenport}, J. R.~A., {Kowalski}, A.~F., {et~al.} 2014, \apj,
  797, 121, \dodoi{10.1088/0004-637X/797/2/121}

\bibitem[{{Hawley} \& {Fisher}(1992)}]{hawley1992}
{Hawley}, S.~L., \& {Fisher}, G.~H. 1992, \apjs, 78, 565,
  \dodoi{10.1086/191640}

\bibitem[{{Hawley} {et~al.}(2003){Hawley}, {Allred}, {Johns-Krull}, {Fisher},
  {Abbett}, {Alekseev}, {Avgoloupis}, {Deustua}, {Gunn}, {Seiradakis}, {Sirk},
  \& {Valenti}}]{hawley2003}
{Hawley}, S.~L., {Allred}, J.~C., {Johns-Krull}, C.~M., {et~al.} 2003, \apj,
  597, 535, \dodoi{10.1086/378351}

\bibitem[{{Herbst} {et~al.}(2021){Herbst}, {Papaioannou}, {Airapetian}, \&
  {Atri}}]{herbst2021}
{Herbst}, K., {Papaioannou}, A., {Airapetian}, V.~S., \& {Atri}, D. 2021, \apj,
  907, 89, \dodoi{10.3847/1538-4357/abcc04}

\bibitem[{{Howard} {et~al.}(2020){Howard}, {Corbett}, {Law}, {Ratzloff},
  {Galliher}, {Glazier}, {Gonzalez}, {Vasquez Soto}, {Fors}, {del Ser}, \&
  {Haislip}}]{howard2020}
{Howard}, W.~S., {Corbett}, H., {Law}, N.~M., {et~al.} 2020, \apj, 902, 115,
  \dodoi{10.3847/1538-4357/abb5b4}

\bibitem[{Hunter(2007)}]{Hunter:2007}
Hunter, J.~D. 2007, Computing in Science \& Engineering, 9, 90,
  \dodoi{10.1109/MCSE.2007.55}

\bibitem[{{Jess} {et~al.}(2008){Jess}, {Mathioudakis}, {Crockett}, \&
  {Keenan}}]{jess2008}
{Jess}, D.~B., {Mathioudakis}, M., {Crockett}, P.~J., \& {Keenan}, F.~P. 2008,
  \apjl, 688, L119, \dodoi{10.1086/595588}

\bibitem[{{Kerr} \& {Fletcher}(2014)}]{kerr2014}
{Kerr}, G.~S., \& {Fletcher}, L. 2014, \apj, 783, 98,
  \dodoi{10.1088/0004-637X/783/2/98}

\bibitem[{{Kirchner} \& {Allen}(2020)}]{kirchner2020}
{Kirchner}, J.~W., \& {Allen}, S.~T. 2020, Hydrology and Earth System Sciences,
  24, 17, \dodoi{10.5194/hess-24-17-2020}

\bibitem[{{Klagyivik} {et~al.}(2017){Klagyivik}, {Deeg}, {Cabrera},
  {Csizmadia}, \& {Almenara}}]{klagyivik2017}
{Klagyivik}, P., {Deeg}, H.~J., {Cabrera}, J., {Csizmadia}, S., \& {Almenara},
  J.~M. 2017, \aap, 602, A117, \dodoi{10.1051/0004-6361/201628244}

\bibitem[{{Kleint} {et~al.}(2016){Kleint}, {Heinzel}, {Judge}, \&
  {Krucker}}]{kleint2016}
{Kleint}, L., {Heinzel}, P., {Judge}, P., \& {Krucker}, S. 2016, \apj, 816, 88,
  \dodoi{10.3847/0004-637X/816/2/88}

\bibitem[{{Kowalski} {et~al.}(2013){Kowalski}, {Hawley}, {Wisniewski}, {Osten},
  {Hilton}, {Holtzman}, {Schmidt}, \& {Davenport}}]{kowalski2013}
{Kowalski}, A.~F., {Hawley}, S.~L., {Wisniewski}, J.~P., {et~al.} 2013, \apjs,
  207, 15, \dodoi{10.1088/0067-0049/207/1/15}

\bibitem[{{Kowalski} {et~al.}(2016){Kowalski}, {Mathioudakis}, {Hawley},
  {Wisniewski}, {Dhillon}, {Marsh}, {Hilton}, \& {Brown}}]{kowalski2016}
{Kowalski}, A.~F., {Mathioudakis}, M., {Hawley}, S.~L., {et~al.} 2016, \apj,
  820, 95, \dodoi{10.3847/0004-637X/820/2/95}

\bibitem[{{Kowalski} {et~al.}(2019){Kowalski}, {Wisniewski}, {Hawley}, {Osten},
  {Brown}, {Fari{\~n}a}, {Valenti}, {Brown}, {Xilouris}, {Schmidt}, \&
  {Johns-Krull}}]{kowalski2019}
{Kowalski}, A.~F., {Wisniewski}, J.~P., {Hawley}, S.~L., {et~al.} 2019, \apj,
  871, 167, \dodoi{10.3847/1538-4357/aaf058}

\bibitem[{{Kretzschmar}(2011)}]{kretzschmar2011}
{Kretzschmar}, M. 2011, \aap, 530, A84, \dodoi{10.1051/0004-6361/201015930}

\bibitem[{{Law} {et~al.}(2016){Law}, {Fors}, {Ratzloff}, {Corbett}, {del Ser},
  \& {Wulfken}}]{law2016}
{Law}, N.~M., {Fors}, O., {Ratzloff}, J., {et~al.} 2016, in Society of
  Photo-Optical Instrumentation Engineers (SPIE) Conference Series, Vol. 9906,
  Ground-based and Airborne Telescopes VI, ed. H.~J. {Hall}, R.~{Gilmozzi}, \&
  H.~K. {Marshall}, 99061M, \dodoi{10.1117/12.2233349}

\bibitem[{{L{\'e}ger} {et~al.}(2009){L{\'e}ger}, {Rouan}, {Schneider}, {Barge},
  {Fridlund}, {Samuel}, {Ollivier}, {Guenther}, {Deleuil}, {Deeg}, {Auvergne},
  {Alonso}, {Aigrain}, {Alapini}, {Almenara}, {Baglin}, {Barbieri}, {Bruntt},
  {Bord{\'e}}, {Bouchy}, {Cabrera}, {Catala}, {Carone}, {Carpano}, {Csizmadia},
  {Dvorak}, {Erikson}, {Ferraz-Mello}, {Foing}, {Fressin}, {Gandolfi},
  {Gillon}, {Gondoin}, {Grasset}, {Guillot}, {Hatzes}, {H{\'e}brard}, {Jorda},
  {Lammer}, {Llebaria}, {Loeillet}, {Mayor}, {Mazeh}, {Moutou}, {P{\"a}tzold},
  {Pont}, {Queloz}, {Rauer}, {Renner}, {Samadi}, {Shporer}, {Sotin}, {Tingley},
  {Wuchterl}, {Adda}, {Agogu}, {Appourchaux}, {Ballans}, {Baron}, {Beaufort},
  {Bellenger}, {Berlin}, {Bernardi}, {Blouin}, {Baudin}, {Bodin}, {Boisnard},
  {Boit}, {Bonneau}, {Borzeix}, {Briet}, {Buey}, {Butler}, {Cailleau},
  {Cautain}, {Chabaud}, {Chaintreuil}, {Chiavassa}, {Costes}, {Cuna Parrho},
  {de Oliveira Fialho}, {Decaudin}, {Defise}, {Djalal}, {Epstein}, {Exil},
  {Faur{\'e}}, {Fenouillet}, {Gaboriaud}, {Gallic}, {Gamet}, {Gavalda},
  {Grolleau}, {Gruneisen}, {Gueguen}, {Guis}, {Guivarc'h}, {Guterman},
  {Hallouard}, {Hasiba}, {Heuripeau}, {Huntzinger}, {Hustaix}, {Imad},
  {Imbert}, {Johlander}, {Jouret}, {Journoud}, {Karioty}, {Kerjean},
  {Lafaille}, {Lafond}, {Lam-Trong}, {Landiech}, {Lapeyrere}, {Larqu{\'e}},
  {Laudet}, {Lautier}, {Lecann}, {Lefevre}, {Leruyet}, {Levacher}, {Magnan},
  {Mazy}, {Mertens}, {Mesnager}, {Meunier}, {Michel}, {Monjoin}, {Naudet},
  {Nguyen-Kim}, {Orcesi}, {Ottacher}, {Perez}, {Peter}, {Plasson}, {Plesseria},
  {Pontet}, {Pradines}, {Quentin}, {Reynaud}, {Rolland}, {Rollenhagen},
  {Romagnan}, {Russ}, {Schmidt}, {Schwartz}, {Sebbag}, {Sedes}, {Smit},
  {Steller}, {Sunter}, {Surace}, {Tello}, {Tiph{\`e}ne}, {Toulouse}, {Ulmer},
  {Vandermarcq}, {Vergnault}, {Vuillemin}, \& {Zanatta}}]{leger2009}
{L{\'e}ger}, A., {Rouan}, D., {Schneider}, J., {et~al.} 2009, \aap, 506, 287,
  \dodoi{10.1051/0004-6361/200911933}

\bibitem[{{Lin} \& {Hudson}(1976)}]{lin1976}
{Lin}, R.~P., \& {Hudson}, H.~S. 1976, \solphys, 50, 153,
  \dodoi{10.1007/BF00206199}

\bibitem[{{Maehara} {et~al.}(2012){Maehara}, {Shibayama}, {Notsu}, {Notsu},
  {Nagao}, {Kusaba}, {Honda}, {Nogami}, \& {Shibata}}]{maehara2012}
{Maehara}, H., {Shibayama}, T., {Notsu}, S., {et~al.} 2012, \nat, 485, 478,
  \dodoi{10.1038/nature11063}

\bibitem[{{Maehara} {et~al.}(2021){Maehara}, {Notsu}, {Namekata}, {Honda},
  {Kowalski}, {Katoh}, {Ohshima}, {Iida}, {Oeda}, {Murata}, {Yamanaka},
  {Takagi}, {Sasada}, {Akitaya}, {Ikuta}, {Okamoto}, {Nogami}, \&
  {Shibata}}]{maehara2021}
{Maehara}, H., {Notsu}, Y., {Namekata}, K., {et~al.} 2021, \pasj, 73, 44,
  \dodoi{10.1093/pasj/psaa098}

\bibitem[{{McQuillan} {et~al.}(2013){McQuillan}, {Aigrain}, \&
  {Mazeh}}]{mcquillan2013}
{McQuillan}, A., {Aigrain}, S., \& {Mazeh}, T. 2013, \mnras, 432, 1203,
  \dodoi{10.1093/mnras/stt536}

\bibitem[{{McQuillan} {et~al.}(2014){McQuillan}, {Mazeh}, \&
  {Aigrain}}]{mcquillan2014}
{McQuillan}, A., {Mazeh}, T., \& {Aigrain}, S. 2014, \apjs, 211, 24,
  \dodoi{10.1088/0067-0049/211/2/24}

\bibitem[{{Mochnacki} \& {Zirin}(1980)}]{mochnacki1980}
{Mochnacki}, S.~W., \& {Zirin}, H. 1980, \apjl, 239, L27,
  \dodoi{10.1086/183285}

\bibitem[{{Namekata} {et~al.}(2017){Namekata}, {Sakaue}, {Watanabe}, {Asai},
  {Maehara}, {Notsu}, {Notsu}, {Honda}, {Ishii}, {Ikuta}, {Nogami}, \&
  {Shibata}}]{namekata2017}
{Namekata}, K., {Sakaue}, T., {Watanabe}, K., {et~al.} 2017, \apj, 851, 91,
  \dodoi{10.3847/1538-4357/aa9b34}

\bibitem[{{Namekata} {et~al.}(2020){Namekata}, {Maehara}, {Sasaki}, {Kawai},
  {Notsu}, {Kowalski}, {Allred}, {Iwakiri}, {Tsuboi}, {Murata}, {Niwano},
  {Shiraishi}, {Adachi}, {Iida}, {Oeda}, {Honda}, {Tozuka}, {Katoh}, {Onozato},
  {Okamoto}, {Isogai}, {Kimura}, {Kojiguchi}, {Wakamatsu}, {Tampo}, {Nogami},
  \& {Shibata}}]{namekata2020}
{Namekata}, K., {Maehara}, H., {Sasaki}, R., {et~al.} 2020, \pasj, 72, 68,
  \dodoi{10.1093/pasj/psaa051}

\bibitem[{{Namekata} {et~al.}(2021){Namekata}, {Maehara}, {Honda}, {Notsu},
  {Okamoto}, {Takahashi}, {Takayama}, {Ohshima}, {Saito}, {Katoh}, {Tozuka},
  {Murata}, {Ogawa}, {Niwano}, {Adachi}, {Oeda}, {Shiraishi}, {Isogai}, {Seki},
  {Ishii}, {Ichimoto}, {Nogami}, \& {Shibata}}]{namekata2021}
{Namekata}, K., {Maehara}, H., {Honda}, S., {et~al.} 2021, Nature Astronomy, 6,
  241, \dodoi{10.1038/s41550-021-01532-8}

\bibitem[{{N{\`e}mec} {et~al.}(2020){N{\`e}mec}, {I{\c{s}}{\i}k}, {Shapiro},
  {Solanki}, {Krivova}, \& {Unruh}}]{nemec2020}
{N{\`e}mec}, N.~E., {I{\c{s}}{\i}k}, E., {Shapiro}, A.~I., {et~al.} 2020, \aap,
  638, A56, \dodoi{10.1051/0004-6361/202038054}

\bibitem[{{Newville} {et~al.}(2014){Newville}, {Stensitzki}, {Allen}, \&
  {Ingargiola}}]{lmfit2014}
{Newville}, M., {Stensitzki}, T., {Allen}, D.~B., \& {Ingargiola}, A. 2014,
  {LMFIT: Non-Linear Least-Square Minimization and Curve-Fitting for Python},
  0.8.0, Zenodo,  Zenodo, \dodoi{10.5281/zenodo.11813}

\bibitem[{{Notsu} {et~al.}(2013){Notsu}, {Shibayama}, {Maehara}, {Notsu},
  {Nagao}, {Honda}, {Ishii}, {Nogami}, \& {Shibata}}]{notsu2013}
{Notsu}, Y., {Shibayama}, T., {Maehara}, H., {et~al.} 2013, \apj, 771, 127,
  \dodoi{10.1088/0004-637X/771/2/127}

\bibitem[{{Okamoto} {et~al.}(2021){Okamoto}, {Notsu}, {Maehara}, {Namekata},
  {Honda}, {Ikuta}, {Nogami}, \& {Shibata}}]{okamoto2021}
{Okamoto}, S., {Notsu}, Y., {Maehara}, H., {et~al.} 2021, \apj, 906, 72,
  \dodoi{10.3847/1538-4357/abc8f5}

\bibitem[{{Ollivier} {et~al.}(2016){Ollivier}, {Deru}, {Chaintreuil},
  {Ferrigno}, {Baglin}, {Almenara}, {Auvergne}, {Barros}, {Baudin}, {Boumier},
  {Chabaud}, {Deeg}, {Guterman}, {Jorda}, {Samadi}, {Tuna}, \& {CoRot
  Team}}]{ollivier2016}
{Ollivier}, M., {Deru}, A., {Chaintreuil}, S., {et~al.} 2016, {II.2 Description
  of processes and corrections from observation to delivery} (EDP sciences),
  41, \dodoi{10.1051/978-2-7598-1876-1.c022}

\bibitem[{{Pickles}(1998)}]{pickles1998}
{Pickles}, A.~J. 1998, \pasp, 110, 863, \dodoi{10.1086/316197}

\bibitem[{{Rauer} {et~al.}(2016){Rauer}, {Aerts}, {Cabrera}, \& {PLATO
  Team}}]{rauer2016}
{Rauer}, H., {Aerts}, C., {Cabrera}, J., \& {PLATO Team}. 2016, Astronomische
  Nachrichten, 337, 961, \dodoi{10.1002/asna.201612408}

\bibitem[{{Rauer} {et~al.}(2014){Rauer}, {Catala}, {Aerts}, {Appourchaux},
  {Benz}, {Brandeker}, {Christensen-Dalsgaard}, {Deleuil}, {Gizon}, {Goupil},
  {G{\"u}del}, {Janot-Pacheco}, {Mas-Hesse}, {Pagano}, {Piotto}, {Pollacco},
  {Santos}, {Smith}, {Su{\'a}rez}, {Szab{\'o}}, {Udry}, {Adibekyan}, {Alibert},
  {Almenara}, {Amaro-Seoane}, {Eiff}, {Asplund}, {Antonello}, {Barnes},
  {Baudin}, {Belkacem}, {Bergemann}, {Bihain}, {Birch}, {Bonfils}, {Boisse},
  {Bonomo}, {Borsa}, {Brand{\~a}o}, {Brocato}, {Brun}, {Burleigh}, {Burston},
  {Cabrera}, {Cassisi}, {Chaplin}, {Charpinet}, {Chiappini}, {Church},
  {Csizmadia}, {Cunha}, {Damasso}, {Davies}, {Deeg}, {D{\'\i}az}, {Dreizler},
  {Dreyer}, {Eggenberger}, {Ehrenreich}, {Eigm{\"u}ller}, {Erikson}, {Farmer},
  {Feltzing}, {de Oliveira Fialho}, {Figueira}, {Forveille}, {Fridlund},
  {Garc{\'\i}a}, {Giommi}, {Giuffrida}, {Godolt}, {Gomes da Silva}, {Granzer},
  {Grenfell}, {Grotsch-Noels}, {G{\"u}nther}, {Haswell}, {Hatzes},
  {H{\'e}brard}, {Hekker}, {Helled}, {Heng}, {Jenkins}, {Johansen},
  {Khodachenko}, {Kislyakova}, {Kley}, {Kolb}, {Krivova}, {Kupka}, {Lammer},
  {Lanza}, {Lebreton}, {Magrin}, {Marcos-Arenal}, {Marrese}, {Marques},
  {Martins}, {Mathis}, {Mathur}, {Messina}, {Miglio}, {Montalban}, {Montalto},
  {Monteiro}, {Moradi}, {Moravveji}, {Mordasini}, {Morel}, {Mortier},
  {Nascimbeni}, {Nelson}, {Nielsen}, {Noack}, {Norton}, {Ofir}, {Oshagh},
  {Ouazzani}, {P{\'a}pics}, {Parro}, {Petit}, {Plez}, {Poretti}, {Quirrenbach},
  {Ragazzoni}, {Raimondo}, {Rainer}, {Reese}, {Redmer}, {Reffert},
  {Rojas-Ayala}, {Roxburgh}, {Salmon}, {Santerne}, {Schneider}, {Schou},
  {Schuh}, {Schunker}, {Silva-Valio}, {Silvotti}, {Skillen}, {Snellen}, {Sohl},
  {Sousa}, {Sozzetti}, {Stello}, {Strassmeier}, {{\v{S}}vanda}, {Szab{\'o}},
  {Tkachenko}, {Valencia}, {Van Grootel}, {Vauclair}, {Ventura}, {Wagner},
  {Walton}, {Weingrill}, {Werner}, {Wheatley}, \& {Zwintz}}]{rauer2014}
{Rauer}, H., {Catala}, C., {Aerts}, C., {et~al.} 2014, Experimental Astronomy,
  38, 249, \dodoi{10.1007/s10686-014-9383-4}

\bibitem[{{Ricker}(2014)}]{ricker2014}
{Ricker}, G.~R. 2014, \jaavso, 42, 234

\bibitem[{{Riello} {et~al.}(2021){Riello}, {De Angeli}, {Evans}, {Montegriffo},
  {Carrasco}, {Busso}, {Palaversa}, {Burgess}, {Diener}, {Davidson}, {Rowell},
  {Fabricius}, {Jordi}, {Bellazzini}, {Pancino}, {Harrison}, {Cacciari}, {van
  Leeuwen}, {Hambly}, {Hodgkin}, {Osborne}, {Altavilla}, {Barstow}, {Brown},
  {Castellani}, {Cowell}, {De Luise}, {Gilmore}, {Giuffrida}, {Hidalgo},
  {Holland}, {Marinoni}, {Pagani}, {Piersimoni}, {Pulone}, {Ragaini}, {Rainer},
  {Richards}, {Sanna}, {Walton}, {Weiler}, \& {Yoldas}}]{riello2021}
{Riello}, M., {De Angeli}, F., {Evans}, D.~W., {et~al.} 2021, \aap, 649, A3,
  \dodoi{10.1051/0004-6361/202039587}

\bibitem[{{Rodr{\'\i}guez Mart{\'\i}nez} {et~al.}(2020){Rodr{\'\i}guez
  Mart{\'\i}nez}, {Lopez}, {Shappee}, {Schmidt}, {Jayasinghe}, {Kochanek},
  {Auchettl}, \& {Holoien}}]{martinez2020}
{Rodr{\'\i}guez Mart{\'\i}nez}, R., {Lopez}, L.~A., {Shappee}, B.~J., {et~al.}
  2020, \apj, 892, 144, \dodoi{10.3847/1538-4357/ab793a}

\bibitem[{{Rouan} {et~al.}(1999){Rouan}, {Baglin}, {Barge}, {Copet}, {Deleuil},
  {Leger}, {Schneider}, {Toublanc}, \& {Vuillemin}}]{rouan1999}
{Rouan}, D., {Baglin}, A., {Barge}, P., {et~al.} 1999, Physics and Chemistry of
  the Earth C, 24, 567, \dodoi{10.1016/S1464-1917(99)00093-8}

\bibitem[{{Schaefer} {et~al.}(2000){Schaefer}, {King}, \&
  {Deliyannis}}]{schaefer2000}
{Schaefer}, B.~E., {King}, J.~R., \& {Deliyannis}, C.~P. 2000, \apj, 529, 1026,
  \dodoi{10.1086/308325}

\bibitem[{{Shibayama} {et~al.}(2013){Shibayama}, {Maehara}, {Notsu}, {Notsu},
  {Nagao}, {Honda}, {Ishii}, {Nogami}, \& {Shibata}}]{shibayama2013}
{Shibayama}, T., {Maehara}, H., {Notsu}, S., {et~al.} 2013, \apjs, 209, 5,
  \dodoi{10.1088/0067-0049/209/1/5}

\bibitem[{{Silverman}(1986)}]{silverman1986}
{Silverman}, B.~W. 1986, {Density estimation for statistics and data analysis}
  (Chapman and Hall)

\bibitem[{{Toriumi} {et~al.}(2017){Toriumi}, {Schrijver}, {Harra}, {Hudson}, \&
  {Nagashima}}]{toriumi2017}
{Toriumi}, S., {Schrijver}, C.~J., {Harra}, L.~K., {Hudson}, H., \&
  {Nagashima}, K. 2017, \apj, 834, 56, \dodoi{10.3847/1538-4357/834/1/56}

\bibitem[{{Venuti} {et~al.}(2015){Venuti}, {Bouvier}, {Irwin}, {Stauffer},
  {Hillenbrand}, {Rebull}, {Cody}, {Alencar}, {Micela}, {Flaccomio}, \&
  {Peres}}]{venuti2015}
{Venuti}, L., {Bouvier}, J., {Irwin}, J., {et~al.} 2015, \aap, 581, A66,
  \dodoi{10.1051/0004-6361/201526164}

\bibitem[{Virtanen {et~al.}(2020)Virtanen, Gommers, Oliphant, Haberland, Reddy,
  Cournapeau, Burovski, Peterson, Weckesser, Bright, {van der Walt}, Brett,
  Wilson, Millman, Mayorov, Nelson, Jones, Kern, Larson, Carey, Polat, Feng,
  Moore, {VanderPlas}, Laxalde, Perktold, Cimrman, Henriksen, Quintero, Harris,
  Archibald, Ribeiro, Pedregosa, {van Mulbregt}, \& {SciPy 1.0
  Contributors}}]{scipy2020}
Virtanen, P., Gommers, R., Oliphant, T.~E., {et~al.} 2020, Nature Methods, 17,
  261, \dodoi{10.1038/s41592-019-0686-2}

\bibitem[{{Woods} {et~al.}(2006){Woods}, {Kopp}, \& {Chamberlin}}]{woods2006}
{Woods}, T.~N., {Kopp}, G., \& {Chamberlin}, P.~C. 2006, Journal of Geophysical
  Research (Space Physics), 111, A10S14, \dodoi{10.1029/2005JA011507}

\end{thebibliography}
\bibliographystyle{aasjournal}



\end{document}